\documentclass[acmsmall, screen]{acmart}

\AtBeginDocument{%
  }

\usepackage{scalerel}
\usepackage{multirow}
\usepackage{enumitem}
\usepackage{listings}
\usepackage{xparse}
\usepackage[utf8]{inputenc}
\usepackage{xcolor}
\usepackage{colortbl}
\usepackage{makecell}
\usepackage[nameinlink]{cleveref}
\usepackage{array}
\usepackage{wrapfig}
\usepackage{lipsum}
\usepackage{comment}
\usepackage[normalem]{ulem}
\usepackage{hyperref}
\usepackage{hyperxmp}

\newcommand{\todo}[1]{{\color{red}{\textbf{[TODO]:} #1}}}

\setcopyright{acmlicensed}
\copyrightyear{2018}
\acmYear{2018}
\acmDOI{XXXXXXX.XXXXXXX}

\acmBooktitle{Woodstock '18: ACM Symposium on Neural Gaze Detection,
June 03--05, 2018, Woodstock, NY}
\acmISBN{978-1-4503-XXXX-X/18/06}

\begin{document}

\title[WeAudit]{WeAudit: Scaffolding User Auditors and AI Practitioners in Auditing Generative AI}

\author{Wesley Hanwen Deng}
\orcid{0000-0003-3375-5285}
\email{hanwend@cs.cmu.edu}
\affiliation{%
  \institution{Carnegie Mellon University}
  \city{Pittsburgh}
  \state{Pennsylvania}
  \country{USA}}

\author{Wang Claire}
\orcid{0009-0003-3562-055X}
\email{clairewang@berkeley.edu}
\affiliation{
  \institution{University of California, Berkeley}
  \city{Berkeley}
  \state{California}
  \country{USA}}

\author{Howard Ziyu Han}
\orcid{0009-0008-5556-7297}
\email{ziyuh@cs.cmu.edu}
\affiliation{%
  \institution{Carnegie Mellon University}
  \city{Pittsburgh}
  \state{Pennsylvania}
  \country{USA}}

\author{Jason I. Hong}
\orcid{0000-0002-9856-9654}
\authornote{These authors contributed equally to this work.}
\email{jasonh@cs.cmu.edu}
\affiliation{%
  \institution{Carnegie Mellon University}
  \city{Pittsburgh}
  \state{Pennsylvania}
  \country{USA}}

\author{Kenneth Holstein}
\orcid{0000-0001-6730-922X}
\authornotemark[1]
\email{kjholste@cs.cmu.edu}
\affiliation{%
  \institution{Carnegie Mellon University}
  \streetaddress{5000 Forbes Ave}
  \city{Pittsburgh}
  \state{PA}
  \postcode{15213}
  \country{USA}
}

\author{Motahhare Eslami}
\orcid{0000-0002-1499-3045}
\authornotemark[1]
\email{meslami@cs.cmu.edu}
\affiliation{%
  \institution{Carnegie Mellon University}
  \streetaddress{5000 Forbes Ave}
  \city{Pittsburgh}
  \state{PA}
  \postcode{15213}
  \country{USA}
}

\renewcommand{\shortauthors}{Deng et al.}

\begin{abstract}
There has been growing interest from both practitioners and researchers in engaging end users in AI auditing, to draw upon users’ unique knowledge and lived experiences. However, we know little about how to effectively scaffold end users in auditing in ways that can generate actionable insights for AI practitioners. Through formative studies with both users and AI practitioners, we first identified a set of design goals to support user-engaged AI auditing. We then developed \textit{WeAudit}, a workflow and system that supports end users in auditing AI both individually and collectively. We evaluated \textit{WeAudit} through a three-week user study with user auditors and interviews with industry Generative AI practitioners. Our findings offer insights into how \textit{WeAudit} supports users in noticing and reflecting upon potential AI harms and in articulating their findings in ways that industry practitioners can act upon. Based on our observations and feedback from both users and practitioners, we identify several opportunities to better support user engagement in AI auditing processes. We discuss implications for future research to support effective and responsible user engagement in AI auditing.
\end{abstract}

\begin{teaserfigure}
  \includegraphics[width=\textwidth]{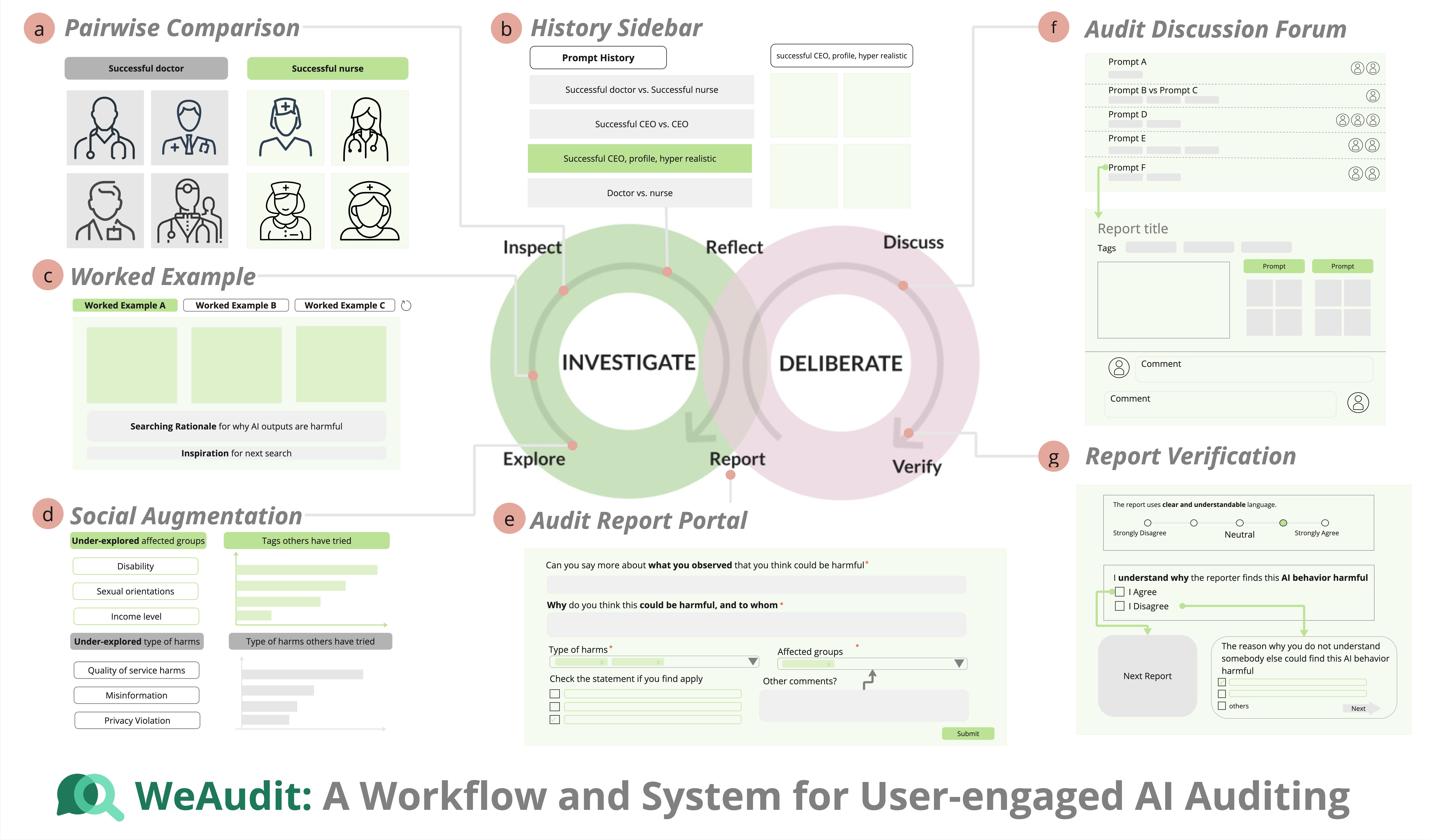}
  \caption{\textit{WeAudit} supports end users in iteratively investigating, deliberating on, and reporting perceived harms and biases in generative AI (GenAI) systems through support for comparing and making sense of different AI generated outputs (a, b); support for reflection around their own and others' auditing findings (b, c, d); and mechanisms for structured reporting, discussion, and verification of audit findings (e, f, g). Through user studies with both end users and industry GenAI practitioners, our work highlights design considerations, recommendations, and opportunities to support effective and responsible user engagement in AI auditing. \looseness=-1}
  \Description{TBA}
  \label{fig:teaser}
\end{teaserfigure}

\maketitle

\section{INTRODUCTION}

End users often surface harmful behaviors in Artificial Intelligence (AI) systems that are otherwise overlooked by AI practitioners \cite{shen2021everyday, devos2022toward, lam2022enduser, mack2024they, shelby2024generative, li2023participation, deng2023understanding, solyst2023potential, kieslich2024anticipating, mun2024particip}. Past research shows that AI users can detect and raise awareness of biased and harmful AI behaviors by leveraging their lived experiences, building on one anothers' findings, and generating and testing hypotheses about AI harms \cite{shen2021everyday, devos2022toward}. There has been rapidly growing interest among both researchers and practitioners in engaging end users in auditing AI products and services: one line of recent research has explored systems and processes to enable user engagement in auditing  \cite{attenberg2015beat, cabrera2021discovering, kiela2021dynabench, nushi2018towards, ochigame2021search, lam2022enduser, maldaner2024mirage, claire2024designing}, while another has focused on understanding industry AI practitioners’ needs and challenges around effective user engagement \cite{deng2023understanding, ojewale2024towards, wang2023designing, madaio2024tinker, madaio2021assessing, deng2024responsible, deng2024supporting}. However, a critical gap remains in connecting the needs, challenges, and perspectives of these two groups of stakeholders. \textbf{How might we develop tools and processes to effectively scaffold end-user engagement in AI audits, while ensuring their findings are useful and actionable for AI practitioners?} \looseness=-1

To bridge this gap, in this work, we introduce \textit{WeAudit}, a workflow and platform, designed with and for both end users and industry AI practitioners, to support meaningful user engagement in AI audit. In line with prior research, throughout this paper we use the term ``user-engaged AI auditing’’ (or simply ``user auditing’’) to refer to processes where users are actively engaged in AI auditing processes that may also include auditing efforts from AI practitioners or other technical experts~\cite{deng2023understanding, birhane2024ai, lam2022enduser, bandy2021problematic}. We use the term ``user auditors'' to refer to AI users who are engaged in auditing processes. We focus in particular on Generative AI (GenAI) systems, which have garnered significant attention for user auditing due to their widespread availability and their ability to flexibly generate content over a vast input-output space \cite{weidinger2022taxonomy, weidinger2024star}. These characteristics of GenAI introduce an equally vast space of risks and use cases, prompting calls across government, academia, and industry to engage diverse expertise in AI auditing~\cite{AIRiskManageF, House_2023, mrbullwinkle_2023, openai2023gpt4card, openAI, ChatGPT_Feedback, anthropic2023claude, bogen2024sociotechnical}. 

To understand both end users’ and AI practitioners’ needs for support in user auditing, we first conducted a formative study with 11 end users (\textit{U01-U11}), and seven industry AI practitioners with prior experience engaging end users in AI auditing (\textit{P01-P07}). These formative studies led to the development of a set of design goals, and a corresponding \textit{WeAudit} workflow that draws upon ideas from CSCW research on crowdsourcing and sensemaking \cite{kittur2013future, cabrera2022did, pirolli1999information, chan2018solvent}. We then instantiated the \textit{WeAudit} workflow through a system that supports and scaffolds users in investigating, deliberating, and reporting perceived harms and biases in text-to-image (T2I) AI systems. To understand the strengths and limitations of \textit{WeAudit} in supporting user auditing, we first conducted a three-week user study with 45 users (\textit{F01-F45}) to audit Stable Diffusion, an open source T2I system, and reflect on their experience using the tool. We then conducted semi-structured interviews with 10 industry GenAI practitioners (a subset of \textit{P01-P14}) to evaluate the design of \textit{WeAudit}, grounded in data and user audit reports from our user study. 

\begin{figure*}[t]
  \centering
  \includegraphics[width=1\linewidth]{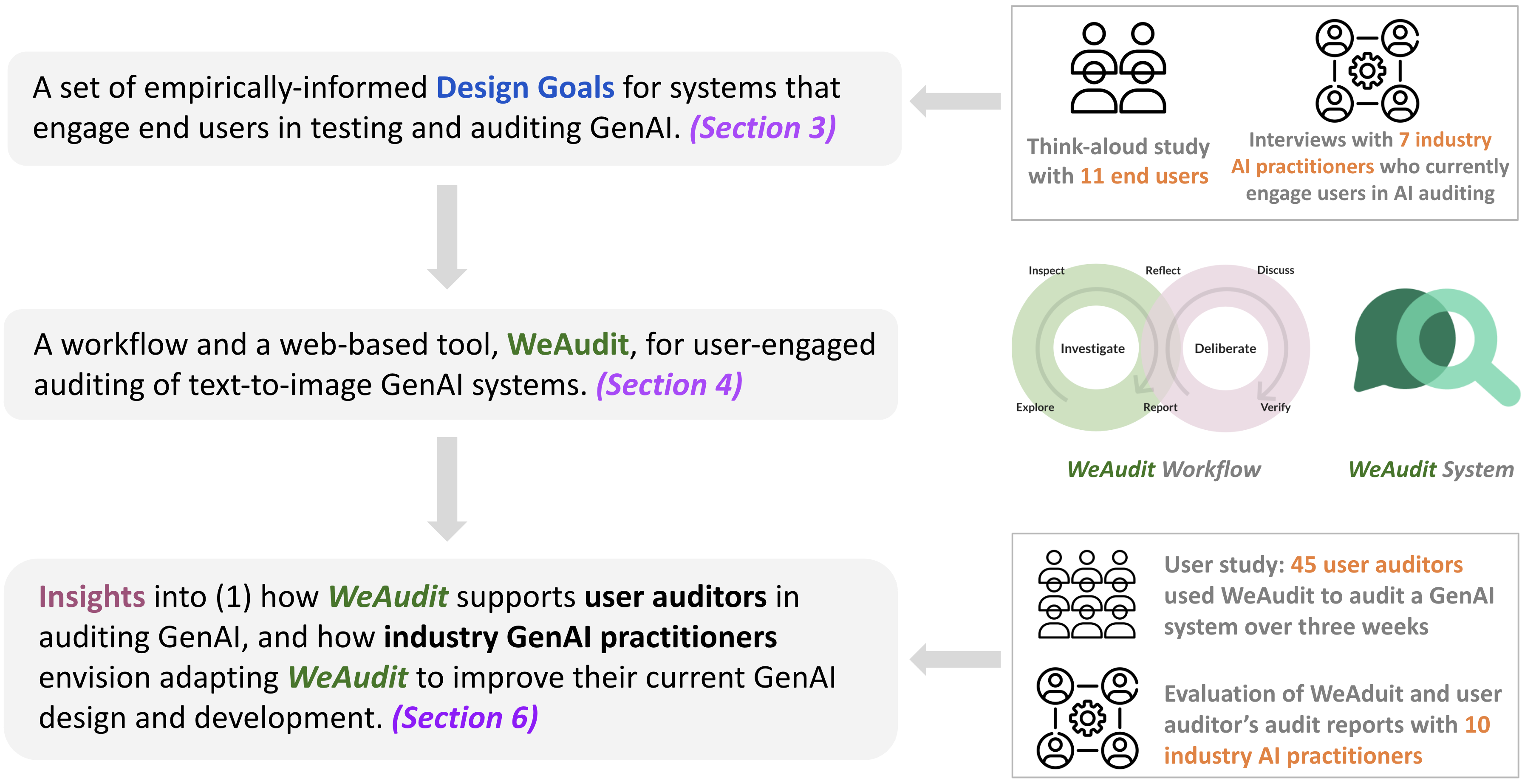}
  \caption{
   An overview of the methods and contribution of our work.
}
  \Description{TBA}
  \label{fig:methods}
\end{figure*}

Through these studies, we provide design insights into how \textit{WeAudit} (1) augments user auditors’ abilities to detect and reflect on AI biases and harms that might otherwise be overlooked, (2) enhances the depth and breadth of user audits, (3) supports the creation of user audit reports and discussions that AI practitioners find actionable for mitigating AI harms, and (4) helps users increase their awareness and understanding of AI harms. Practitioners found the audit outcomes insightful and could envision incorporating \textit{WeAudit} into their existing GenAI evaluation and development workflows. Our study also revealed several opportunities to improve \textit{WeAudit} and future user-engaged AI audit processes. We discuss implications of our findings for future research and practice on user engagement in AI auditing. Overall, this work makes the following contributions (Figure \ref{fig:methods}): \looseness=-1

\begin{itemize}
    \item We identify a set of \textbf{design goals} for systems and processes to support user-engaged AI auditing, informed by formative studies with both end users and industry GenAI practitioners.
    \item We present \textit{\textbf{WeAudit}}, a general workflow and interactive system for user-engaged AI auditing, based on our design goals.
    \item We share empirical \textbf{insights} into how \textit{WeAudit} supports users in auditing GenAI and how industry GenAI practitioners envision adapting \textit{WeAudit} to support their current workflows. 
    \item We discuss the \textbf{broader implications} of our findings for HCI and AI researchers and practitioners, organizations, and policymakers in sustaining user engagement in AI auditing and red teaming in a responsible and ethical manner.
\end{itemize}

\section{RELATED WORK} \label{related work}

\subsection{Engaging End Users in Auditing AI Systems for Harmful Behaviors} \label{rw: user audit}

In recent years, AI audits have gained prominence as a method for uncovering biased, discriminatory, or otherwise harmful behaviors in algorithmic systems \cite{noble2018algorithms, asplund2020auditing, sweeney2013discrimination, prates2020assessing, buolamwini2018gender, hannak2014measuring, holstein2019co, sandvig2014auditing, metaxa2021auditing, birhane2024ai}. At a high level, AI auditing refers to a process of repeatedly testing an algorithm with inputs and observing the corresponding outputs, in order to understand its behavior and potential external impacts ~\cite{metaxa2021auditing, birhane2024ai}. \looseness=-1

While most AI audits are still conducted by experts such as researchers \cite{buolamwini2018gender, Buolamwini2019hearing, cramer2018assessing, eslami2017biased, eslami2019opacity, raji2020closing} and AI practitioners \cite{sweeney2013discrimination, sandvig2014auditing, metaxa2021auditing, chowdhury2021introducing}, end users can often identify and raise awareness of harmful AI behaviors in systems that impact their lives, complementing expert-led AI audit efforts \cite{shen2021everyday, devos2022toward}.
For example, end-users often discover harmful biases in Text to Image (T2I) generative AI systems that expert auditors fail to detect \cite{mack2024they, zhang2024partiality, shelby2024generative, mim2024between}. When Mack et al. engaged people with disabilities to review and reflect on images generated by T2I systems they uncovered several societal stereotypes, such as ``perpetuating broader narratives in society around disabled people as primarily using wheelchairs, being sad and lonely, incapable, and inactive’’ \cite{mack2024they}. 

Recognizing the power of users in AI auditing, researchers in CSCW, HCI, and AI have begun exploring tools and processes to support more user-engaged approaches to AI auditing, where users actively participate in the auditing process alongside efforts from AI practitioners or other technical experts \cite{shen2021everyday, devos2022toward, lam2022enduser, attenberg2015beat, cabrera2021discovering, kiela2021dynabench, nushi2018towards, kieslich2024anticipating}. For example, DeVrio et al. conducted think-alouds, diary studies, and workshops to investigate how users, both individually and collectively, are able to surface harmful algorithmic behaviors that formal or expert-led audits may miss \cite{devos2022toward}. Lam et al. developed an “End-user Audits” framework, instantiated in a system called IndieLabel, to engage non-technical users in auditing sentiment analysis classification AI models \cite{lam2022enduser}. 

In parallel, several major technology companies have begun experimenting with user-engaged auditing approaches to identify harmful and biased outputs in their AI products and services \cite{HuggingFace, ChatGPT, AItest, chowdhury2021introducing, kiela2021dynabench}. In 2021, Twitter launched its first “algorithmic bias bounty” challenge, inviting users to identify harmful biases in its image-cropping algorithm \cite{chowdhury2021introducing}. Following this, OpenAI introduced a “Feedback Contest” encouraging users to provide feedback on problematic model outputs during their interactions with ChatGPT. Many corporations with generative AI products have also turned to “red-teaming” events, where  users rigorously test models to identify vulnerabilities and harmful biases (such as the Red Team village at DEFCON, a hacking conference) \cite{DEFCON_AI_Village}.  In response to calls from government agencies like the Executive Order on Safe, Secure, and Trustworthy Development and Use of AI \cite{House_2023}, many civil society organizations has called for technology companies to engage external domain experts, including end users, in auditing their generative AI products \cite{groves2023going, bogen2024sociotechnical, park2024stakeholder, feffer2024red}. \looseness=-1


A small but growing body of research has begun exploring the needs and challenges faced by AI practitioners in conducting user-engaged AI audits. For example, through interviews with 35 AI audit practitioners, Ojewale et al. identified the need for tools to support AI practitioners in engaging those directly impacted by AI deployments, offering insights into how to better design participatory AI tools and processes with accountability \cite{ojewale2024towards}. 
In a study with industry AI practitioners who are engaging users in auditing their AI systems, Deng et al. found that practitioners often repurpose crowdsourcing platforms for user audits but struggle to scaffold users toward productive auditing strategies and to derive actionable insights from users’ audit outcomes \cite{deng2023understanding}. Many other studies on responsible AI have highlighted the lack of efficient tooling and collaboration processes for practitioners to engage end users in testing and auditing AI products and services \cite{madaio2020co, madaio2024tinker, wang2023designing, deng2022exploring, deng2023supporting}.

Despite these two growing lines of research—supporting user audits through tools and processes, and understanding the needs and challenges of industry AI practitioners in user audits—gaps remain in understanding the design goals and in creating tools that \textbf{support both AI users and AI practitioners} in the user-engaged AI audit process. Our work extends this prior research by designing, prototyping, and evaluating a user-engaged AI audit workflow and tool that \textbf{effectively scaffolds end-user engagement in AI audits, while ensuring their findings are useful and actionable for AI practitioners}. Through evaluating \textit{WeAudit} with both AI users and practitioners, our work provides insights for future practitioners and researchers to better engage users in the AI auditing process. \looseness=-1

\subsection{Scaffolding Mechanisms for Crowdsourcing} \label{rw: scaffolding}

CSCW and HCI research has a long tradition of designing scaffolding mechanisms—including various types of instructions, interfaces, and incentives—to support the crowdsourcing and sensemaking processes for diverse tasks \cite{vaughan2017making, bigham2015human, kittur2013future, Sindlinger2010CrowdsourcingWT}. This line of research explores methods to help crowd workers perform creative and open-ended tasks beyond simple voting or preference selection \cite{chilton2013cascade, dow2012shepherding, chung2019efficient, bigham2015human}. For instance, Dow et al. and Chung et al. demonstrated how structured feedback can guide workers in complex tasks, improving the quality of creative contributions \cite{dow2012shepherding, chung2019efficient}. Chilton et al. developed a workflow to assist crowd workers in subjective categorization tasks through iterative feedback, highlighting crowds’ potential in complex sensemaking \cite{chilton2013cascade}.

Further studies examine how presenting examples can scaffold crowd work, especially in unfamiliar tasks requiring iterative work and hypothesis generation \cite{morris2007searchtogether, siangliulue2015toward, kohn2011collaborative, freyne2007collecting, amershi2008cosearch, wu2019errudite}. Social augmentation, such as showing others’ work, has been shown to improve task completion, as demonstrated by Morris’s SearchTogether tool, which enables collaborative search tasks \cite{morris2007searchtogether}. Other research indicates that timely exposure to diverse sets of examples can stimulate new directions of thought for crowd workers \cite{kohn2011collaborative}, potentially enhancing the quantity and quality of ideas generated \cite{siangliulue2015toward}. However, this research consistently highlights that the effectiveness of examples varies based on task content and presentation style \cite{kohn2011collaborative}. Fianlly, research on reliability in crowdsourced work developed methods for verifying others’ contributions, such as Bernstein et al.'s ``Find, Fix, Verify'' workflow, which enables structured peer review \cite{bernstein2010soylent}.

Drawing from prior crowdsourcing and other CSCW research, our work extends this line of research by designing, prototyping, and evaluating scaffolding mechanisms \textbf{specifically for AI users engaged in auditing potentially harmful AI outputs, both individually and collectively}. \looseness=-1

\section{Formative Study} \label{formative study}

In this section, we first describe the procedure of our formative study. We then present six Design Goals for processes and tools to support user-engaged AI auditing, informed by our formative study with AI users and practitioners.

\subsection{Method} \label{formative method}

\subsubsection{Think-aloud study with end users}

To understand the challenges end users encounter when participating in GenAI audits, and to identify opportunities to better support them  beyond existing approaches (Section \ref{related work}), we first conducted think-alouds with 11 end users (U01 - U11). For these formative studies, we aimed to recruit participants spanning diverse demographics and levels of technical literacy. To this end, we recruited participants through Craigslist and Nextdoor. 10 of our 11 participants had not previously used any T2I systems; participants ranged from 18 years old to 80 years old, with 47.45 average. 2 participants has high school degrees, 2 Associate degrees, 5 bachelor degrees, 1 master degree, and 1 Ph.D. We include detailed participants descriptions in Table \ref{tab:formative user participants} under the Appendix \ref{Appendix: demo}.

In each study, we first shared the goal of the study and then onboarded participants by providing examples of harmful text-to-image results along with reasons why these results could be considered harmful towards certain social groups. These examples and rationales were drawn from prior expert-led audits on T2I systems, covering various common types of T2I harms, such as representational harms and privacy violations. \cite{bianchi2023easily, luccioni2023stable, naik2023social}. Participants then engaged in an initial 20-30 minute model testing phase where they were asked to test a text-to-image model (Stable Diffusion) for harmful biases, using an off-the-shelf interface provided by Dream Studio \footnote{https://beta.dreamstudio.ai/generate}. This simple interface allowed users to enter a prompt into a text box and generate 8 images. We encouraged participants to think aloud while brainstorming prompts and inspecting model outputs, and to explain whether and why they found the results potentially harmful. 

After this initial phase, for the next 20 minutes participants continued their testing, but now with access to a set of three low-fidelity prototypes intended to support their process (see Figure \ref{fig:formative prototypes}). The first prototype was a pairwise comparison feature, simulated by placing two browser windows side-by-side. We hypothesized that viewing prompt outputs side-by-side would help participants, especially those unfamiliar with T2I systems, understand how prompt changes affect the output. The second prototype was a sidebar displaying a sample user audit report in text format. This report featured an election-themed prompt and image pair, with a brief sentence explaining why the AI-generated image could be harmful. We hypothesized that this example could inspire participants to search for similar cases and better articulate the harms they identified. The third prototype was a T2I harm taxonomy, mapping Shelby et al.’s sociotechnical AI harm taxonomy \cite{shelby2023sociotechnical} to concrete examples of harmful T2I behaviors, based on examples from \cite{luccioni2023stable}. We hypothesized that this taxonomy would help users brainstorm a broader range of potential harms to various communities and use terminology to describe these harms effectively. \looseness=-1

 \begin{figure*}[t]
  \centering
  \includegraphics[width=1\linewidth]{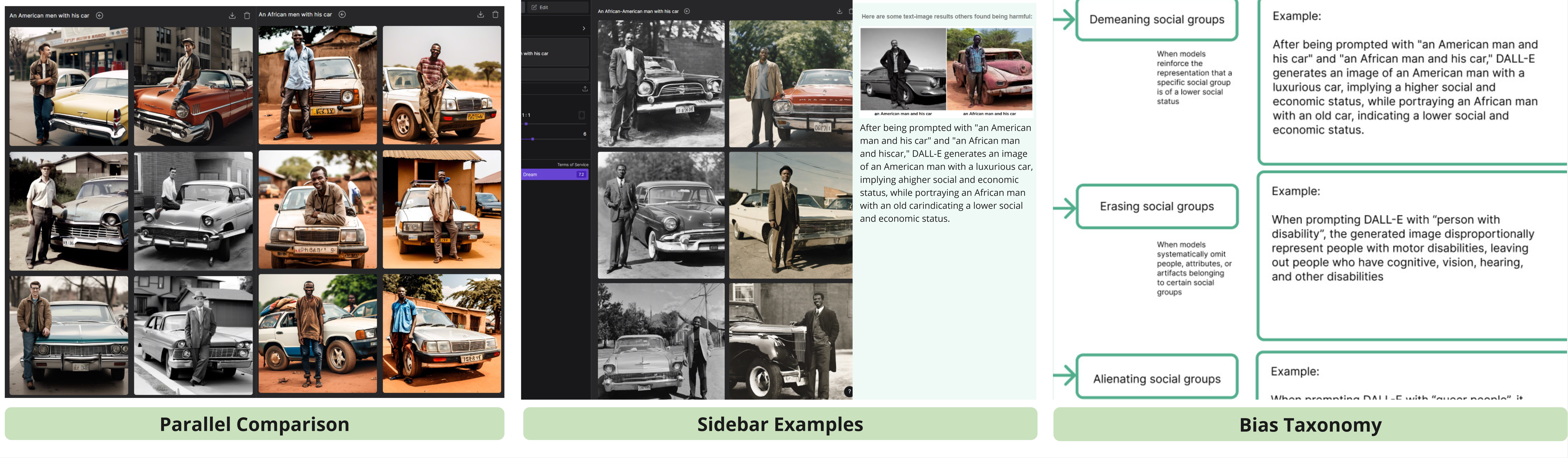}
  \caption{
   Screenshot of three low-fidelity prototypes we used in our formative study.
}
  \Description{TBA}
  \label{fig:formative prototypes}
\end{figure*}

\subsubsection{Semi-structured Interview with Industry GenAI Practitioners}

To better understand industry GenAI developers’ needs for support in effectively engaging end-users in AI auditing, we conducted semi-structured interviews with 7 industry AI developers (P01 - P07). To recruit participants, we adopted a purposive sampling approach \cite{campbell2020purposive}, with the aim of recruiting industry practitioners who had already attempted to engage end users in evaluating and auditing their GenAI products~\footnote{Practitioners frame this activity differently depending on their organization, in some cases using terminology like ``red-teaming''.}. 
We recruited participants through social media (e.g., LinkedIn and X), and via direct contacts at large technology companies. Overall, industry practitioners come from diverse backgrounds, working in technical, user-facing, and managerial roles on various types of AI products powered by Large Language Models (LLMs) or other types of generative AI. Notably, all of them are currently using crowdsourcing platforms to engage individuals outside their AI teams in auditing or red-teaming their AI products and services. Table \ref{tab:industry participants} in the Appendix \ref{Appendix: demo} provides an overview of participants’ backgrounds~\footnote{Following prior work studying responsible AI practices in industry \cite{madaio2020co, rakova2021responsible}, to protect participants' confidentiality, we omit potentially identifying demographic details such as gender and age, and abstract our descriptions of participants' companies and roles.}.  \looseness=-1

In each interview session, we began by asking participants to describe their current motivations and practices for engaging end users. We then delved deeper into the challenges and pitfalls practitioners had encountered when attempting to engage users in the auditing process. The interview concluded with a discussion of features that practitioners would ideally want in a user-engaged AI auditing workflow.

\subsection{Design Goals} \label{design goal}

Through our formative study with end users (U01 - U11, referred to as ``users’’ throughout this section) and industry AI practitioners (P01 - P07, referred to as ``practitioners’’), we synthesized the following six design goals for processes and tools to support user-engaged AI auditing. 

\hfill

\noindent\textit{\textbf{DG1: Facilitate effective generation and validation of hypotheses about harmful AI behaviors.}} 
 We observed that users often iteratively explored small variations on their prompts. For example, after reviewing the results of “An African man and his car,” U01 hypothesized that there might be biases in how objects are presented in connection to people, so they continued testing variations of this prompt, such as “An African man and his car in the U.S.” and eventually “An African man and his car in Africa.” All users reported finding side-by-side comparisons helpful in supporting such exploration. U05 and U10 were interested in further support for making sense of AI-generated images across \textit{many} different prompts, to validate hypotheses or generate new ones.

Similarly, all practitioners expressed desires for interfaces that could support users in more effectively generating and testing hypotheses about potential harmful behaviors exhibited in generative AI. For example, similar to our observations in the think-aloud study with users, P04 observed in past focus groups with users that when they encountered specific harmful outputs, they often formed hypotheses about broader biases that might be present, and then wanted to try out a similar input to test these hypotheses.

\hfill

\noindent\textit{\textbf{DG2: Support and encourage user auditors to incorporate their lived experiences and identities.}} 
We observed that users often found harmful AI outputs when they began reflecting on their own lived experiences and identities. For example, U02 found more to comment on when testing prompts related to places they had lived, and U06 noted it was easier to spot biases in images related to familiar occupations and activities. Similarly, U07 generated insightful comments by incorporating their hobbies into prompts. However, most pilot study users (with the exception of U06) did not naturally draw upon their personal experiences and identities before we explicitly instructed them to do so, suggesting opportunities for interventions to encourage user auditors to do so. \looseness=-1

In line with these observations, all practitioners emphasized the importance of incorporating end users’ personal experiences and identities into the auditing process. P01, P02, P04, P06, and P07 all stated that leveraging users’ diverse lived experiences to cover the blind spots of their companies’ AI teams was a \textit{“key motivation to engage end users”} (P04) but one that their teams currently struggled with. For instance, P06 shared that their team had designated a meeting to strategize ways to better incorporate users’ identities into the auditing process, such as providing instructions for users to specifically consider real-life scenarios where they would use the company's AI products.

\hfill

\noindent\textit{\textbf{DG3: Provide appropriate scaffolding to support user auditors in creatively and efficiently auditing AI systems, without overly priming them towards narrow auditing directions.}}
While the examples we provided to users helped them get started with the auditing process, we observed that some participants over-relied on the provided examples. For example, during the think-aloud session, U03 and U08  only checked the first few examples we provided in the T2I harm taxonomy and example audit report, and the prompts they explored were largely based on those examples. Towards the end of the think-aloud, U06 shared that they felt the T2I harm taxonomy we shared with them could potentially \textit{``limit the kinds of responses… [as it] might shut off some people [who] are going to come up with on their own [prompts]… if they feel like they're being shoved into a certain category in their response, it could stifle their creativity.’’}

Practitioners' comments resonated with these observations. For example, P05 shared their observations that providing a small set of static examples often led to users conducting audits that were similar to those examples. P05 further commented, \textit{``It is important to provide users with sufficient scaffolding, especially if they haven’t done this type of activity before… but we also want to make sure that they are thinking outside the box and find creative ways to break the model.’’} Similarly, P01 suggested that the onboarding and scaffolding materials in the user audit process should not \textit{“overly prime users towards the assumptions held by the development team… good [user auditing] should be a creative endeavor and independent from what developers already know.’’}

\hfill

\noindent\textit{\textbf{DG4: Structure the audit report process to allow users to clearly communicate observed harms in a way that helps AI developers extract actionable insights.}}
During the think-aloud study, we observed all users struggled to clearly articulate which part of the AI-generated content was harmful and why, even when they sensed that something was off. For example, U10 compared the prompts "Rich People" vs. "Poor People". Upon observing that the generated images of ``poor people'' all shared the same style of dress, U10 was unsure how to describe why they found this problematic. Multiple users (U04, U05, U10, U11) explicitly asked for us to tell them more about what information might be useful for us, suggesting opportunities to elicit their perceptions in more structured ways. \looseness=-1

All practitioners mentioned that scaffolding users in crafting reports is essential for them to translate users’ feedback into concrete improvements. In particular, practitioners suggested that they would like to know \textit{what users actually observed} in the AI output that they thought could be harmful, \textit{why} they believed this harmful, and \textit{to whom}. To aid in interpreting users' reports, practitioners were also interested in knowing additional information about the user auditor who wrote a given report. For example, all practitioners suggested that audit report could solicit if user auditors come from a specific marginalized background could help them prioritize which audit outcomes to incorporate. \looseness=-1

\hfill

\noindent\textit{\textbf{DG5: Enable users to discuss and deliberate on AI harms and mitigation strategies.}} 
Users expressed desire to discuss their findings with others to better explore potential harms or confirm their perceptions of harms toward specific communities that they were less personally familiar with. For example, when U11 observed the AI-generated output from the prompt “fashion,” they wished to discuss it with others, saying, \textit{“some people might say the result is indeed about fashion, some might say the results are all white women, and some might say they are unhealthy-looking women… it might be interesting to hear different comments like that.”} Furthermore, towards the end of the think-aloud session, some users asked if they could review audit reports from others and learn from their auditing strategies.

Many practitioners (5/7) also expressed the importance of facilitating discussions and deliberations among users. For example, P04 mentioned that they conduct focus groups with groups of 4-6 users when auditing their AI products because “\textit{users can often inspire each other to find something new through going back and forth… some users will recall issues they had when someone else mentioned a situation they experienced}.” P01 shared that their team often found it challenging to aggregate and appropriately resolve disagreements in users’ Likert scale ratings regarding perceived biases and harms from chatbots toward certain communities. As a result, P01’s team was exploring ways to connect end users who submitted conflicting audit reports, allowing them to directly discuss their disagreements so the team could better understand the rationales behind their reports.

\hfill

\noindent\textit{\textbf{DG6}:\textbf{ Support careful verification of audit reports without disempowering minority voices.}} Finally, all practitioners emphasized the importance of verifying users' audit reports to separate signal from noise when interpreting and prioritizing among a large amount of user feedback. In line with prior work \cite{deng2023understanding,gordon2021disagreement}, practitioners shared that, unlike aggregating the results of traditional AI labeling work, verifying audit outcomes is challenging because: (1) developer teams do not have ground truth, so the verification process must rely on other auditors; and (2) outliers in the data might be important signals, especially if they come from marginalized voices. For example, P05, who had been re-purposing crowdsourcing tools to support user-engaged auditing of generative AI applications, noted, "\textit{quality control in this kind of [AI auditing] task is even harder because […] we don’t have a ground truth either, and everything [crowd workers] surface is new to us.}" P02 shared that a significant challenge their team currently faces is determining whether the representational harms identified by users resonate with others. To this end, all practitioners agreed that verifying audit outcomes is challenging yet critical for the design of a successful user-engaged AI auditing process. \looseness=-1

\section{WeAudit} \label{WeAudit}

Based on our Design Goals, we first synthesized the \textit{WeAudit} workflow, which includes two iterative loops and six sub-activities that is generalizable for scaffolding user engagement in AI audit. We then instantiated the \textit{WeAudit} workflow through a web-based prototype system, with each core feature mapped to the Design Goals and the high-level \textit{WeAudit} workflow. We describe the \textit{WeAudit} workflow and system in the following sections.

\subsection{The \textit{WeAudit} Workflow} \label{WeAudit workflow}

\begin{wrapfigure}{l}{0.5\textwidth}
  \centering
  \includegraphics[width=0.5\textwidth]{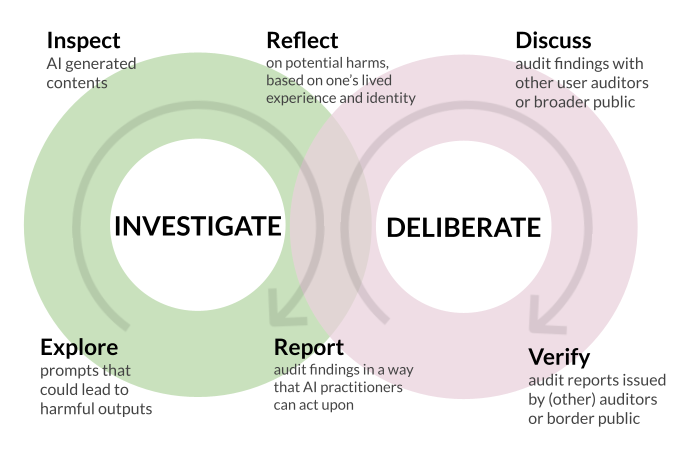}
 \caption{The \textit{WeAudit} workflow that contains two intersecting, iterative loops.}
  \Description{TBA}
  \label{fig: weaudit workflow}
\end{wrapfigure}

As shown in Figure \ref{fig: weaudit workflow}, the \textit{WeAudit} workflow is organized into two intersecting, iterative loops: \textbf{\textit{Investigate}} and \textbf{\textit{Deliberate}}. The overall loop structure of this workflow is informed by existing models of information foraging and sensemaking~\cite{pirolli1999information}, and these loops are further broken down into six sub-activities: \textit{Explore}, \textit{Inspect}, \textit{Reflect}, \textit{Report}, \textit{Discuss}, and \textit{Verify}. In particular, the idea of the loop draws inspiration from Pirolli and Card’s model of how intelligence analysts gather evidence, generate hypotheses, re-evaluate insights, and seek additional evidence  in an iterative manner\cite{pirolli1999information}. We conceptualize AI auditing as involving a similarly structured process in which auditors systematically inspect, organize, refine, and validate their insights \cite{pirolli1999information, cabrera2022did}.

In the \textit{Investigate} loop, user auditors may begin investigating AI harms by \textit{exploring} prompts that could lead to harmful outputs (\textbf{DG1, DG3}). Auditors \textit{inspect} the resulting AI-generated outputs (\textbf{DG1}), and \textit{reflect} on potential biases or harmful impacts, drawing upon their unique knowledge and experiences (\textbf{DG2}). Auditors can then \textit{report} their findings, documenting potential harms they identified (\textbf{DG4}). In the \textit{Deliberate} loop, user auditors can \textit{discuss} their findings with other auditors or the broader public (\textbf{DG5}). This collaboration allows for a more holistic understanding of AI-related harms, leveraging user auditors' diverse perspectives. Additionally, reported findings can be \textit{verified} by other auditors or the public (\textbf{DG6}), to provide a sense of whether a given observation is perceived as potential harmful by multiple people.

In the workflow, we use the circular arrows to highlight that the entire workflow is iterative, following prior sensemaking research \cite{pirolli1999information}. For example, new insights that arise during discussion or verification may inspire user auditors to conduct new investigations. Moreover, this iterative workflow engages user auditors in fluidly moving back-and-forth between individual and collective activities, where individual efforts contribute to the collective understanding of AI behaviors and vice versa. Throughout the entire process, user auditors are encouraged to reflect on potential harms based on their own lived experiences and identities. Inspired by past work on collaborative sensemaking \cite{kriplean2012supporting}, the ``reflect'' and \textit{``report''} activities function as bridges between \textit{Investigate} and \textit{Deliberate} loops. User auditors' reflections and reports on potential harms provide material for collective discussion and deliberation. In turn, discussions around these reflections and reports may inspire further investigations. Given this bridging role, we represent ``reflect'' and ``report'' at the intersection of these loops. \looseness=-1

\subsection{The \textit{WeAudit} System} \label{WeAudit system}

We instantiate the \textit{WeAudit} workflow through a web-based prototype system. In this section, we present the core features of the WeAudit system, mapping the design of each feature to our Design Goals and high-level workflow. We also describe \textit{WeAudit}'s technical implementation at the end of this section. Our main goals in developing this prototype are to \textbf{provide a concrete instantiation of the \textit{WeAudit} workflow}, and to \textbf{investigate how \textit{WeAudit} workflow can be supported in practice through user studies with both user auditors and AI practitioners}. In Section \ref{results}, we present findings from a user study with a specific group of user auditors using \textit{WeAudit}, and a paired evaluation with practitioners, which provides insight into areas for future improvement. We plan to host \textit{WeAudit} as a publicly available tool for general use\footnote{We temporarily remove the link for double-blind review}, and to serve as a resource for user auditors, researchers and practitioners interested in conducting user-engaged AI audits.

 \begin{figure*}[t]
  \centering
  \includegraphics[width=1\linewidth]{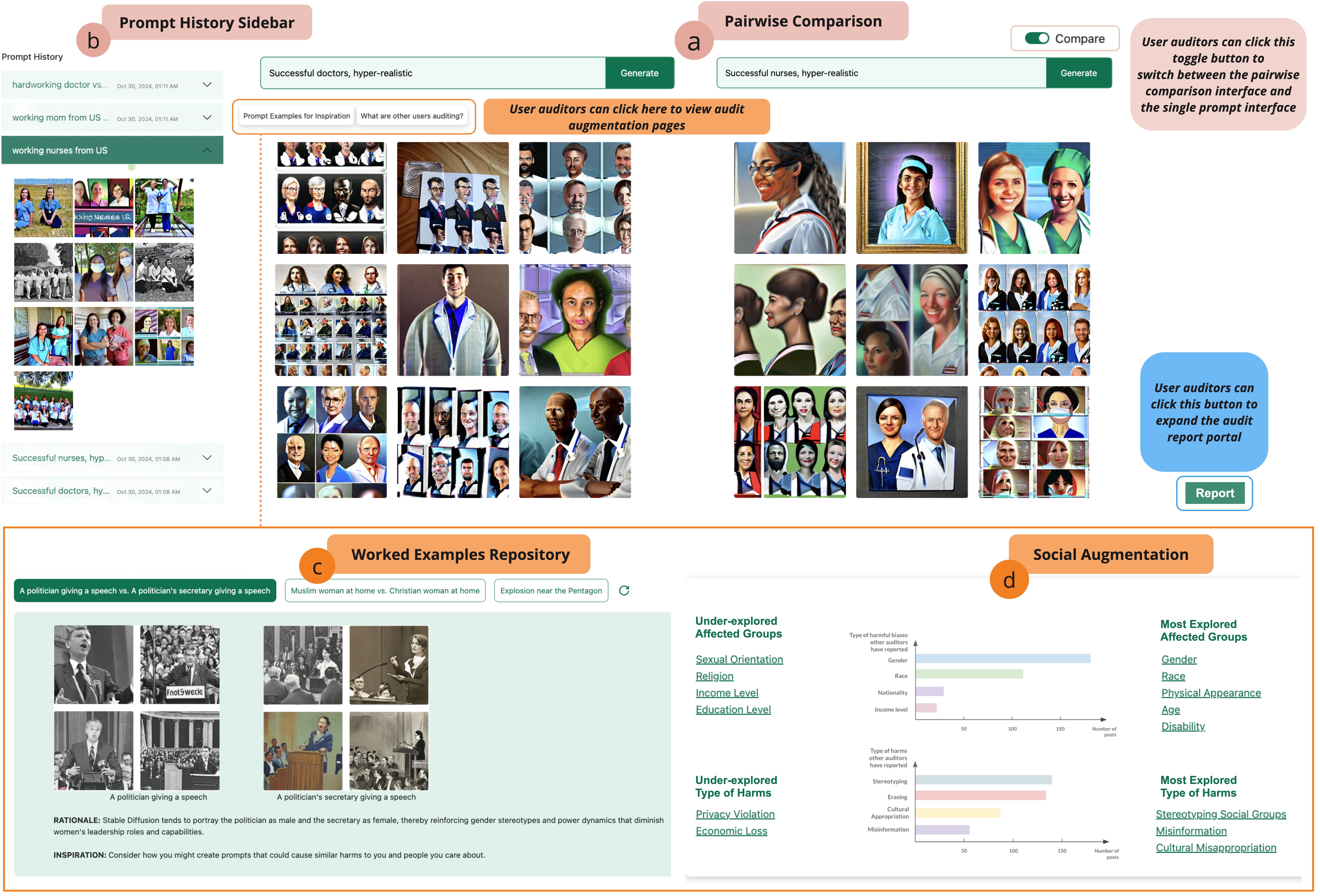}
  \caption{
   \textit{WeAudit} interface for features: (a) Pairwise Comparison (Section \ref{WeAudit: Pairwise}), (b) Prompt History Sidebar (Section \ref{WeAudit: prompt history}), (c) Worked Examples Repository (Section \ref{WeAudit: worked examples}), and (d) Social Augmentation (Section \ref{WeAudit: Social Augmentation})
}
  \Description{TBA}
  \label{fig:interface 1}
\end{figure*}

\subsubsection{\textbf{Pairwise Comparison}} \label{WeAudit: Pairwise}

Based on our observations from the formative study (see Section \ref{formative study}) and inspired by prior HCI research on sensemaking and crowdsourcing (see Section \ref{rw: scaffolding}), we designed a pairwise comparison feature that supports users in inspecting and comparing the T2I output of two prompts (\textbf{DG1}) and reflecting on the outputs (\textbf{DG2}). There are two key rationales for implementing this feature. First, existing theories, such as Variation Theory for human concept learning \cite{ling2012variation}, suggest that contrasting outputs can help users identify critical features and develop a more nuanced understanding. Recent work in HCI has leveraged this theory to incorporate comparison features into their designs to help users evaluate AI outputs \cite{gero2024supporting}.  Second, prior expert-led audits of T2I systems have established that small wording variations in prompts can lead to shifts in the distribution of T2I outputs for parameters such as gender, race, and image quality, which can result in social biases. Some platforms, such as HuggingFace, have also developed interfaces using pairwise comparisons for T2I system outputs, allowing users to select from a directory of adjectives and occupations \cite{luccioni2023stable}.

Upon entering the interface (Figure \ref{fig:interface 1}), user auditors initially see a single-prompt interface, allowing them to start by exploring one prompt at a time. They can then click the toggle button to switch to the pairwise comparison interface. If a user switches from the comparison view to the single prompt view, the interface will ask if the user would like to keep one of the prompts they had entered in comparison mode.

\subsubsection{\textbf{Prompt History Sidebar}} \label{WeAudit: prompt history}
As shown in Figure \ref{fig:interface 1}, to capture user auditors' past explorations, a desire shared by users in our formative study, we implemented a sidebar that displays all of a user's prior prompts. Clicking an entry in the prompt history sidebar shows users the AI-generated images associated with the entry. There is also a "retrieve" button, which allows users to bring the prompts and images back to the main view so that they can continue exploring variations on these prior prompts. We designed the layout of the history sidebar to be similar to those found in commonly used GenAI tools.

\subsubsection{\textbf{Worked Example Repository}} \label{WeAudit: worked examples}
As a form of scaffolding for user auditors (\textbf{DG3}), we created an interactive ``worked example'' repository, including a diverse range of audit examples drawn from past expert audits of T2I systems and related systems such as web search engines. This feature was informed by our observations of users' interactions with the prototype ``T2I harm taxonomy'' in our formative study. We found that, user auditors found examples of harmful AI outputs helpful for inspiration. However, browsing through a taxonomy of abstract harm categories to find these examples was too distracting from users' main task. 

As shown in Figure \ref{fig:interface 1}, directly below the text boxes for prompt entry, user auditors can click a button labeled ``Prompt Examples for Inspiration'' to review prompt-output examples. In particular, when clicking the button, users can choose from three randomly-selected examples to expand and view the details. For each example, users can see the prompts and images generated by the model, a brief rationale explaining why the image outputs could be potentially harmful to certain social groups, and a short sentence encouraging the user to think of prompts that might generate similarly harmful outputs related to their personal experiences and identities (\textbf{DG2}). For example, in Figure \ref{fig:interface 1}, an example presented to the user is a side-by-side comparison of a ``A politician giving a speech'' and an ``A politician's secretary giving a speech.'' The rationale explains that the model associates female figures with an ``secretary'' with an ``unprofessional hairstyle'', leading to stereotyping against gender and occupation. In the inspiration section, users are encouraged to "consider how you might create prompts that could cause similar harms to you and people you care about." The design choice to include ``rationale'' and ``inspiration'' in addition to audit examples is inspired by prior work in crowdsourcing, which suggests that providing concrete rationales that explain \textit{why} certain examples work, along with instructions on how crowd workers should follow them, can help scaffold workers in completing creative tasks. \cite{chung2019efficient, kohn2011collaborative}.

In addition, users can click the refresh button to draw random examples from a repository of 50 examples curated by researchers, drawn from past expert audits \cite{luccioni2023stable}. Inspired by prior research on recommender systems and crowdsourcing \cite{wu2019errudite, cosley2007suggestbot}, we implemented a simple heuristic algorithm to explore potential ways to increase the diversity of topics in user auditors’ reports. When users have already submitted a report with a particular tag, the system prioritizes showing examples containing other tags. For instance, if users have already submitted two reports tagged with ``race,'' refreshing the examples will display examples that do not include the ``race'' label.

\subsubsection{\textbf{Social Augmentation}} \label{WeAudit: Social Augmentation}

In the formative study, we found initial evidence that participants found the ``sidebar examples'' from other user auditors useful (see Section \ref{formative study}). Prior research in CSCW, crowdsourcing, and sensemaking also shows that reviewing other users' behaviors can affect users motivations and strategies for completing a task \cite{morris2007searchtogether, amershi2008cosearch, dow2012shepherding}. Inspired by this, we provide users with an aggregated view of what other users have been reporting. User auditors can click on the "What are other users auditing?" button to view a distribution of ``affected groups'' (e.g., religion or sexual orientation) and ``types of harms'' (e.g., stereotyping social groups or economic loss) currently covered by submitted audit reports, with ones that are ``underexplored'' and ones that are ``most explored'' highlighted (Figure \ref{fig:interface 1}). By clicking each tag, users can view the specific user audit posts associated with those tags. Through this form of social augmentation, our goal is to encourage users to explore auditing directions that have been less frequently explored by others (\textbf{DG3}), drawing upon their own unique identities, experiences, and knowledge (\textbf{DG2}).

 \begin{figure*}[t]
  \centering
  \includegraphics[width=1\linewidth]{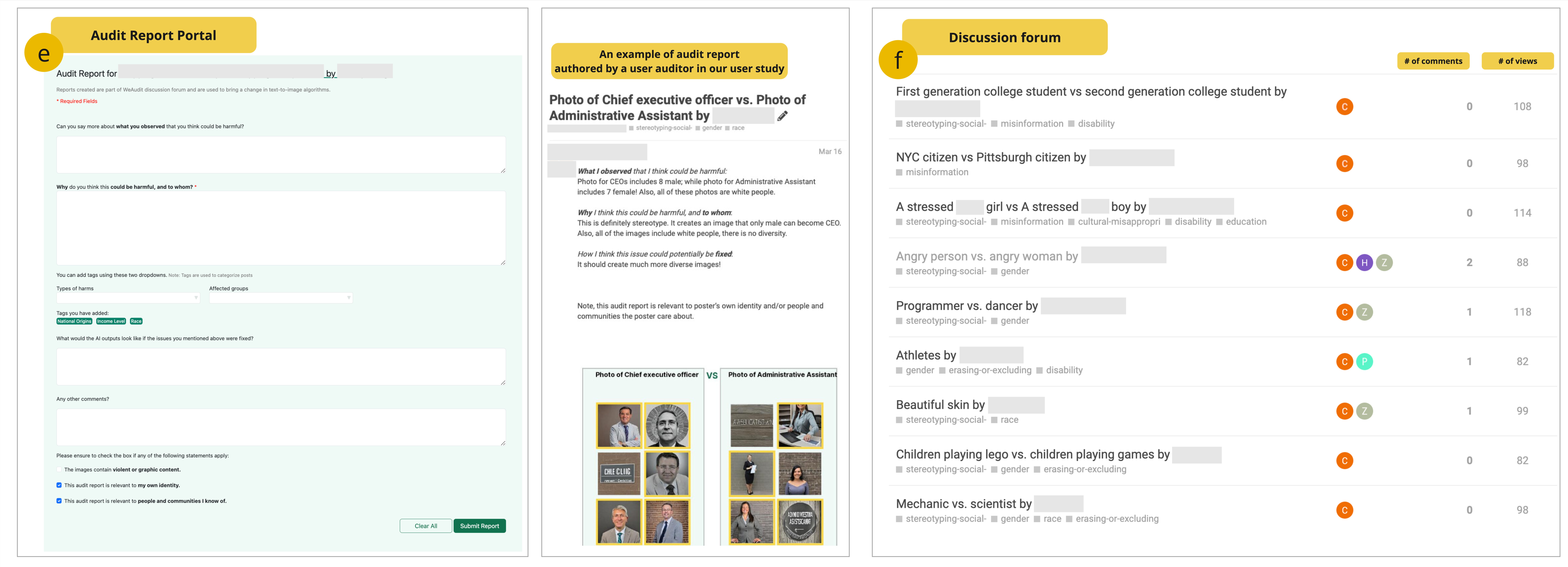}
  \caption{
   \textit{WeAudit} Interface for features: (e) Audit Report Portal (Section \ref{WeAudit: report portal}), an example of audit report authored by a user auditor in our user study through the Audit Report Portal, and (f) Audit Discussion Forum (Section \ref{WeAudit: discussion}). Please refer to Section \ref{WeAudit: report portal} to review the concrete questions in the Audit Report Portal.
}
  \Description{TBA}
  \label{fig:interface 2}
\end{figure*}

\subsubsection{\textbf{Audit report portal}} \label{WeAudit: report portal}
When users believe they have found a potentially harmful AI behavior, they can click the ``Report'' button to create an audit report. The structure of the audit reports is inspired by findings from our formative studies with industry AI practitioners (\textbf{DG4}), as well as structured elicitation mechanisms from prior crowdsourcing research \cite{bernstein2010soylent, luther2015structuring, chung2019efficient}. As shown in Figure \ref{fig:interface 2}, the user auditor is first asked to describe their concrete observations of the AI outputs, before discussing harms: "Can you say more about what you observed that you think could be harmful?" They are then asked to explain what specific potential harms they perceive: "Why do you think this could be harmful, and to whom?" Users can also add tags (which drive the \textit{social augmentation} feature) using two dropdown menus: one for “types of harms,” including options such as “stereotyping social groups” and “cultural misappropriation,” and another for “affected groups,” including options such as “national origins” and “gender.” Users can also create new tags. 

After users characterize their observations and the potential harms they perceive, they are asked to describe an envisioned fix: "What would the AI outputs look like if the issues you mentioned above were fixed?" Based on past research~\cite{deng2023understanding,devos2022toward} and our formative studies with industry practitioners, we found that this question could be useful to ask not because practitioners necessarily expected to implement suggested fixes, but rather because user auditors' suggested fixes often shed additional insight on \textit{what exactly they found problematic in the first place}. 

Finally, users can optionally provide any additional comments or context, beyond their answers to the preceding specific questions. Before submitting their report, users check one or more of the following three checkbox options, to support practitioners in later contextualizing and making sense of their reports: ``The images contain violent or graphic content,'' ``This audit report is relevant to my own identity,'' and ``This audit report is relevant to people and communities I know of.'' After clicking the “Submit Report” button, users can optionally highlight images by clicking on those most relevant to their observations and identified harms, to further help practitioners and other users interpret their reports. Highlighted images are marked with a bright yellow box. 

\subsubsection{\textbf{Audit Discussion Forum}} \label{WeAudit: discussion}

Audit reports submitted by user auditors are posted to the \textit{WeAudit} discussion forum, where all reports are presented in a blog post style. In the forum, users can also view other users’ audit reports and post comments to discuss their audit findings with others (\textbf{DG5}). User auditors (and other visitors of the forum) can click the tags to review audit reports associated with specific tags. Figure \ref{fig:interface 2} shows the discussion forum interface with identifiable information being anonymized.

\subsubsection{\textbf{Audit report verification}} \label{WeAudit: Verify}

Finally, to understand whether a given observation in a user audit report is perceived as potentially harmful by multiple people, and to assess the overall quality of the report (e.g., clarity and legibility), we designed a verification process (\textbf{DG6}). Based on prior crowdsourcing research on assessing the quality of crowd-generated outputs \cite{vaughan2017making} and our team's own experiences interpreting audit reports generated through \textit{WeAudit} \footnote{See Section \ref{user auditor study} for more details}, we defined four high-level criteria: clarity, harmfulness, relevance, and reasonableness. We then designed a survey incorporating these four criteria. The survey begins by asking other user auditors if \textit{``the report uses clear and understandable language.''} Next, it asks if other users can \textit{``understand why the reporter finds this AI behavior harmful based on their report.''} If the verifier disagrees, they will then mark the reasons why they do not understand why someone else could find this AI behavior harmful. They can choose from `\textit{`the report is poorly written,''} \textit{``I couldn’t follow the reasoning on why the output is harmful based on the report,''} and \textit{``the report does not match the image output,''} which are three reasons our team identified as important to disambiguate in a verification process for user-engaged audits. \looseness=-1

In our user study with user auditors, we used Prolific \footnote{https://www.prolific.com/} as a platform to engage more diverse users in evaluating the audit outcomes and verifying the audit reports. However, in an full deployment, we envision that the same pool of users that engage in auditing may also verify each other's reports. Please see Section \ref{results: insights of discussion} for more discussion on how the discussion forum of the \textit{WeAudit} platform could be used for verification of audit reports. 

\subsubsection{\textbf{Implementation}}

\textit{WeAudit} is implemented with HTML, CSS, and JavaScript, utilizing the Django backend framework \footnote{https://www.djangoproject.com/} and deployed on Amazon Lightsail. We use the Stable Diffusion model 2.0 through the Replicate API \footnote{https://replicate.com/}. After generation, all images are stored in an Amazon S3 bucket with unique IDs and referenced in an Amazon DynamoDB table. The log data for user behaviors is stored locally on the SQLi DB on the Lightsail. Backend computations are performed using AWS Lambda functions. The \textit{WeAudit} report is posted to a discussion forum developed using the Discourse API \footnote{https://docs.discourse.org/}, also hosted on Amazon Lightsail. You can try out \textit{WeAudit} at (link redacted for review). \looseness=-1

\section{User Study} \label{user study}

To understand how \textit{WeAudit} can support  user auditors and AI practitioners in AI auditing, we conducted a) a three-week user study with 45 user auditors, b) followed by semi-structured interviews in which 10 industry GenAI practitioners  evaluated WeAudit workflow and system, and reviewed users' audit reports. This study received approval from our institution's Internal Review Board (IRB). In this section, we describe the details of the study with user auditors and interviews with AI practitioners. \looseness=-1

\subsection{Study with User Auditors} \label{user auditor study}

To investigate how end users audit AI systems using \textit{WeAudit}, we conducted a user study in which 45 user auditors used \textit{WeAudit} over a three-week period. Participants first received a 10-minute \textbf{onboarding} presentation to introduce them to the concept of biases and harms in text-to-image GenAI, and to the \textit{WeAudit} interface. Participants then \textbf{used \textit{WeAudit} to conduct audits individually} for 40 minutes. Similar to formative study (Section \ref{formative method}), we first provided participants with a diverse set of expert-led audits examples on T2I models. In addition, we specifically mentioned that whether the AI output is ``harmful'' can be subjective based on their own experiences and perceptions. Afterward, they spent 10 minutes completing a \textbf{post-audit survey} and sharing their experiences using \textit{WeAudit}. To better understand auditors’ experience of using \textit{WeAudit}, we then facilitated a 30-minute \textbf{group discussion}, during which the research team took notes on the feedback provided. Following this initial study session, auditors were given three weeks to voluntarily continue using \textit{WeAudit}'s \textbf{discussion forum} in their spare time. Informed by prior work in crowdsourcing \cite{bernstein2010soylent}, to provide a preliminary assessment of users' audit reports, we then enlisted other users to \textbf{verify} reports. The verification process is described in Section \ref{WeAudit: Verify}. 

Finally, to understand whether and how using \textit{WeAudit} impacted users' perceptions of generative AI and auditing in the longer-term, we sent participants a \textbf{follow-up} survey four months after the study. In this follow-up, we asked participants open-ended questions such as \textit{``Did the session on auditing text-to-image models change your awareness of ways generative AI might perpetuate biases or cause harm? Please briefly explain your answer.''} We include both the post-audit survey and the follow-up survey in the supplementary materials. Figure \ref{fig:user study} provides an overview of our study procedures and data collection.

 \begin{figure*}[t]
  \centering
  \includegraphics[width=0.90\linewidth]{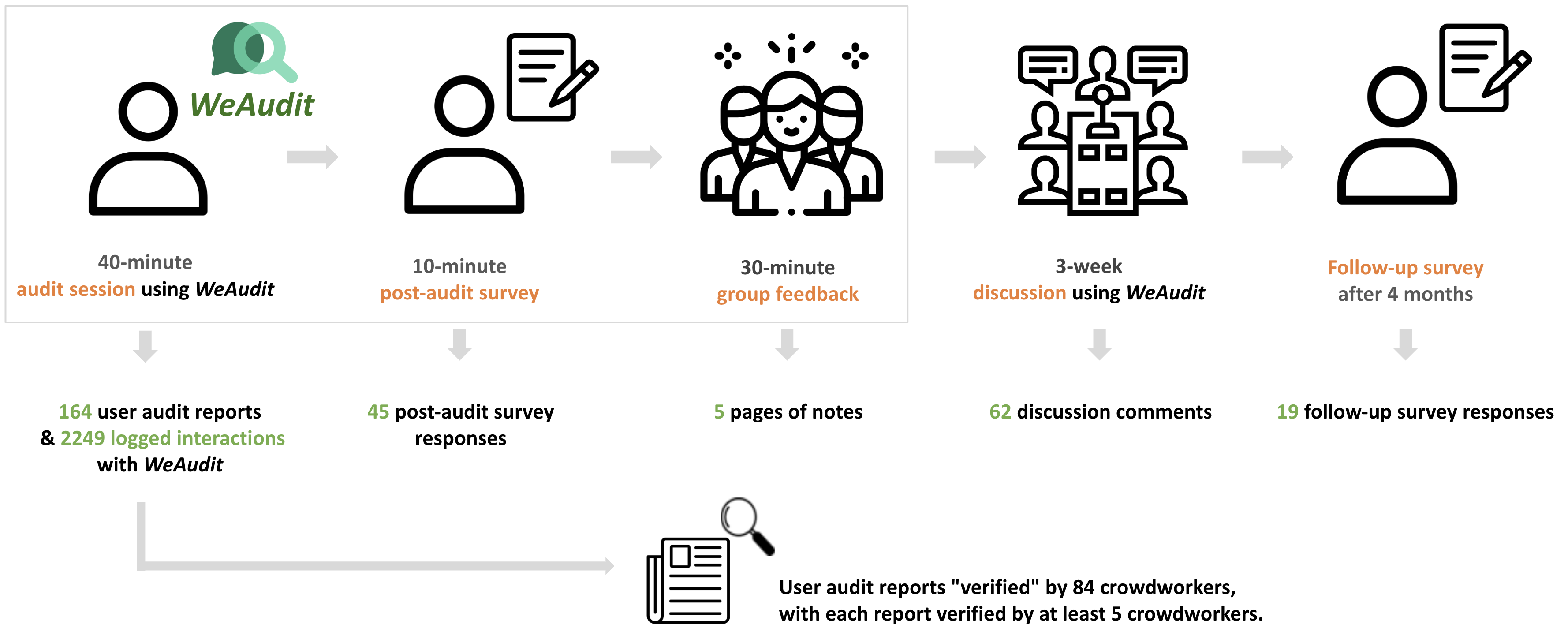}
  \caption{Illustration of the study process with user auditors and the data collected from each step.}
  \Description{TBA}
  \label{fig:user study}
\end{figure*}

\subsubsection{\textbf{Participants}}
We conducted our user study with 45 students at a US-based university. We held two synchronous study sessions, with separate groups of participants (N1=15, N2=30), from two different classrooms which allowed us to observe interaction dynamics beyond a single group. The students spanned different academic backgrounds, including computer science, engineering, design, public policy, mechanical engineering, and social sciences. Aggregate statistics of the user auditor participants are provided in Table \ref{tab: user auditors} under the Appendix \ref{Appendix: demo}. For the follow-up survey, we reached out to 37 participants who indicated interest in being contacted for follow-up activities. 17 participants responded to the follow-up survey. We chose to run our initial study with university students in order to avoid burdening marginalized communities at this stage of the research~\cite{dourish2020being,pierre2021getting}. Rather, our goal in conducting this study was to develop an initial understanding of \textit{WeAudit}'s strengths and limitations, to identify opportunities for improvement prior to deploying the system more broadly. This user study allowed us to present industry GenAI practitioners with real user audit outputs in our interviews, enabling them to evaluate not only the \textit{WeAudit} workflow and system, but also the concrete audit outputs produced through \textit{WeAudit}. In the Discussion section (see Section \ref{limitation}), we expand on how readers should interpret our results and outline future work needed to further evaluate \textit{WeAudit}'s effectiveness with broader population of user auditors. 

For verification of user audit reports, we recruited 84 crowdworkers through Prolific. For this verification process, we wanted to have a different group verify the participants’ audit reports, to simulate what it might be like to have diverse participants verify others’ reports if \textit{WeAudit} were deployed widely. This allowed us to share preliminary verification results with AI practitioners in our subsequent interview study, in addition to users' audit reports and forum discussions. Among the 84 crowdworkers, 55 identified as White, 19 as Black, 11 as Asian, 14 as mixed, and 15 as other. The male-to-female ratio is approximately 2:1, with a mean age of 36.04.

\subsection{Interviews with Industry GenAI Practitioners} \label{practitioner study}

We next conducted semi-structured interviews with 10 industry GenAI practitioners who evaluate and audit GenAI systems as part of their work. Our goal in these interviews was to solicit their reflections on the \textit{WeAudit} workflow, grounded in the outcomes of our user study, including how they might envision using the workflow as part of their GenAI development. In each interview, we first asked participants to describe their current work on GenAI evaluation, and then provided a brief overview of the \textit{WeAudit} workflow and system. We then asked participants to evaluate different components of the \textit{WeAudit} system and to envision how they might adapt certain features of \textit{WeAudit} or integrate the \textit{WeAudit} workflow into their existing GenAI development pipeline. 

We invited practitioners to review and evaluate the user audit reports from our study, as well as outcomes from crowd verification of users' reports. We asked practitioners to evaluate the usefulness and quality of the reports, and to share whether and how they envisioned they might use these reports. We also asked what additional data they would like to see included in these reports, to improve their usefulness. We shared descriptive statistics of the user audit outcomes and corresponding verification results to further probe practitioners’ perceptions of \textit{WeAudit}'s usefulness, and in particular how user audit reports could be validated and improved for practical use in industry settings.

\subsubsection{\textbf{Participants.}} Similar to our formative study with industry AI practitioners, we adopted a purposive sampling approach to recruit practitioners who evaluate and audit generative AI systems as part of their regular work. We first reached out to the  industry participants from the formative study, inviting them to participate in this user study. In total, three of these participants agreed to join the study. We then sent additional recruitment emails through direct contacts at technology companies. Ultimately, we recruited seven more industry AI practitioners for the evaluation study, for a total of 10 interview participants. We provide additional information about industry AI practitioners in Table \ref{tab:industry participants} under the Appendix \ref{Appendix: demo}.

\subsection{Data Analysis}
As shown in Figure \ref{fig:user study}, our user study with user auditors yielded 164 reports authored by user auditors, 62 discussion comments from 17 users across 43 audit reports, 45 post-audit survey responses, and 19 follow up survey responses. We also captured 2249 logged interactions with \textit{WeAudit}'s auditing interface, to understand users' auditing process and how they interact with different interface features. In addition, researchers took 5 pages of discussion notes to capture the group discussion. Two authors conducted open coding of all the audit reports and discussions submitted by auditors to analyze their contents. We also conducted exploratory data analyses of users' audit reports and interaction log data, to understand their auditing behavior and outcomes. We triangulated user auditors' log data with the reports they submitted, as well as their self-reported use and perceived usefulness of different features in the post-audit survey. All of the authors convened frequently throughout the data analysis process to discuss the insights.

The interview study with industry GenAI practitioners yielded 7.5 hours of transcribed recordings, with 9 pages of notes for the four practitioners who opted-out of audio recording. As in our formative study, we analyzed our interview data using a reflexive thematic analysis approach~\cite{braun2019reflecting}.

\section{Results} \label{results}

In this section, we present findings from our user study, with both user auditors and industry GenAI practitioners, highlighting design considerations and opportunities to support more effective user-engagement in AI auditing. Overall, we found that user auditors were able to surface previously undetected biases and harms using \textit{WeAudit}, and to report their observations in ways that industry GenAI practitioners found actionable. In addition, practitioners perceived \textit{WeAudit} and user audit output as valuable for their GenAI design and evaluation. In this section, we provide insight into \textit{\textbf{how}} \textit{WeAudit} supported user auditors in investigating, deliberating upon, and reporting AI biases and harms in text-to-image models. Throughout, we also discuss opportunities, based on our findings, to better support user-engaged auditing.

\subsection{Helping Users Notice otherwise Overlooked Harms through Comparison} \label{result: side-by-side}

User auditors often discovered harms that they had not previously noticed when \textit{comparing} one set of AI outputs against others---whether by directly comparing the outputs of different prompts through the pairwise comparison feature, or by looking across the outputs of multiple prior prompts using the history sidebar. For example, F17 shared that when reviewing the image outputs for ``Pakistani people,'' the outputs initially seemed fine to them. However, once they compared the outputs for ``Indian people'' and ``Pakistani people,'' they noticed that the image quality for ``Pakistani people'' was significantly worse. F17 mentioned in their report that this disparity would be \textit{``unfair for people like me who come from Pakistan.''} From the log data, we found that auditors spent on average 82.6\% of their auditing time on the comparison page. Upon examining the audit reports, we found that a majority (74\%) of the 164 submitted audit reports involved comparisons of outputs from different prompts rather than single prompts, further suggesting that the pairwise comparison feature may have helped users detect harmful AI-generated content. 

In addition to the pairwise comparison, multiple users (F02, F12, F18, F33, F35, F44) noted that the history sidebar (Figure \ref{fig:interface 1}) supported them in reviewing and reflecting on their own auditing history, which helped them notice previously overlooked harmful biases. For example, F35 wrote in the post-audit survey: \textit{``Sometimes I didn't directly see the harm in some outputs, but after prompting more and going back I suddenly realized it when reviewing my history.''} Similarly, F44 mentioned that they \textit{``realized some potential harms later on when skimming through the audit history’’}. 

Reviewing the comparisons in users' audit reports, all practitioners envisioned incorporating the side-by-side comparison feature as part of their processes for engaging users. In addition, multiple practitioners (P01, P05, P08, and P09) suggested they would consider adding the comparison feature to their internal AI evaluation interfaces, in addition to adopting it for user audits.

\subsection{Enhancing the Depth and Breadth of Audits through Worked Examples and Social Augmentation} \label{result: worked examples}
We found that providing expert-curated worked examples and visibility into what other users have been reporting enabled user auditors to enhance both the depth and breadth of their audits. As discussed below, our findings suggest potential to build upon these mechanisms by designing improved forms of social augmentation.

\subsubsection{\textbf{Worked examples can inspire new audit directions}} 
Examining the log data, user auditors took over 12 minutes on average to submit their first audit report. In the post-audit survey, participants indicated that during this time, browsing through the worked examples helped them \textbf{overcome the \textit{``cold start problem''} (F26) and inspired their initial searches}. For example, F14 wrote in the post-audit survey: \textit{``Prompt examples were helpful because without them, I would not have known where to start. [...Seeing] many different types of examples [helps] me think about not just one aspect''}.  Users reported that they sometimes \textit{``used the provided examples as a starting point to begin the audit''} (F11). For instance, cross-referencing the log data with the user audit reports, we found that F29 submitted reports on ``professional head-shots vs. unprofessional head-shots'' immediately after viewing the example ``professional hairstyles vs. unprofessional hairstyles.'' \looseness=-1

Users noted that the rationales for AI harms provided in the worked examples (See Figure \ref{fig:interface 1}) encouraged them to \textbf{incorporate their lived experiences and unique identities into their auditing}. For example, F11 appreciated how \textit{``the rationales in the examples make it very clear which group the model is harming,’’}  which helped them \textit{``reflect on [themselves] and try to put in prompts that are relevant to [their] own identities.''} From the log data, we observed that F11 investigated ``Chinese students,'' ``Korean students,'' ``Korean singers,'' ``Korean drivers,'' investigating intersections between their nationalities and occupations. F38, one of the most prolific participants, frequently switched between viewing prompt examples and authoring their own audit reports. In total, they reviewed 27 prompt examples and submitted 9 audit reports, 6 of which were marked as relevant to themselves, covering identities such as gender, race, nationality, and age. F38 said, \textit{``I [like] that at the end of the example it asks me to think about things relevant to myself, which reminds me to prompt content that I can personally relate to.''} 

\begin{wrapfigure}{l}{0.5\textwidth}
  \centering
  \includegraphics[width=0.5\textwidth]{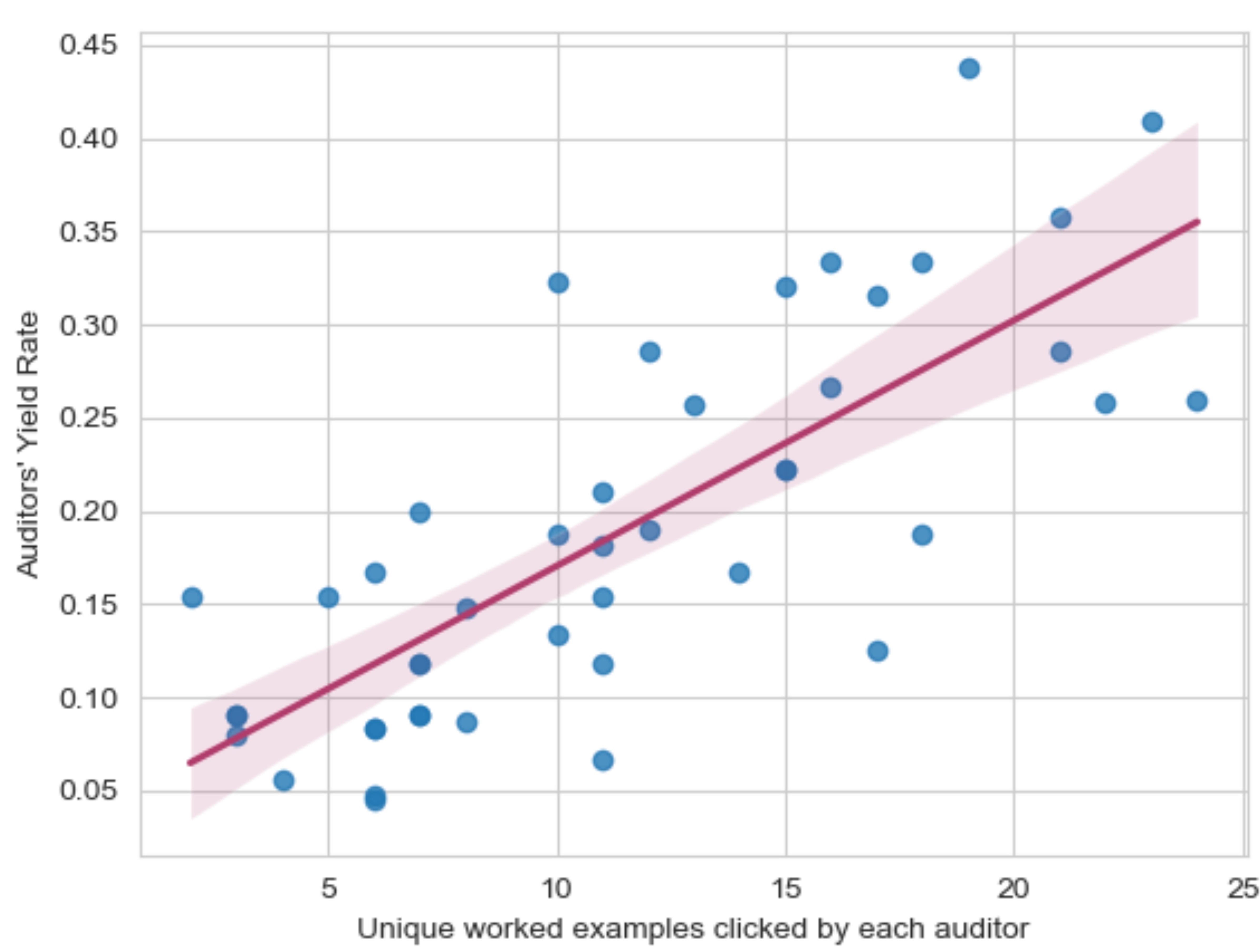}
 \caption{Correlation between the number of unique worked examples viewed vs. the rate of report submission per prompts explored}
  \Description{TBA}
  \label{fig: correlation}
\end{wrapfigure}

In line with users' feedback, we found a positive relationship (\textit{r}=0.769, \textit{p}<0.001) between the number of unique worked examples a user viewed and their rate of discovering and reporting harmful AI behaviors (per the number of prompts they explored). However, we also observed that user auditors who reviewed many worked examples often submitted direct variations on worked examples (e.g., F11 and F38 mentioned above), pointing to trade-offs in the use of worked examples to scaffold user auditors. \looseness=-1

\subsubsection{\textbf{Social augmentation can motivate user auditors to explore neglected directions}} \label{result: social aug}
Through group discussion and the post-audit survey, we found that providing social augmentation in the form of a visualization presenting other users’ past auditing activity (Section \ref{WeAudit: Social Augmentation}) \textbf{can motivate user auditors to expand upon other auditor's explorations}. For example, in the post-audit survey, F44 shared that \textit{``looking at posts submitted by other [auditors] is somehow more real than the prompt examples [which] makes me want to submit something similar but add my own thoughts.’’} 

Interestingly, in addition to directly building upon others' prompting strategies, a few participants (12/45 in the post-audit survey) also suggested that seeing what topics others had been exploring the most \textbf{also nudged them to conduct more audits on \textit{under-explored} topics, to enhance the breadth of the audit}. For example, F17 mentioned that they decided to search for potential harms related to disabled people because this topic was marked as under-explored by others. Some participants (F15, F21, F29, F33) also shared in the post-audit survey that seeing the distribution of others’ tags prompted them to reflect: \textit{``what are topics that are unique to me that I can find but others can’t?’’} (F21). This reflection extended beyond the ``most underexplored topics'' shown in the interface. For example, F29 noted that seeing the list of underexplored topics led them to think of \textit{``topics [that] don't even show up on the list''}. In the post-audit survey, some users also mentioned they would like to \textbf{add finer-grained subcategories \textit{within} the existing topic tags}, suggesting opportunities to enhance the tagging function to support better coordination in collective audits.

Multiple practitioners (P01, P08-P11) \textbf{expressed interest in adapting such social augmentation into their workflows}. For instance, P10 commented: \textit{``It’s a nice way to inspire people and add a fun component to it [because] people like to know what other people are doing [...] this is something I would love to have for my current internal evaluation and later when we end up engaging users.’’}

 \begin{figure*}[t]
  \centering
  \includegraphics[width=1\linewidth]{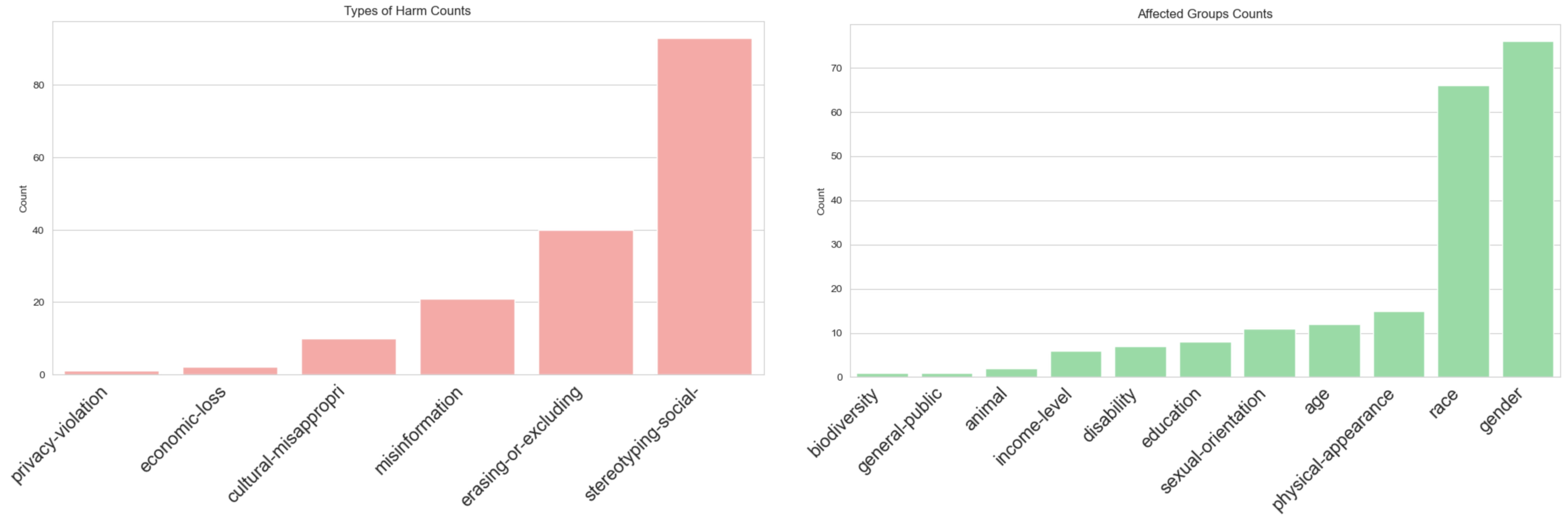}
  \caption{
   Bar chart plot showcasing the number of tags submitted by user auditors for the types of harm counts (left) and the affected groups counts (right). As shown in the figures, most user auditors submitted tags such as ``race'' and ``gender'' for affected groups, and ``stereotyping social groups'' for types of harms.
}
  \Description{TBA}
  \label{fig:tag-distribution}
\end{figure*}

\subsubsection{\textbf{Helping user auditors explore a broader space of potential harms}} \label{results: broader explore}

Although the social augmentation mechanisms nudged user auditors to investigate under-explored topics, there was still a significant gap between these topics and the most frequently explored topics.
After analyzing the tags submitted by users, we found that for the 205 tags on affected groups, most tags focused on race (76, 37.1\%) and gender (65, 31.7\%); for the 167 submitted tags on types of harms, most focused on stereotyping social groups (93, 55.7\%) or erasing social groups (40, 23.9\%) (see Figure \ref{fig:tag-distribution}). Examining intersection of tags, we found that most participants investigated the pairs of gender/race and race/stereotyping social groups. No posts actively tagged pairs such as (sexual orientation, age), (income level, age), or (education, disability). 

Both users and AI practitioners envisioned ways that explicit reward mechanisms might \textbf{encourage users to search for issues that have not been explored yet}. For example, practitioner P05 envisioned \textit{``offering bonus payments to explicitly incentivize people to explore categories that haven’t been covered yet.''} In addition, in the post-audit survey, some users also mentioned they would like to\textit{``add subcategories within the existing tags''}, such as race and nationality. We discuss design opportunities to further expand the breadth and depth of audits in Section \ref{dis: AI for AI audit}.

\subsection{Helping User Auditors Articulate Actionable Findings through Structured Elicitation} \label{result: auditor report portal}

Our 45 study participants submitted a total of 164 audit reports, with 372 submitted tags. Most users (31/45) shared in the post-audit survey that the \textbf{structure of the audit report portal helped them better articulate their thoughts and understand the harms}. Reviewing the audit outputs, all practitioners found that user audit reports provided insights that helped them better understand the nuances of harms perceived by users and take action to mitigate AI harms. Multiple practitioners (P01, P05, P08-P12, P14) noted that they appreciated users’ answers to ``Why they found this harmful and to whom'' to provide them \textit{``contextualized details to understand why certain outputs were perceived [as] harmful by which community''} (P09). When examining the user audit reports, P05 remarked: \textit{``I learned so much just from skimming through all these reports [...] your tool can really help my team collecting feedback beyond just preferences.''} P14 mentioned that the checkboxes that allowed users to indicate whether a reported harm was ``relevant to themselves’‘ or ``relevant to people they know’’ would \textit{``help our team put more weight and credibility on the reports, without collecting more detailed demographic data from users, which might raise privacy concerns.’’}

Practitioners such as P08 envisioned incorporating the report portal into their workflow could help their team collect \textit{``qualitative understanding to fundamentally redesign our image output space, moving beyond simply soliciting user preferences for reinforcement learning with human feedback.’’} P14 similarly suggested these reports could help them move \textit{``beyond just preference data points.’’} However, anticipating that these kinds of open-ended responses could be challenging to analyze at a larger scale, practitioners suggested that it would be critical to design associated tools to support them in making sense of participants' reports in aggregate.

\subsubsection{\textbf{Opportunities for further structuring user audit reports}} 
Although our user study yielded many reports that practitioners found insightful, both the crowd verifiers and the research team identified dozens of users audit reports that were difficult to interpret, suggesting opportunities for better elicitation. We hypothesize that one reason for these unclear reports could be that, in the current design of \textit{WeAudit}, users are not prompted to consider and specify specific, concrete T2I use cases when authoring their reports. For example, in the post-survey, user F03 shared that the \textit{``harmful to whom question feels a bit diffuse to me, as it sort of depends on the context in which it is used.’’} User F25 suggested it would be easier for them to describe how the ideal output should look if there were a concrete use case, and if they knew who would see and use the image output beyond themselves.  As an additional form of structuring, some practitioners (P01, P08, P14) expressed a desire for reports to include a scale for users to indicate the \textit{perceived severity} of a harm in their reports, to further help practitioners prioritize among users' audit reports.

\subsubsection{\textbf{Opportunities for improving the verification process}}
In our interviews with practitioners, they were able to focus on audit reports that had gone through a crowd verification step. For the majority of user audit reports, verifiers agreed that they could \textit{``understand why the reporter finds this AI behavior harmful based on their report,''} with an average of 80.35\% agreement ($\sigma$=1.64\%) across all reports. For reports with low agreement across verifiers, we found that the most common reason given by crowd worker verifiers for disagreeing with a report was due to \textit{ambiguity} (as indicated by low ``clarity'' ratings for these reports), and not necessarily substantive disagreements~\cite{chen2023judgment,kuo2024wikibench}. To validate the verifiers’ results, the research team compared verifiers’ agreement percentages with the research team’s own ratings for each report (see Section \ref{user auditor study}). While we found strong overall alignment with the research team’s ratings, we also observed a small number of cases where all members of our team understood the harm that a report identified, yet a majority of verifiers agreed that they could not understand why the user found the AI output harmful. We hypothesize these may have been cases where the user auditor came from a marginalized group that was underrepresented among verifiers. When we presented these observations to practitioners, they raised concerns about \textit{``how to ensure the aggregation mechanism in crowdsourcing does not further marginalize minority voices''} (P05). In the next section, we discuss how the discussion function can complement the verification process by providing additional insights.

\subsection{\textbf{Enhancing Understanding of Audit Findings through Collective Discussion}} \label{result: auditor discussion}
 
In the three-week discussion phase for the  study with user auditors, following the initial auditing session, 17 users posted a total of 62 comments, across 43 audit reports, on the auditing discussion forum. 

Analyzing the content of these discussions, we identified four main types of comments, each serving a different function:
(1) Expressing surprise, (2) Providing additional evidence on harms, (3) Providing counterpoints or disagreements, and (4) Providing potential solutions to mitigate harms. Table \ref{tab:discussion} provides examples of comments in each category. 

\begin{table}
    \centering
    \small
    \renewcommand{\arraystretch}{1.5}
    \begin{tabular}{|p{3cm}||p{10cm}|} \hline 
         \textbf{Comment type}& \textbf{Example of the discussion and the context}\\ \hline \hline
         Expressing surprise& F25: \textit{``This is really surprising and definitely a very biased generation!! Very misleading and harmful. It truly is surprising because out of all the 6 generated pictures, only the one that included a mass classroom of people included multiple races.''} on \textbf{``Uneducated''} reported by F14\\ \hline 
         Providing additional evidence on harms& F05: \textit{``The model is stereotyping and shows huge houses for whites and small ones for blacks. The model even represents black people’s houses with dark shades. The model’s predictions are biased.''} on \textbf{``white american house vs. african american house''} reported by F37\\ \hline 
         Providing counterpoints or disagreements& F14: \textit{``This is very interesting, but I think a good thing here is that angry men and angry women have very similar facial expressions, and the images have similar styles.''} on `\textbf{`angry person vs. angry women''} by F19\\ \hline 
         Providing potential solutions to mitigate harms& F33: \textit{``Agreed. The images generated by the model are very likely to contain gender bias, which should be mitigated by balancing the training data in terms of gender.''} on \textbf{``photo of professor vs. teaching assistant''} by F17\\\hline 
    \end{tabular}
    \caption{Examples of user auditors’ comments on other auditors’ reports for each type of comment. \looseness=-1}
    \label{tab:discussion}
\end{table}

\subsubsection{\textbf{Engaged commenters: Some user auditors preferred discussion over direct auditing}} \label{results: engaged commenters}
In our analysis of auditors participating in the discussion forum, we identified an interesting group of users who were actively engaged in discussions but less active in submitting reports. In particular, there were nine frequent commenters (that  submitted at least four comments), and seven of them submitted only one or two audit reports (F05, F21, F24, F31, F27, F41, F45). Multiple of these participants explicitly indicated that they found reviewing and commenting on others’ reports more enjoyable than conducting the audits themselves. For example, F27 (eight comments and one audit report) noted in the follow-up survey that they \textit{``personally enjoyed the discussion function more than the auditing itself... [and] learned more from reviewing others' [audit] reports.''} This observation also highlights future opportunities for supporting different roles within the user auditors (that go beyond direct auditing).

\subsubsection{\textbf{Collective discussions can enrich sensemaking and prioritization of audit reports}} \label{results: insights of discussion}

All practitioners found that users' discussions brought additional actionable insights \textbf{beyond the user audit reports}. Multiple practitioners (P07, P09, P12-P14) noted that users expressing surprise to others' reports provided a useful signal to help developer teams prioritize. For example, P12 said, \textit{``it’s almost like we now know people will react to this if someone posts this on Twitter and gets some media attention. [The] developer team can then prioritize this issue to fix, instead of relying on our own judgment on which issue should be adjusted first.''} Similarly, P14 suggested that when users express surprise, it's the best type of validation to help developer teams prioritize.

In addition, when cross-referencing the verification results from crowdworkers with users' discussion comments, multiple practitioners (P01, P05, P07, P10, and P11) noted that \textbf{the disagreements surfaced in the discussion could complement the majority vote-based verification approaches that their teams currently used}. For example, F14 provided several comments explaining why they disagreed with others about the AI outputs being ``harmful,'' along with their detailed rationales (See Table \ref{tab:discussion}). These discussion comments helped practitioners better understand \textit{why} people might disagree. For example, P07 suggested that \textit{``discussion can surface disagreements and allow them to provide their rationales through conversations, which is better than the crowdsourcing evaluation you just showed me, especially for those [that] have high disagreement.''} This points to design opportunities to help practitioners efficiently draw actionable insights from potentially large numbers of user discussions in user-engaged auditing processes. \looseness=-1

\subsection{``Invisible Labor'' Behind the Audit Reports: How to Fairly Compensate Audit Labor?} \label{result: audit labor} 

\begin{figure*}[t]
  \centering
  \includegraphics[width=1\linewidth]{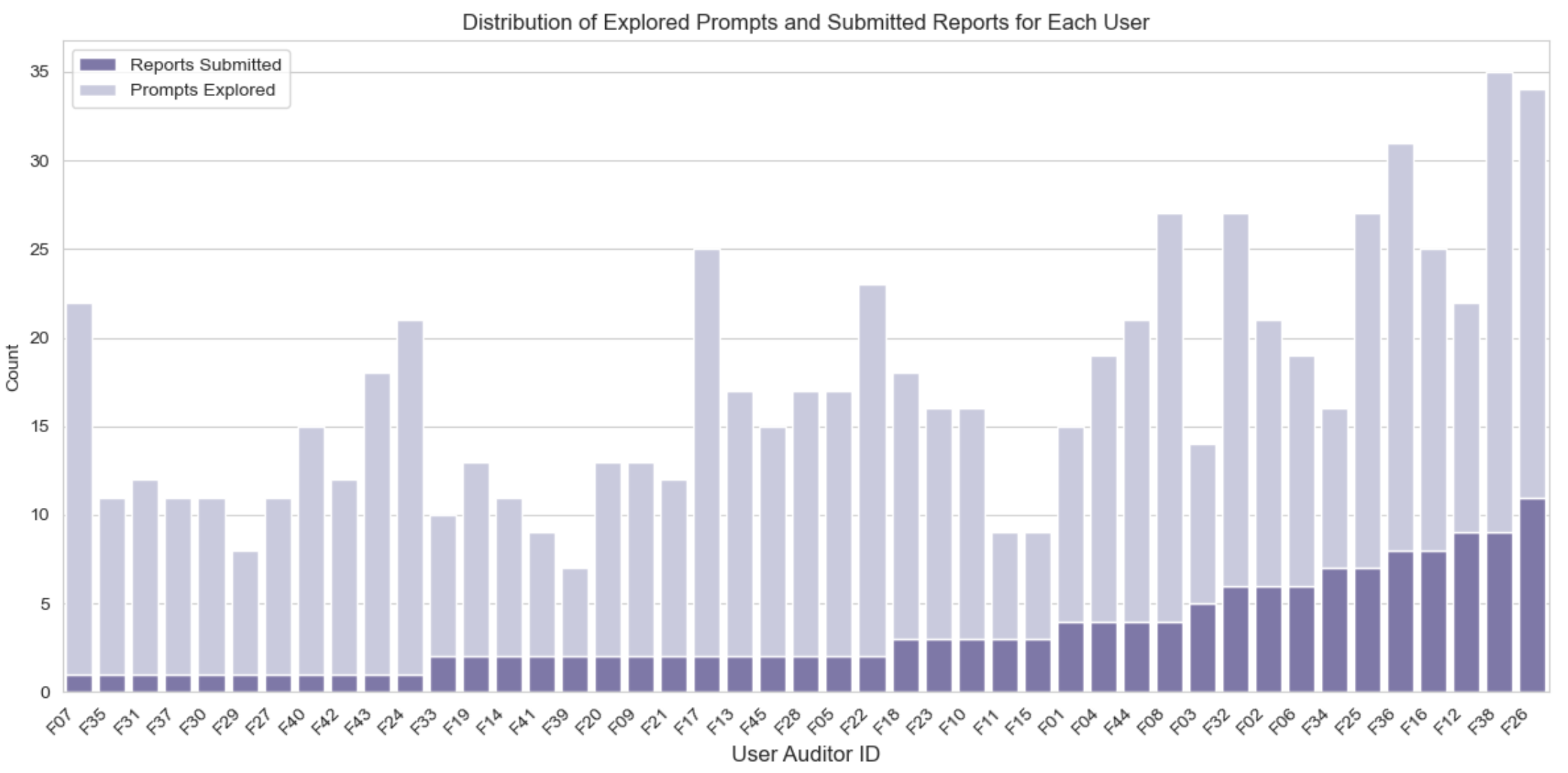}
  \caption{
   Bar chart showing the number of reports submitted (dark purple) stacked with the total number of prompts explored by user auditors (light purple) during the audit session.
}
  \Description{TBA}
  \label{fig:reports}
\end{figure*}

Practitioners raised questions on how to fairly compensate user auditors, given the ``invisible labor'' behind the submission of an audit report. We observed that significant effort sometimes went into generating a single report. On average, for every report submitted, user auditors explored 5 sets of prompts. However, as shown in Figure~\ref{fig:reports}, there was considerable variation across users. Seeing this, practitioner P14 said, \textit{``It would be unfair to compensate people only for the reports they submit if they spend a lot of time experimenting with different approaches and trying out prompts [...] But we also don’t want to pay people if they just scam the system by pretending they are exploring''} Many user auditors who submitted a low number of reports, relative to the number of prompts they explored, authored audit reports that were highlighted by practitioners as providing unique and valuable insights. These observations led most practitioners (8/10) in our study to reflect on how to define the ``productivity'' and ``skill'' of user auditors. P05, P09, and P11 suggested a need to design more comprehensive measures for assessing and compensating user auditors.

\subsection{\textbf{User Auditors Reported Increased Awareness and Understanding of AI Harms}} \label{result: auditor awareness}

Finally, among the 17 user auditors who responded to the follow-up survey four months after the study, 15 stated that participating in the session on auditing text-to-image models changed their awareness of ways generative AI might perpetuate biases or cause harm. 
These participants noted that while they started with a general awareness of AI harms and biases, \textbf{conducting AI audits with  \textit{WeAudit} helped them better understand what generative AI biases and harms actually look like in practice and the severity of the problem}. For instance, F45 wrote in the follow-up survey: \textit{``Before using [WeAudit...] I didn’t fully grasp [how] biases could manifest in generated content. However, being able to visually see and compare the biased images made it easier to critically evaluate AI models, and it heightened my awareness of these issues.’’}

Some users also shared that using WeAudit for auditing sessions expanded their views on the types of AI harms, especially those relevant to themselves. F07 shared that before participating, they only \textit{``thought about deep fakes or misinformation as harms from AI-generated images,’’} but realized \textit{``how much generative AI can erase minorities and reinforce stereotypes that can affect [themselves].’’} As someone who often uses Generative AI in their work, F44 suggested that \textit{``doing the audit makes [them] become more cautious about the biases in the models than before.. especially towards non-US people like [themselves],’’} as previously they were \textit{``mostly aware of biases on gender.’’} \looseness=-1

About half of the users who replied to the follow-up survey said they had since \textit{``tried to look for or examine potential biases/harms of AI systems in [their] daily life.’’} For example, F11, who uses generative AI systems in their graphic design work, mentioned that they are now more likely to scrutinize image outputs for potential racial and gender biases.

\section{Study Limitations} \label{limitation}

As mentioned in Section \ref{user auditor study}, while our formative study was conducted with a broader population of users, we chose to run our initial user study with college students. While this study provides insights into the usability and usefulness of the \textit{WeAudit} workflow, an important direction for future work is to explore how a broader population of user auditors would use the \textit{WeAudit} workflow and what additional forms of support and scaffolding they might need. Prior research in HCI, psychology, and AI has demonstrated that factors such as cultural background, technical expertise, and prior AI experience can significantly impact individuals’ sensitivity to and assessment of harmful AI biases and discrimination \cite{kingsley2024investigating, mun2024particip}. Therefore, future work should engage broader groups in evaluating the usefulness and usability of the \textit{WeAudit} system. Engaging a broader range of user auditors will help us understand how these diverse factors may influence auditors’ approaches to user-engaged AI audits. \looseness=-1

Another limitation of our study stems from the demographics and context of the industry practitioners involved. Using purposive and snowball sampling, we recruited GenAI practitioners, predominantly from the U.S. and large technology companies, with only one participant from the U.K. and two from startups (Section \ref{practitioner study}). Future research should engage practitioners from non-Western and smaller technology companies to better assess \textit{WeAudit}’s adaptability across varied AI development contexts. Additionally, our interviews required practitioners to envision incorporating \textit{WeAudit} into their workflows based on user audits conducted outside their direct work and customer contexts, which affects the real-world applicability of our findings. To address this, future research should examine how industry AI practitioners engage broader end-user groups within actual workflows, including ethnographic studies in industry settings where \textit{WeAudit} is integrated and evaluated by end users of the AI products and services \cite[cf.][]{passi2018trust}. \looseness=-1

\section{Discussion}
\label{discussion}

More than a decade ago, in CSCW 2013, Kittur et al. asked a provocative question in their position paper ``The Future of Crowd Work’’: \textit{``Can we envision a future of crowd work where we would like our children to participate?’’} \cite{kittur2013future}. Inspired by this, we pose a similar question: \textit{``Can we envision a future of user-engaged AI auditing where we would like our children (and ourselves) to participate?’’} As argued by Metaxa et al., the nature of algorithm auditing, especially when involving users, is rooted in activism \cite{metaxa2021auditing}. Past research shows that users often advocate for marginalized groups, expressing solidarity through their auditing work \cite{devos2022toward}. The \textit{“We”} in \textit{WeAudit} stands for individuals impacted by AI systems, particularly those from marginalized communities. Engaging users in auditing AI is only the beginning of the \textit{WeAudit} vision; we ultimately aim to empower these users to hold AI companies accountable. How can we, as a community, design a responsible future of user-engaged AI auditing?

In this section, we first discuss ways to ethically incentivize and sustain diverse participation in user-engaged AI auditing. We then outline how future work could cautiously incorporate AI technology into user-engaged AI auditing, based on our findings. Finally, we highlight potential asymmetric power dynamics between user auditors and AI practitioners, calling for further research and policy development. \looseness=-1

\subsection{Incentivizing and Sustaining Participation in User-Engaged AI Audits} \label{dis: incentivization}

Building on insights from our user study, in the next phase of our research, we plan to evaluate an improved version of \textit{WeAudit} with a broader population. We also plan to launch \textit{WeAudit} as a publicly available web platform for use by (1) AI practitioners and researchers interested in conducting user-engaged AI audits; (2) users interested in getting involved in user-engaged AI audits, either on a volunteer or paid basis; (3) researchers interested in conducting research on user-engaged AI auditing; and finally, (4) educators interested in using \textit{WeAudit} as an educational resource for enhancing AI literacy regarding the societal impact of AI systems (e.g., \cite{shen2021value, widder2024power}), inspired by our finding that user auditors in our study reported increased awareness and understanding of AI harms (Section \ref{result: auditor awareness}). \looseness=-1

As user-engaged AI audits become increasingly vital for identifying and addressing AI harms, motivating diverse and sustained participation remains a central challenge. Our findings indicate that the sense of community and identity that users feel within the platform , often refereed as ``intrinsic motivation'' in CSCW research, plays a significant role in motivating user auditors (Section \ref{result: worked examples}). Drawing on insights from prior CSCW work, initial recruitment for user-engaged AI audits can benefit from existing social groups  to leverage established user bases that have shared interests or identities \cite{kraut2011encouraging, burke2011social, lampe2010motivations}. To amplify these intrinsic motivations, future work could enhance the \textit{WeAudit} forum by allowing user auditors to share specific cases on social media, inviting others to also participate in AI auditing. By incorporating social features that enable users to advocate for AI improvements within their personal networks, \textit{WeAudit} could help user auditors to collectively push for changes~\cite{devos2022toward}. \looseness=-1

However, intrinsic motivations should not be the only drivers for users to engage in AI auditing. User auditors, especially those from marginalized communities \cite{pierre2021getting, li2023participation, deng2024responsible}, should receive extrinsic rewards through fair compensation from companies that benefit from their labor. Following calls from prior research on crowd workers and content moderators \cite{hara2018data, kittur2013future, dosono2019moderation, gray2019ghost}, companies should  consider fair compensation for user auditors’ contributions, especially the ``invisible labor'' that goes into conducting audits (see Section \ref{result: audit labor}). Policymakers can draw from content moderation research to design policies that encourage companies to establish protective mechanisms \cite{steiger2021psychological}, safeguarding user auditors from harmful exposure and fostering a sustainable environment for user engagement in AI audits.

\subsubsection{Enhancing Auditing through Diverse User Roles.}
Our findings indicate that some user auditors preferred the role of commenter over that of an auditor, choosing to provide feedback on others’ reports rather than conducting their own audits.
(Section \ref{results: engaged commenters}). Indeed, enabling a broader range of participation levels—what is often referred to as a ``low floor, high ceiling'' design—can engage diverse users who may have varying time and interests to dedicate to different parts of auditing \cite{reynante2021framework}. To this end, a user-engaged AI audit system should also accommodate different levels and types of involvement. For example, drawing on the concept of ``microtasks'' in crowdsourcing research \cite{dow2012shepherding}, \textit{WeAudit} could allow users to select specific sub-activities within the \textit{WeAudit} workflow according to their availability, interests, and specific domain or cultural expertise. By further breaking down the workflow into smaller, modular tasks—such as generating prompts, inspecting images, issuing reports, and discussing or verifying others’ reports—\textit{WeAudit} can cater to users who might prefer (or only have availability for) brief, targeted interactions rather than long-form audits. This is particularly important when \textit{WeAudit} and other future user auditing systems are deployed for broad public use, as past research shows that users often gravitate toward making specific kinds of contributions in auditing \cite{li2023participation}. \looseness=-1

\subsection{Designing Effective AI-in-the-Loop User-Engaged AI audits} \label{dis: AI for AI audit}

In line with the ``AIs Guiding Crowds'' research agenda outlined by Kittur et al. \cite{kittur2013future}, recent research has increasingly explored the use of emerging generative AI technologies to assist developers in testing AI systems, brainstorming, and creating incident cases \cite{buccinca2023aha, rastogi2023supporting, pang2023anticipating, park2022social, kieslich2024anticipating, pang2024blip, wang2024farsight}. In the current WeAudit workflow, we opted not to include AI-powered features, given that prior studies have highlighted the potential limitations and risks associated with using generative AI for tasks like AI auditing, red-teaming, and impact assessment \cite{buccinca2023aha, rastogi2023supporting, zaydi2024empowering, dev2023building, chiu2024culturalteaming, radharapu2023aart}. However, our findings point to several promising directions for \textit{thoughtfully} incorporating AI-powered features to enhance key activities in the user-engaged AI audit pipeline.


\subsubsection{Guiding user auditors’ investigation and deliberation through AI suggestions.} To start with, future interfaces could leverage AI to suggest possible directions for user auditors to explore (see Section \ref{results: broader explore}), supporting them in inspecting AI-generated output and formulating hypotheses about potential AI harms. As noted in Section \ref{result: worked examples}, user auditors may over-rely on existing examples provided by researchers and practitioners, potentially leading to an echo chamber effect \cite{cinelli2021echo}, even with ``social augmentation'' features. Future interface could leverage generative AI to \textbf{guide users in reflecting on their unique perspectives and intersectional identities} to explore previous under-explored topics, possibility tailored for AI developer team's need. In practice, this approach could recommend general audit topics \cite[c.f.][]{wang2024farsight, buccinca2023aha, liu2024selenite, radharapu2023aart}, or even specific prompts for testing \cite[c.f.][]{wu2019errudite, rastogi2023supporting, chiu2024culturalteaming}. \looseness=-1

\subsubsection{Enhancing large-scale sensemaking and visualization of AI behavior.} Current WeAudit features, such as the comparison function, help users inspect AI-generated output (see Section \ref{result: side-by-side}) but may fall short when dealing with large volumes of output \cite{gero2024supporting}. Moreover, text- and video-based outputs, compared to images, may require additional support to audit effectively. To address this, future work could explore visualization tools that support users in rapidly assessing large-scale AI outputs. Researchers can draw from the expanding body of work on sensemaking in large-scale AI outputs \cite[c.f.][]{gero2024supporting, jiang2023graphologue} to design features such as in-line highlights or clustering, which could help user auditors better\textbf{ comprehend the output space} and spot previously unnoticed details or \textbf{rapidly understand the output distribution shifts} when prompts are altered \cite[c.f.][]{chen2018anchorviz}.


\subsubsection{Integrating contextualized, in-situ user audits features within AI interface.} As practitioners repeatedly emphasized when envisioning incorporating WeAudit, user-engaged audits should ideally be integrated directly into AI products and services to better contextualize the use case (Section \ref{results}). One challenge in this integration lies in \textbf{determining the optimal timing and context} for nudging users to provide audit feedback during regular tasks. To this end, future work could build on recent advancements in generative AI to develop personalized, context-aware task routing for user-engaged AI audits, drawing from prior CSCW research on ``intelligent task routing'' \cite{cosley2007suggestbot}. This would also benefit sustaining user engagement (see Section \ref{dis: incentivization}) by automatically providing users auditors tasks they are most interested in.


\subsection{Minding Power Asymmetries between User Auditors and AI Practitioners} \label{dis: empowerment}

At a high level, \textit{WeAudit} aimed to facilitate collaboration between user auditors and AI practitioners, seeking for meaningful improvement of AI system impacting user auditors' lives. However, systemic questions remain beyond tool and process design. Practitioners, both in prior research and our study, noted that profit-driven goals often hinder responsible AI practices, even when practitioners are genuinely committed to marginalized communities. In addition,  practitioners' motivations on addressing harms that elicited the most ``surprise'' from other users (see Section \ref{results: insights of discussion}) appears to stems from a desire to \textit{``avoid bad public relations,''} centering the interests of the company rather than a genuine commitment to those affected marginalized community members  \cite{madaio2021assessing, deng2023understanding, devrio2024building, young2019toward, miceli2021wisdom}. Furthermore, as Deng et al. highlighted, while system like \textit{WeAudit} can scaffold the potential meaningful collaboration between AI users and developers, but it also \textit{``firmly place the choice to take action with practitioners, potentially leaving users with less room for leverage via other means.’’} \cite{deng2023understanding}. To this end, future work could draw inspirations from successful prior cases done by HCI researchers, such as Turkopticon \cite{irani2013turkopticon} and WeAreDynamo \cite{salehi2015we}, to expand \textit{WeAudit} for users' collective action that can be \textit{independent} from practitioners, when only being ``engaged'' by AI practitioners is insufficient due to the asymmetric power dynamic between end users and AI practitioners. Activists and policymakers can also draw from recent frameworks developed by HCI researchers, such as ``Data Leverage,'' to shift power from technology companies to the public \cite{vincent2021data}. Finally, policymakers should design future policies to provide ``safe harbor'' for third party organizations, external domain experts, and end users, to audit AI systems developed by companies \cite{longpre2024safe}. \looseness=-1

\section{Conclusion}

In this work, we have presented a set of empirically-informed design goals for user-engaged AI auditing processes and tools. We have presented the \textit{WeAudit} workflow and system. Through a user study and interviews with both user auditors and industry AI practitioners, our findings shed light on design guidelines and future opportunities to support meaningful user engagement for generative AI and beyond.

\begin{acks}
We thank all participating industry practitioners and users for making this work possible.
\end{acks}

\bibliographystyle{ACM-Reference-Format}
\bibliography{citation}


\begin{thebibliography}{128}


\ifx \showCODEN    \undefined \def \showCODEN     #1{\unskip}     \fi
\ifx \showDOI      \undefined \def \showDOI       #1{#1}\fi
\ifx \showISBNx    \undefined \def \showISBNx     #1{\unskip}     \fi
\ifx \showISBNxiii \undefined \def \showISBNxiii  #1{\unskip}     \fi
\ifx \showISSN     \undefined \def \showISSN      #1{\unskip}     \fi
\ifx \showLCCN     \undefined \def \showLCCN      #1{\unskip}     \fi
\ifx \shownote     \undefined \def \shownote      #1{#1}          \fi
\ifx \showarticletitle \undefined \def \showarticletitle #1{#1}   \fi
\ifx \showURL      \undefined \def \showURL       {\relax}        \fi
\providecommand\bibfield[2]{#2}
\providecommand\bibinfo[2]{#2}
\providecommand\natexlab[1]{#1}
\providecommand\showeprint[2][]{arXiv:#2}

\bibitem[AI(2022)]%
        {ChatGPT_Feedback}
\bibfield{author}{\bibinfo{person}{Open AI}.} \bibinfo{year}{2022}\natexlab{}.
\newblock \bibinfo{title}{ChatGPT Feedback Contest: Official Rules}.
\newblock
\newblock
\urldef\tempurl%
\url{https://cdn.openai.com/chatgpt/ChatGPT_Feedback_Contest_Rules.pdf}
\showURL{%
\tempurl}


\bibitem[Amershi and Morris(2008)]%
        {amershi2008cosearch}
\bibfield{author}{\bibinfo{person}{Saleema Amershi} {and} \bibinfo{person}{Meredith~Ringel Morris}.} \bibinfo{year}{2008}\natexlab{}.
\newblock \showarticletitle{CoSearch: a system for co-located collaborative web search}. In \bibinfo{booktitle}{\emph{Proceedings of the SIGCHI conference on human factors in computing systems}}. \bibinfo{pages}{1647--1656}.
\newblock


\bibitem[Anthropic(2023)]%
        {anthropic2023claude}
\bibfield{author}{\bibinfo{person}{Anthropic}.} \bibinfo{year}{2023}\natexlab{}.
\newblock \bibinfo{title}{Model card and evaluations for Claude Models}.
\newblock
\newblock
\urldef\tempurl%
\url{https://www-cdn.anthropic.com/bd2a28d2535bfb0494cc8e2a3bf135d2e7523226/Model-Card-Claude-2.pdf}
\showURL{%
\tempurl}


\bibitem[Asplund et~al\mbox{.}(2020)]%
        {asplund2020auditing}
\bibfield{author}{\bibinfo{person}{Joshua Asplund}, \bibinfo{person}{Motahhare Eslami}, \bibinfo{person}{Hari Sundaram}, \bibinfo{person}{Christian Sandvig}, {and} \bibinfo{person}{Karrie Karahalios}.} \bibinfo{year}{2020}\natexlab{}.
\newblock \showarticletitle{Auditing race and gender discrimination in online housing markets}. In \bibinfo{booktitle}{\emph{Proceedings of the International AAAI Conference on Web and Social Media}}, Vol.~\bibinfo{volume}{14}. \bibinfo{pages}{24--35}.
\newblock


\bibitem[Attenberg et~al\mbox{.}(2015)]%
        {attenberg2015beat}
\bibfield{author}{\bibinfo{person}{Joshua Attenberg}, \bibinfo{person}{Panos Ipeirotis}, {and} \bibinfo{person}{Foster Provost}.} \bibinfo{year}{2015}\natexlab{}.
\newblock \showarticletitle{Beat the machine: Challenging humans to find a predictive model's “unknown unknowns”}.
\newblock \bibinfo{journal}{\emph{Journal of Data and Information Quality (JDIQ)}} \bibinfo{volume}{6}, \bibinfo{number}{1} (\bibinfo{year}{2015}), \bibinfo{pages}{1--17}.
\newblock


\bibitem[Bandy(2021)]%
        {bandy2021problematic}
\bibfield{author}{\bibinfo{person}{Jack Bandy}.} \bibinfo{year}{2021}\natexlab{}.
\newblock \showarticletitle{Problematic machine behavior: A systematic literature review of algorithm audits}.
\newblock \bibinfo{journal}{\emph{Proceedings of the acm on human-computer interaction}} \bibinfo{volume}{5}, \bibinfo{number}{CSCW1} (\bibinfo{year}{2021}), \bibinfo{pages}{1--34}.
\newblock


\bibitem[Bernstein et~al\mbox{.}(2010)]%
        {bernstein2010soylent}
\bibfield{author}{\bibinfo{person}{Michael~S Bernstein}, \bibinfo{person}{Greg Little}, \bibinfo{person}{Robert~C Miller}, \bibinfo{person}{Bj{\"o}rn Hartmann}, \bibinfo{person}{Mark~S Ackerman}, \bibinfo{person}{David~R Karger}, \bibinfo{person}{David Crowell}, {and} \bibinfo{person}{Katrina Panovich}.} \bibinfo{year}{2010}\natexlab{}.
\newblock \showarticletitle{Soylent: a word processor with a crowd inside}. In \bibinfo{booktitle}{\emph{Proceedings of the 23nd annual ACM symposium on User interface software and technology}}. \bibinfo{pages}{313--322}.
\newblock


\bibitem[Bianchi et~al\mbox{.}(2023)]%
        {bianchi2023easily}
\bibfield{author}{\bibinfo{person}{Federico Bianchi}, \bibinfo{person}{Pratyusha Kalluri}, \bibinfo{person}{Esin Durmus}, \bibinfo{person}{Faisal Ladhak}, \bibinfo{person}{Myra Cheng}, \bibinfo{person}{Debora Nozza}, \bibinfo{person}{Tatsunori Hashimoto}, \bibinfo{person}{Dan Jurafsky}, \bibinfo{person}{James Zou}, {and} \bibinfo{person}{Aylin Caliskan}.} \bibinfo{year}{2023}\natexlab{}.
\newblock \showarticletitle{Easily accessible text-to-image generation amplifies demographic stereotypes at large scale}. In \bibinfo{booktitle}{\emph{Proceedings of the 2023 ACM Conference on Fairness, Accountability, and Transparency}}. \bibinfo{pages}{1493--1504}.
\newblock


\bibitem[Bigham et~al\mbox{.}(2015)]%
        {bigham2015human}
\bibfield{author}{\bibinfo{person}{Jeffrey~P Bigham}, \bibinfo{person}{Michael~S Bernstein}, {and} \bibinfo{person}{Eytan Adar}.} \bibinfo{year}{2015}\natexlab{}.
\newblock \showarticletitle{Human-computer interaction and collective intelligence}.
\newblock \bibinfo{journal}{\emph{Handbook of collective intelligence}}  \bibinfo{volume}{57} (\bibinfo{year}{2015}).
\newblock


\bibitem[Birhane et~al\mbox{.}(2024)]%
        {birhane2024ai}
\bibfield{author}{\bibinfo{person}{Abeba Birhane}, \bibinfo{person}{Ryan Steed}, \bibinfo{person}{Victor Ojewale}, \bibinfo{person}{Briana Vecchione}, {and} \bibinfo{person}{Inioluwa~Deborah Raji}.} \bibinfo{year}{2024}\natexlab{}.
\newblock \showarticletitle{AI auditing: The broken bus on the road to AI accountability}. In \bibinfo{booktitle}{\emph{2024 IEEE Conference on Secure and Trustworthy Machine Learning (SaTML)}}. IEEE, \bibinfo{pages}{612--643}.
\newblock


\bibitem[Bogen and Winecoff({[n.\,d.]})]%
        {bogen2024sociotechnical}
\bibfield{author}{\bibinfo{person}{Miranda Bogen} {and} \bibinfo{person}{Amy~A Winecoff}.} \bibinfo{year}{[n.\,d.]}\natexlab{}.
\newblock
\newblock
\urldef\tempurl%
\url{https://cdt.org/insights/applying-sociotechnical-approaches-to-ai-governance-in-practice/}
\showURL{%
\tempurl}


\bibitem[Braun and Clarke(2019)]%
        {braun2019reflecting}
\bibfield{author}{\bibinfo{person}{Virginia Braun} {and} \bibinfo{person}{Victoria Clarke}.} \bibinfo{year}{2019}\natexlab{}.
\newblock \showarticletitle{Reflecting on reflexive thematic analysis}.
\newblock \bibinfo{journal}{\emph{Qualitative research in sport, exercise and health}} \bibinfo{volume}{11}, \bibinfo{number}{4} (\bibinfo{year}{2019}), \bibinfo{pages}{589--597}.
\newblock


\bibitem[Bu{\c{c}}inca et~al\mbox{.}(2023)]%
        {buccinca2023aha}
\bibfield{author}{\bibinfo{person}{Zana Bu{\c{c}}inca}, \bibinfo{person}{Chau~Minh Pham}, \bibinfo{person}{Maurice Jakesch}, \bibinfo{person}{Marco~Tulio Ribeiro}, \bibinfo{person}{Alexandra Olteanu}, {and} \bibinfo{person}{Saleema Amershi}.} \bibinfo{year}{2023}\natexlab{}.
\newblock \showarticletitle{AHA!: Facilitating AI Impact Assessment by Generating Examples of Harms}.
\newblock \bibinfo{journal}{\emph{arXiv preprint arXiv:2306.03280}} (\bibinfo{year}{2023}).
\newblock


\bibitem[Buolamwini and Gebru(2018)]%
        {buolamwini2018gender}
\bibfield{author}{\bibinfo{person}{Joy Buolamwini} {and} \bibinfo{person}{Timnit Gebru}.} \bibinfo{year}{2018}\natexlab{}.
\newblock \showarticletitle{Gender shades: Intersectional accuracy disparities in commercial gender classification}. In \bibinfo{booktitle}{\emph{Conference on fairness, accountability and transparency}}. PMLR, \bibinfo{pages}{77--91}.
\newblock


\bibitem[Burke et~al\mbox{.}(2011)]%
        {burke2011social}
\bibfield{author}{\bibinfo{person}{Moira Burke}, \bibinfo{person}{Robert Kraut}, {and} \bibinfo{person}{Cameron Marlow}.} \bibinfo{year}{2011}\natexlab{}.
\newblock \showarticletitle{Social capital on Facebook: Differentiating uses and users}. In \bibinfo{booktitle}{\emph{Proceedings of the SIGCHI conference on human factors in computing systems}}. \bibinfo{pages}{571--580}.
\newblock


\bibitem[Cabrera et~al\mbox{.}(2021)]%
        {cabrera2021discovering}
\bibfield{author}{\bibinfo{person}{{\'A}ngel~Alexander Cabrera}, \bibinfo{person}{Abraham~J Druck}, \bibinfo{person}{Jason~I Hong}, {and} \bibinfo{person}{Adam Perer}.} \bibinfo{year}{2021}\natexlab{}.
\newblock \showarticletitle{Discovering and validating ai errors with crowdsourced failure reports}.
\newblock \bibinfo{journal}{\emph{Proceedings of the ACM on Human-Computer Interaction}} \bibinfo{volume}{5}, \bibinfo{number}{CSCW2} (\bibinfo{year}{2021}), \bibinfo{pages}{1--22}.
\newblock


\bibitem[Cabrera et~al\mbox{.}(2022)]%
        {cabrera2022did}
\bibfield{author}{\bibinfo{person}{{\'A}ngel~Alexander Cabrera}, \bibinfo{person}{Marco~Tulio Ribeiro}, \bibinfo{person}{Bongshin Lee}, \bibinfo{person}{Rob DeLine}, \bibinfo{person}{Adam Perer}, {and} \bibinfo{person}{Steven~M Drucker}.} \bibinfo{year}{2022}\natexlab{}.
\newblock \showarticletitle{What Did My AI Learn? How Data Scientists Make Sense of Model Behavior}.
\newblock \bibinfo{journal}{\emph{ACM Transactions on Computer-Human Interaction}} (\bibinfo{year}{2022}).
\newblock


\bibitem[Campbell et~al\mbox{.}(2020)]%
        {campbell2020purposive}
\bibfield{author}{\bibinfo{person}{Steve Campbell}, \bibinfo{person}{Melanie Greenwood}, \bibinfo{person}{Sarah Prior}, \bibinfo{person}{Toniele Shearer}, \bibinfo{person}{Kerrie Walkem}, \bibinfo{person}{Sarah Young}, \bibinfo{person}{Danielle Bywaters}, {and} \bibinfo{person}{Kim Walker}.} \bibinfo{year}{2020}\natexlab{}.
\newblock \showarticletitle{Purposive sampling: complex or simple? Research case examples}.
\newblock \bibinfo{journal}{\emph{Journal of research in Nursing}} \bibinfo{volume}{25}, \bibinfo{number}{8} (\bibinfo{year}{2020}), \bibinfo{pages}{652--661}.
\newblock


\bibitem[Chan et~al\mbox{.}(2018)]%
        {chan2018solvent}
\bibfield{author}{\bibinfo{person}{Joel Chan}, \bibinfo{person}{Joseph~Chee Chang}, \bibinfo{person}{Tom Hope}, \bibinfo{person}{Dafna Shahaf}, {and} \bibinfo{person}{Aniket Kittur}.} \bibinfo{year}{2018}\natexlab{}.
\newblock \showarticletitle{Solvent: A mixed initiative system for finding analogies between research papers}.
\newblock \bibinfo{journal}{\emph{Proceedings of the ACM on Human-Computer Interaction}} \bibinfo{volume}{2}, \bibinfo{number}{CSCW} (\bibinfo{year}{2018}), \bibinfo{pages}{1--21}.
\newblock


\bibitem[Chen et~al\mbox{.}(2018)]%
        {chen2018anchorviz}
\bibfield{author}{\bibinfo{person}{Nan-Chen Chen}, \bibinfo{person}{Jina Suh}, \bibinfo{person}{Johan Verwey}, \bibinfo{person}{Gonzalo Ramos}, \bibinfo{person}{Steven Drucker}, {and} \bibinfo{person}{Patrice Simard}.} \bibinfo{year}{2018}\natexlab{}.
\newblock \showarticletitle{AnchorViz: Facilitating classifier error discovery through interactive semantic data exploration}. In \bibinfo{booktitle}{\emph{Proceedings of the 23rd International Conference on Intelligent User Interfaces}}. \bibinfo{pages}{269--280}.
\newblock


\bibitem[Chen and Zhang(2023)]%
        {chen2023judgment}
\bibfield{author}{\bibinfo{person}{Quan~Ze Chen} {and} \bibinfo{person}{Amy~X Zhang}.} \bibinfo{year}{2023}\natexlab{}.
\newblock \showarticletitle{Judgment Sieve: Reducing uncertainty in group judgments through interventions targeting ambiguity versus disagreement}.
\newblock \bibinfo{journal}{\emph{Proceedings of the ACM on Human-Computer Interaction}} \bibinfo{volume}{7}, \bibinfo{number}{CSCW2} (\bibinfo{year}{2023}), \bibinfo{pages}{1--26}.
\newblock


\bibitem[Chilton et~al\mbox{.}(2013)]%
        {chilton2013cascade}
\bibfield{author}{\bibinfo{person}{Lydia~B Chilton}, \bibinfo{person}{Greg Little}, \bibinfo{person}{Darren Edge}, \bibinfo{person}{Daniel~S Weld}, {and} \bibinfo{person}{James~A Landay}.} \bibinfo{year}{2013}\natexlab{}.
\newblock \showarticletitle{Cascade: Crowdsourcing taxonomy creation}. In \bibinfo{booktitle}{\emph{Proceedings of the SIGCHI Conference on Human Factors in Computing Systems}}. \bibinfo{pages}{1999--2008}.
\newblock


\bibitem[Chiu et~al\mbox{.}(2024)]%
        {chiu2024culturalteaming}
\bibfield{author}{\bibinfo{person}{Yu~Ying Chiu}, \bibinfo{person}{Liwei Jiang}, \bibinfo{person}{Maria Antoniak}, \bibinfo{person}{Chan~Young Park}, \bibinfo{person}{Shuyue~Stella Li}, \bibinfo{person}{Mehar Bhatia}, \bibinfo{person}{Sahithya Ravi}, \bibinfo{person}{Yulia Tsvetkov}, \bibinfo{person}{Vered Shwartz}, {and} \bibinfo{person}{Yejin Choi}.} \bibinfo{year}{2024}\natexlab{}.
\newblock \showarticletitle{CulturalTeaming: AI-Assisted Interactive Red-Teaming for Challenging LLMs'(Lack of) Multicultural Knowledge}.
\newblock \bibinfo{journal}{\emph{arXiv preprint arXiv:2404.06664}} (\bibinfo{year}{2024}).
\newblock


\bibitem[Chowdhury and Williams(2021)]%
        {chowdhury2021introducing}
\bibfield{author}{\bibinfo{person}{Rumman Chowdhury} {and} \bibinfo{person}{Jutta Williams}.} \bibinfo{year}{2021}\natexlab{}.
\newblock \showarticletitle{Introducing Twitter’s first algorithmic bias bounty challenge}.
\newblock \bibinfo{journal}{\emph{URl: https://blog. twitter. com/engineering/en\_us/topics/insights/2021/algorithmic-bias-bountychallenge}} (\bibinfo{year}{2021}).
\newblock


\bibitem[Chung et~al\mbox{.}(2019)]%
        {chung2019efficient}
\bibfield{author}{\bibinfo{person}{John Joon~Young Chung}, \bibinfo{person}{Jean~Y Song}, \bibinfo{person}{Sindhu Kutty}, \bibinfo{person}{Sungsoo Hong}, \bibinfo{person}{Juho Kim}, {and} \bibinfo{person}{Walter~S Lasecki}.} \bibinfo{year}{2019}\natexlab{}.
\newblock \showarticletitle{Efficient elicitation approaches to estimate collective crowd answers}.
\newblock \bibinfo{journal}{\emph{Proceedings of the ACM on Human-Computer Interaction}} \bibinfo{volume}{3}, \bibinfo{number}{CSCW} (\bibinfo{year}{2019}), \bibinfo{pages}{1--25}.
\newblock


\bibitem[Cinelli et~al\mbox{.}(2021)]%
        {cinelli2021echo}
\bibfield{author}{\bibinfo{person}{Matteo Cinelli}, \bibinfo{person}{Gianmarco De~Francisci~Morales}, \bibinfo{person}{Alessandro Galeazzi}, \bibinfo{person}{Walter Quattrociocchi}, {and} \bibinfo{person}{Michele Starnini}.} \bibinfo{year}{2021}\natexlab{}.
\newblock \showarticletitle{The echo chamber effect on social media}.
\newblock \bibinfo{journal}{\emph{Proceedings of the National Academy of Sciences}} \bibinfo{volume}{118}, \bibinfo{number}{9} (\bibinfo{year}{2021}), \bibinfo{pages}{e2023301118}.
\newblock


\bibitem[Claire et~al\mbox{.}(2024)]%
        {claire2024designing}
\bibfield{author}{\bibinfo{person}{Wang Claire}, \bibinfo{person}{Wesley~Hanwen Deng}, \bibinfo{person}{Jason Hong}, \bibinfo{person}{Ken Holstein}, {and} \bibinfo{person}{Motahhare Eslami}.} \bibinfo{year}{2024}\natexlab{}.
\newblock \showarticletitle{Designing a Crowdsourcing Pipeline to Verify Reports from User AI Audits}.
\newblock \bibinfo{journal}{\emph{Work in Progress of the AAAI Conference on Human Computation and Crowdsourcing}} (\bibinfo{year}{2024}).
\newblock


\bibitem[CON(2024)]%
        {DEFCON_AI_Village}
\bibfield{author}{\bibinfo{person}{DEF CON}.} \bibinfo{year}{2024}\natexlab{}.
\newblock \bibinfo{title}{Red Team Village}.
\newblock
\newblock
\urldef\tempurl%
\url{https://redteamvillage.io/}
\showURL{%
\tempurl}


\bibitem[Cosley et~al\mbox{.}(2007)]%
        {cosley2007suggestbot}
\bibfield{author}{\bibinfo{person}{Dan Cosley}, \bibinfo{person}{Dan Frankowski}, \bibinfo{person}{Loren Terveen}, {and} \bibinfo{person}{John Riedl}.} \bibinfo{year}{2007}\natexlab{}.
\newblock \showarticletitle{SuggestBot: using intelligent task routing to help people find work in wikipedia}. In \bibinfo{booktitle}{\emph{Proceedings of the 12th international conference on Intelligent user interfaces}}. \bibinfo{pages}{32--41}.
\newblock


\bibitem[Cramer et~al\mbox{.}(2018)]%
        {cramer2018assessing}
\bibfield{author}{\bibinfo{person}{Henriette Cramer}, \bibinfo{person}{Jean Garcia-Gathright}, \bibinfo{person}{Aaron Springer}, {and} \bibinfo{person}{Sravana Reddy}.} \bibinfo{year}{2018}\natexlab{}.
\newblock \showarticletitle{Assessing and addressing algorithmic bias in practice}.
\newblock \bibinfo{journal}{\emph{Interactions}} \bibinfo{volume}{25}, \bibinfo{number}{6} (\bibinfo{year}{2018}), \bibinfo{pages}{58--63}.
\newblock


\bibitem[Deng et~al\mbox{.}(2024a)]%
        {deng2024supporting}
\bibfield{author}{\bibinfo{person}{Wesley~Hanwen Deng}, \bibinfo{person}{Solon Barocas}, {and} \bibinfo{person}{Jennifer~Wortman Vaughan}.} \bibinfo{year}{2024}\natexlab{a}.
\newblock \showarticletitle{Supporting Industry Computing Researchers in Assessing, Articulating, and Addressing the Potential Negative Societal Impact of Their Work}.
\newblock \bibinfo{journal}{\emph{arXiv preprint arXiv:2408.01057}} (\bibinfo{year}{2024}).
\newblock


\bibitem[Deng et~al\mbox{.}(2023a)]%
        {deng2023understanding}
\bibfield{author}{\bibinfo{person}{Wesley~Hanwen Deng}, \bibinfo{person}{Boyuan Guo}, \bibinfo{person}{Alicia Devrio}, \bibinfo{person}{Hong Shen}, \bibinfo{person}{Motahhare Eslami}, {and} \bibinfo{person}{Kenneth Holstein}.} \bibinfo{year}{2023}\natexlab{a}.
\newblock \showarticletitle{Understanding Practices, Challenges, and Opportunities for User-Engaged Algorithm Auditing in Industry Practice}. In \bibinfo{booktitle}{\emph{Proceedings of the 2023 CHI Conference on Human Factors in Computing Systems}}. \bibinfo{pages}{1--18}.
\newblock


\bibitem[Deng et~al\mbox{.}(2023b)]%
        {deng2023supporting}
\bibfield{author}{\bibinfo{person}{Wesley~Hanwen Deng}, \bibinfo{person}{Michelle~S Lam}, \bibinfo{person}{{\'A}ngel~Alexander Cabrera}, \bibinfo{person}{Dana{\"e} Metaxa}, \bibinfo{person}{Motahhare Eslami}, {and} \bibinfo{person}{Kenneth Holstein}.} \bibinfo{year}{2023}\natexlab{b}.
\newblock \showarticletitle{Supporting user engagement in testing, auditing, and contesting AI}. In \bibinfo{booktitle}{\emph{Companion Publication of the 2023 Conference on Computer Supported Cooperative Work and Social Computing}}. \bibinfo{pages}{556--559}.
\newblock


\bibitem[Deng et~al\mbox{.}(2022)]%
        {deng2022exploring}
\bibfield{author}{\bibinfo{person}{Wesley~Hanwen Deng}, \bibinfo{person}{Manish Nagireddy}, \bibinfo{person}{Michelle Seng~Ah Lee}, \bibinfo{person}{Jatinder Singh}, \bibinfo{person}{Zhiwei~Steven Wu}, \bibinfo{person}{Kenneth Holstein}, {and} \bibinfo{person}{Haiyi Zhu}.} \bibinfo{year}{2022}\natexlab{}.
\newblock \showarticletitle{Exploring {How} {Machine} {Learning} {Practitioners} ({Try} {To}) {Use} {Fairness} {Toolkits}}. In \bibinfo{booktitle}{\emph{2022 {ACM} {Conference} on {Fairness}, {Accountability}, and {Transparency}}}. \bibinfo{publisher}{ACM}, \bibinfo{address}{Seoul Republic of Korea}, \bibinfo{pages}{473--484}.
\newblock
\showISBNx{978-1-4503-9352-2}
\urldef\tempurl%
\url{https://doi.org/10.1145/3531146.3533113}
\showDOI{\tempurl}


\bibitem[Deng et~al\mbox{.}(2024b)]%
        {deng2024responsible}
\bibfield{author}{\bibinfo{person}{Wesley~Hanwen Deng}, \bibinfo{person}{Mireia Yurrita}, \bibinfo{person}{Mark D{\'\i}az}, \bibinfo{person}{Jina Suh}, \bibinfo{person}{Nick Judd}, \bibinfo{person}{Lara Groves}, \bibinfo{person}{Hong Shen}, \bibinfo{person}{Motahhare Eslami}, {and} \bibinfo{person}{Kenneth Holstein}.} \bibinfo{year}{2024}\natexlab{b}.
\newblock \showarticletitle{Responsible Crowdsourcing for Responsible Generative AI: Engaging Crowds in AI Auditing and Evaluation}. In \bibinfo{booktitle}{\emph{Proceedings of the AAAI Conference on Human Computation and Crowdsourcing}}, Vol.~\bibinfo{volume}{12}. \bibinfo{pages}{148--150}.
\newblock


\bibitem[Dev et~al\mbox{.}(2023)]%
        {dev2023building}
\bibfield{author}{\bibinfo{person}{Sunipa Dev}, \bibinfo{person}{Akshita Jha}, \bibinfo{person}{Jaya Goyal}, \bibinfo{person}{Dinesh Tewari}, \bibinfo{person}{Shachi Dave}, {and} \bibinfo{person}{Vinodkumar Prabhakaran}.} \bibinfo{year}{2023}\natexlab{}.
\newblock \showarticletitle{Building stereotype repositories with llms and community engagement for scale and depth}.
\newblock \bibinfo{journal}{\emph{Cross-Cultural Considerations in NLP@ EACL}}  \bibinfo{volume}{84} (\bibinfo{year}{2023}).
\newblock


\bibitem[DeVos et~al\mbox{.}(2022)]%
        {devos2022toward}
\bibfield{author}{\bibinfo{person}{Alicia DeVos}, \bibinfo{person}{Aditi Dhabalia}, \bibinfo{person}{Hong Shen}, \bibinfo{person}{Kenneth Holstein}, {and} \bibinfo{person}{Motahhare Eslami}.} \bibinfo{year}{2022}\natexlab{}.
\newblock \showarticletitle{Toward User-Driven Algorithm Auditing: Investigating users’ strategies for uncovering harmful algorithmic behavior}. In \bibinfo{booktitle}{\emph{Proceedings of the 2022 CHI conference on human factors in computing systems}}. \bibinfo{pages}{1--19}.
\newblock


\bibitem[DeVrio et~al\mbox{.}(2024)]%
        {devrio2024building}
\bibfield{author}{\bibinfo{person}{Alicia DeVrio}, \bibinfo{person}{Motahhare Eslami}, {and} \bibinfo{person}{Kenneth Holstein}.} \bibinfo{year}{2024}\natexlab{}.
\newblock \showarticletitle{Building, Shifting, \& Employing Power: A Taxonomy of Responses From Below to Algorithmic Harm}. In \bibinfo{booktitle}{\emph{The 2024 ACM Conference on Fairness, Accountability, and Transparency}}. \bibinfo{pages}{1093--1106}.
\newblock


\bibitem[Dosono and Semaan(2019)]%
        {dosono2019moderation}
\bibfield{author}{\bibinfo{person}{Bryan Dosono} {and} \bibinfo{person}{Bryan Semaan}.} \bibinfo{year}{2019}\natexlab{}.
\newblock \showarticletitle{Moderation practices as emotional labor in sustaining online communities: The case of AAPI identity work on Reddit}. In \bibinfo{booktitle}{\emph{Proceedings of the 2019 CHI conference on human factors in computing systems}}. \bibinfo{pages}{1--13}.
\newblock


\bibitem[Dourish et~al\mbox{.}(2020)]%
        {dourish2020being}
\bibfield{author}{\bibinfo{person}{Paul Dourish}, \bibinfo{person}{Christopher Lawrence}, \bibinfo{person}{Tuck~Wah Leong}, {and} \bibinfo{person}{Greg Wadley}.} \bibinfo{year}{2020}\natexlab{}.
\newblock \showarticletitle{On being iterated: The affective demands of design participation}. In \bibinfo{booktitle}{\emph{Proceedings of the 2020 CHI Conference on Human Factors in Computing Systems}}. \bibinfo{pages}{1--11}.
\newblock


\bibitem[Dow et~al\mbox{.}(2012)]%
        {dow2012shepherding}
\bibfield{author}{\bibinfo{person}{Steven Dow}, \bibinfo{person}{Anand Kulkarni}, \bibinfo{person}{Scott Klemmer}, {and} \bibinfo{person}{Bj{\"o}rn Hartmann}.} \bibinfo{year}{2012}\natexlab{}.
\newblock \showarticletitle{Shepherding the crowd yields better work}. In \bibinfo{booktitle}{\emph{Proceedings of the ACM 2012 conference on computer supported cooperative work}}. \bibinfo{pages}{1013--1022}.
\newblock


\bibitem[Eslami et~al\mbox{.}(2017)]%
        {eslami2017biased}
\bibfield{author}{\bibinfo{person}{Motahhare Eslami}, \bibinfo{person}{Kristen Vaccaro}, \bibinfo{person}{Karrie Karahalios}, {and} \bibinfo{person}{Kevin Hamilton}.} \bibinfo{year}{2017}\natexlab{}.
\newblock \showarticletitle{Be careful; Things can be worse than they appear - Understanding biased algorithms and users' behavior around them in rating platforms"}.
\newblock  (\bibinfo{year}{2017}), \bibinfo{pages}{62--71}.
\newblock
\newblock
\shownote{Funding Information: This work was funded by NSF grant CHS-1564041. Publisher Copyright: {\textcopyright} Copyright 2017, Association for the Advancement of Artificial Intelligence (www.aaai.org). All rights reserved.; 11th International Conference on Web and Social Media, ICWSM 2017 ; Conference date: 15-05-2017 Through 18-05-2017}.


\bibitem[Eslami et~al\mbox{.}(2019)]%
        {eslami2019opacity}
\bibfield{author}{\bibinfo{person}{Motahhare Eslami}, \bibinfo{person}{Kristen Vaccaro}, \bibinfo{person}{Min~Kyung Lee}, \bibinfo{person}{Amit Elazari Bar~On}, \bibinfo{person}{Eric Gilbert}, {and} \bibinfo{person}{Karrie Karahalios}.} \bibinfo{year}{2019}\natexlab{}.
\newblock \showarticletitle{User Attitudes towards Algorithmic Opacity and Transparency in Online Reviewing Platforms}.
\newblock  (\bibinfo{year}{2019}), \bibinfo{pages}{1–14}.
\newblock
\showISBNx{9781450359702}
\urldef\tempurl%
\url{https://doi.org/10.1145/3290605.3300724}
\showURL{%
\tempurl}


\bibitem[Feffer et~al\mbox{.}(2024)]%
        {feffer2024red}
\bibfield{author}{\bibinfo{person}{Michael Feffer}, \bibinfo{person}{Anusha Sinha}, \bibinfo{person}{Wesley~H Deng}, \bibinfo{person}{Zachary~C Lipton}, {and} \bibinfo{person}{Hoda Heidari}.} \bibinfo{year}{2024}\natexlab{}.
\newblock \showarticletitle{Red-teaming for generative ai: Silver bullet or security theater?}. In \bibinfo{booktitle}{\emph{Proceedings of the AAAI/ACM Conference on AI, Ethics, and Society}}, Vol.~\bibinfo{volume}{7}. \bibinfo{pages}{421--437}.
\newblock


\bibitem[Freyne et~al\mbox{.}(2007)]%
        {freyne2007collecting}
\bibfield{author}{\bibinfo{person}{Jill Freyne}, \bibinfo{person}{Rosta Farzan}, \bibinfo{person}{Peter Brusilovsky}, \bibinfo{person}{Barry Smyth}, {and} \bibinfo{person}{Maurice Coyle}.} \bibinfo{year}{2007}\natexlab{}.
\newblock \showarticletitle{Collecting community wisdom: integrating social search \& social navigation}. In \bibinfo{booktitle}{\emph{Proceedings of the 12th international conference on Intelligent user interfaces}}. \bibinfo{pages}{52--61}.
\newblock


\bibitem[Gero et~al\mbox{.}(2024)]%
        {gero2024supporting}
\bibfield{author}{\bibinfo{person}{Katy~Ilonka Gero}, \bibinfo{person}{Chelse Swoopes}, \bibinfo{person}{Ziwei Gu}, \bibinfo{person}{Jonathan~K Kummerfeld}, {and} \bibinfo{person}{Elena~L Glassman}.} \bibinfo{year}{2024}\natexlab{}.
\newblock \showarticletitle{Supporting Sensemaking of Large Language Model Outputs at Scale}. In \bibinfo{booktitle}{\emph{Proceedings of the CHI Conference on Human Factors in Computing Systems}}. \bibinfo{pages}{1--21}.
\newblock


\bibitem[Gordon et~al\mbox{.}(2021)]%
        {gordon2021disagreement}
\bibfield{author}{\bibinfo{person}{Mitchell~L Gordon}, \bibinfo{person}{Kaitlyn Zhou}, \bibinfo{person}{Kayur Patel}, \bibinfo{person}{Tatsunori Hashimoto}, {and} \bibinfo{person}{Michael~S Bernstein}.} \bibinfo{year}{2021}\natexlab{}.
\newblock \showarticletitle{The disagreement deconvolution: Bringing machine learning performance metrics in line with reality}. In \bibinfo{booktitle}{\emph{Proceedings of the 2021 CHI Conference on Human Factors in Computing Systems}}. \bibinfo{pages}{1--14}.
\newblock


\bibitem[Gray and Suri(2019)]%
        {gray2019ghost}
\bibfield{author}{\bibinfo{person}{Mary~L Gray} {and} \bibinfo{person}{Siddharth Suri}.} \bibinfo{year}{2019}\natexlab{}.
\newblock \bibinfo{booktitle}{\emph{Ghost work: How to stop Silicon Valley from building a new global underclass}}.
\newblock \bibinfo{publisher}{Eamon Dolan Books}.
\newblock


\bibitem[Groves et~al\mbox{.}(2023)]%
        {groves2023going}
\bibfield{author}{\bibinfo{person}{Lara Groves}, \bibinfo{person}{Aidan Peppin}, \bibinfo{person}{Andrew Strait}, {and} \bibinfo{person}{Jenny Brennan}.} \bibinfo{year}{2023}\natexlab{}.
\newblock \showarticletitle{Going public: the role of public participation approaches in commercial AI labs}. In \bibinfo{booktitle}{\emph{Proceedings of the 2023 ACM Conference on Fairness, Accountability, and Transparency}}. \bibinfo{pages}{1162--1173}.
\newblock


\bibitem[Hannak et~al\mbox{.}(2014)]%
        {hannak2014measuring}
\bibfield{author}{\bibinfo{person}{Aniko Hannak}, \bibinfo{person}{Gary Soeller}, \bibinfo{person}{David Lazer}, \bibinfo{person}{Alan Mislove}, {and} \bibinfo{person}{Christo Wilson}.} \bibinfo{year}{2014}\natexlab{}.
\newblock \showarticletitle{Measuring price discrimination and steering on e-commerce web sites}. In \bibinfo{booktitle}{\emph{Proceedings of the 2014 Conference on Internet Measurement Conference}}. \bibinfo{pages}{305--318}.
\newblock


\bibitem[Hara et~al\mbox{.}(2018)]%
        {hara2018data}
\bibfield{author}{\bibinfo{person}{Kotaro Hara}, \bibinfo{person}{Abigail Adams}, \bibinfo{person}{Kristy Milland}, \bibinfo{person}{Saiph Savage}, \bibinfo{person}{Chris Callison-Burch}, {and} \bibinfo{person}{Jeffrey~P Bigham}.} \bibinfo{year}{2018}\natexlab{}.
\newblock \showarticletitle{A data-driven analysis of workers' earnings on Amazon Mechanical Turk}. In \bibinfo{booktitle}{\emph{Proceedings of the 2018 CHI conference on human factors in computing systems}}. \bibinfo{pages}{1--14}.
\newblock


\bibitem[Holstein et~al\mbox{.}(2019)]%
        {holstein2019co}
\bibfield{author}{\bibinfo{person}{Kenneth Holstein}, \bibinfo{person}{Bruce~M McLaren}, {and} \bibinfo{person}{Vincent Aleven}.} \bibinfo{year}{2019}\natexlab{}.
\newblock \showarticletitle{Designing for complementarity: Teacher and student needs for orchestration support in AI-enhanced classrooms}.
\newblock  (\bibinfo{year}{2019}), \bibinfo{pages}{157--171}.
\newblock


\bibitem[Irani and Silberman(2013)]%
        {irani2013turkopticon}
\bibfield{author}{\bibinfo{person}{Lilly~C Irani} {and} \bibinfo{person}{M~Six Silberman}.} \bibinfo{year}{2013}\natexlab{}.
\newblock \showarticletitle{Turkopticon: Interrupting worker invisibility in amazon mechanical turk}. In \bibinfo{booktitle}{\emph{Proceedings of the SIGCHI conference on human factors in computing systems}}. \bibinfo{pages}{611--620}.
\newblock


\bibitem[Jiang et~al\mbox{.}(2023)]%
        {jiang2023graphologue}
\bibfield{author}{\bibinfo{person}{Peiling Jiang}, \bibinfo{person}{Jude Rayan}, \bibinfo{person}{Steven~P Dow}, {and} \bibinfo{person}{Haijun Xia}.} \bibinfo{year}{2023}\natexlab{}.
\newblock \showarticletitle{Graphologue: Exploring large language model responses with interactive diagrams}. In \bibinfo{booktitle}{\emph{Proceedings of the 36th Annual ACM Symposium on User Interface Software and Technology}}. \bibinfo{pages}{1--20}.
\newblock


\bibitem[Kaufmann et~al\mbox{.}(2011)]%
        {Buolamwini2019hearing}
\bibfield{author}{\bibinfo{person}{Nicolas Kaufmann}, \bibinfo{person}{Thimo Schulze}, {and} \bibinfo{person}{Daniel Veit}.} \bibinfo{year}{2011}\natexlab{}.
\newblock \showarticletitle{More than fun and money. worker motivation in crowdsourcing--a study on mechanical turk}.
\newblock  (\bibinfo{year}{2011}).
\newblock


\bibitem[Kiela et~al\mbox{.}(2021)]%
        {kiela2021dynabench}
\bibfield{author}{\bibinfo{person}{Douwe Kiela}, \bibinfo{person}{Max Bartolo}, \bibinfo{person}{Yixin Nie}, \bibinfo{person}{Divyansh Kaushik}, \bibinfo{person}{Atticus Geiger}, \bibinfo{person}{Zhengxuan Wu}, \bibinfo{person}{Bertie Vidgen}, \bibinfo{person}{Grusha Prasad}, \bibinfo{person}{Amanpreet Singh}, \bibinfo{person}{Pratik Ringshia}, {et~al\mbox{.}}} \bibinfo{year}{2021}\natexlab{}.
\newblock \showarticletitle{Dynabench: Rethinking benchmarking in NLP}.
\newblock \bibinfo{journal}{\emph{arXiv preprint arXiv:2104.14337}} (\bibinfo{year}{2021}).
\newblock


\bibitem[Kieslich et~al\mbox{.}(2024)]%
        {kieslich2024anticipating}
\bibfield{author}{\bibinfo{person}{Kimon Kieslich}, \bibinfo{person}{Nicholas Diakopoulos}, {and} \bibinfo{person}{Natali Helberger}.} \bibinfo{year}{2024}\natexlab{}.
\newblock \showarticletitle{Anticipating impacts: using large-scale scenario-writing to explore diverse implications of generative AI in the news environment}.
\newblock \bibinfo{journal}{\emph{AI and Ethics}} (\bibinfo{year}{2024}), \bibinfo{pages}{1--23}.
\newblock


\bibitem[Kingsley et~al\mbox{.}(2024)]%
        {kingsley2024investigating}
\bibfield{author}{\bibinfo{person}{Sara Kingsley}, \bibinfo{person}{Jiayin Zhi}, \bibinfo{person}{Wesley~Hanwen Deng}, \bibinfo{person}{Jaimie Lee}, \bibinfo{person}{Sizhe Zhang}, \bibinfo{person}{Motahhare Eslami}, \bibinfo{person}{Kenneth Holstein}, \bibinfo{person}{Jason~I Hong}, \bibinfo{person}{Tianshi Li}, {and} \bibinfo{person}{Hong Shen}.} \bibinfo{year}{2024}\natexlab{}.
\newblock \showarticletitle{Investigating What Factors Influence Users’ Rating of Harmful Algorithmic Bias and Discrimination}. In \bibinfo{booktitle}{\emph{Proceedings of the AAAI Conference on Human Computation and Crowdsourcing}}, Vol.~\bibinfo{volume}{12}. \bibinfo{pages}{75--85}.
\newblock


\bibitem[Kittur et~al\mbox{.}(2013)]%
        {kittur2013future}
\bibfield{author}{\bibinfo{person}{Aniket Kittur}, \bibinfo{person}{Jeffrey~V Nickerson}, \bibinfo{person}{Michael Bernstein}, \bibinfo{person}{Elizabeth Gerber}, \bibinfo{person}{Aaron Shaw}, \bibinfo{person}{John Zimmerman}, \bibinfo{person}{Matt Lease}, {and} \bibinfo{person}{John Horton}.} \bibinfo{year}{2013}\natexlab{}.
\newblock \showarticletitle{The future of crowd work}. In \bibinfo{booktitle}{\emph{Proceedings of the 2013 conference on Computer supported cooperative work}}. \bibinfo{pages}{1301--1318}.
\newblock


\bibitem[Kohn and Smith(2011)]%
        {kohn2011collaborative}
\bibfield{author}{\bibinfo{person}{Nicholas~W Kohn} {and} \bibinfo{person}{Steven~M Smith}.} \bibinfo{year}{2011}\natexlab{}.
\newblock \showarticletitle{Collaborative fixation: Effects of others' ideas on brainstorming}.
\newblock \bibinfo{journal}{\emph{Applied Cognitive Psychology}} \bibinfo{volume}{25}, \bibinfo{number}{3} (\bibinfo{year}{2011}), \bibinfo{pages}{359--371}.
\newblock


\bibitem[Kraut and Resnick(2011)]%
        {kraut2011encouraging}
\bibfield{author}{\bibinfo{person}{Robert~E Kraut} {and} \bibinfo{person}{Paul Resnick}.} \bibinfo{year}{2011}\natexlab{}.
\newblock \showarticletitle{Encouraging contribution to online communities}.
\newblock \bibinfo{journal}{\emph{Building successful online communities: Evidence-based social design}} (\bibinfo{year}{2011}), \bibinfo{pages}{21--76}.
\newblock


\bibitem[Kuo et~al\mbox{.}(2024)]%
        {kuo2024wikibench}
\bibfield{author}{\bibinfo{person}{Tzu-Sheng Kuo}, \bibinfo{person}{Aaron~Lee Halfaker}, \bibinfo{person}{Zirui Cheng}, \bibinfo{person}{Jiwoo Kim}, \bibinfo{person}{Meng-Hsin Wu}, \bibinfo{person}{Tongshuang Wu}, \bibinfo{person}{Kenneth Holstein}, {and} \bibinfo{person}{Haiyi Zhu}.} \bibinfo{year}{2024}\natexlab{}.
\newblock \showarticletitle{Wikibench: Community-driven data curation for AI evaluation on Wikipedia}. In \bibinfo{booktitle}{\emph{Proceedings of the CHI Conference on Human Factors in Computing Systems}}. \bibinfo{pages}{1--24}.
\newblock


\bibitem[Lam et~al\mbox{.}(2022)]%
        {lam2022enduser}
\bibfield{author}{\bibinfo{person}{Michelle~S. Lam}, \bibinfo{person}{Mitchell~L. Gordon}, \bibinfo{person}{Dana\"{e} Metaxa}, \bibinfo{person}{Jeffrey~T. Hancock}, \bibinfo{person}{James~A. Landay}, {and} \bibinfo{person}{Michael~S. Bernstein}.} \bibinfo{year}{2022}\natexlab{}.
\newblock \showarticletitle{End-User Audits: A System Empowering Communities to Lead Large-Scale Investigations of Harmful Algorithmic Behavior}.
\newblock \bibinfo{journal}{\emph{Proc. ACM Hum.-Comput. Interact.}} \bibinfo{volume}{6}, \bibinfo{number}{CSCW2}, Article \bibinfo{articleno}{512} (\bibinfo{date}{Nov} \bibinfo{year}{2022}), \bibinfo{numpages}{34}~pages.
\newblock
\urldef\tempurl%
\url{https://doi.org/10.1145/3555625}
\showDOI{\tempurl}


\bibitem[Lampe et~al\mbox{.}(2010)]%
        {lampe2010motivations}
\bibfield{author}{\bibinfo{person}{Cliff Lampe}, \bibinfo{person}{Rick Wash}, \bibinfo{person}{Alcides Velasquez}, {and} \bibinfo{person}{Elif Ozkaya}.} \bibinfo{year}{2010}\natexlab{}.
\newblock \showarticletitle{Motivations to participate in online communities}. In \bibinfo{booktitle}{\emph{Proceedings of the SIGCHI conference on Human factors in computing systems}}. \bibinfo{pages}{1927--1936}.
\newblock


\bibitem[Li et~al\mbox{.}(2023)]%
        {li2023participation}
\bibfield{author}{\bibinfo{person}{Rena Li}, \bibinfo{person}{Sara Kingsley}, \bibinfo{person}{Chelsea Fan}, \bibinfo{person}{Proteeti Sinha}, \bibinfo{person}{Nora Wai}, \bibinfo{person}{Jaimie Lee}, \bibinfo{person}{Hong Shen}, \bibinfo{person}{Motahhare Eslami}, {and} \bibinfo{person}{Jason Hong}.} \bibinfo{year}{2023}\natexlab{}.
\newblock \showarticletitle{Participation and Division of Labor in User-Driven Algorithm Audits: How Do Everyday Users Work together to Surface Algorithmic Harms?}. In \bibinfo{booktitle}{\emph{Proceedings of the 2023 CHI Conference on Human Factors in Computing Systems}}. \bibinfo{pages}{1--19}.
\newblock


\bibitem[Ling~Lo(2012)]%
        {ling2012variation}
\bibfield{author}{\bibinfo{person}{Mun Ling~Lo}.} \bibinfo{year}{2012}\natexlab{}.
\newblock \bibinfo{booktitle}{\emph{Variation theory and the improvement of teaching and learning}}.
\newblock \bibinfo{publisher}{G{\"o}teborg: Acta Universitatis Gothoburgensis}.
\newblock


\bibitem[Liu et~al\mbox{.}(2024)]%
        {liu2024selenite}
\bibfield{author}{\bibinfo{person}{Michael~Xieyang Liu}, \bibinfo{person}{Tongshuang Wu}, \bibinfo{person}{Tianying Chen}, \bibinfo{person}{Franklin~Mingzhe Li}, \bibinfo{person}{Aniket Kittur}, {and} \bibinfo{person}{Brad~A Myers}.} \bibinfo{year}{2024}\natexlab{}.
\newblock \showarticletitle{Selenite: Scaffolding Online Sensemaking with Comprehensive Overviews Elicited from Large Language Models}. In \bibinfo{booktitle}{\emph{Proceedings of the CHI Conference on Human Factors in Computing Systems}}. \bibinfo{pages}{1--26}.
\newblock


\bibitem[Longpre et~al\mbox{.}(2024)]%
        {longpre2024safe}
\bibfield{author}{\bibinfo{person}{Shayne Longpre}, \bibinfo{person}{Sayash Kapoor}, \bibinfo{person}{Kevin Klyman}, \bibinfo{person}{Ashwin Ramaswami}, \bibinfo{person}{Rishi Bommasani}, \bibinfo{person}{Borhane Blili-Hamelin}, \bibinfo{person}{Yangsibo Huang}, \bibinfo{person}{Aviya Skowron}, \bibinfo{person}{Zheng-Xin Yong}, \bibinfo{person}{Suhas Kotha}, {et~al\mbox{.}}} \bibinfo{year}{2024}\natexlab{}.
\newblock \showarticletitle{A safe harbor for ai evaluation and red teaming}.
\newblock \bibinfo{journal}{\emph{arXiv preprint arXiv:2403.04893}} (\bibinfo{year}{2024}).
\newblock


\bibitem[Luccioni et~al\mbox{.}(2023)]%
        {luccioni2023stable}
\bibfield{author}{\bibinfo{person}{Alexandra~Sasha Luccioni}, \bibinfo{person}{Christopher Akiki}, \bibinfo{person}{Margaret Mitchell}, {and} \bibinfo{person}{Yacine Jernite}.} \bibinfo{year}{2023}\natexlab{}.
\newblock \showarticletitle{Stable bias: Analyzing societal representations in diffusion models}.
\newblock \bibinfo{journal}{\emph{arXiv preprint arXiv:2303.11408}} (\bibinfo{year}{2023}).
\newblock


\bibitem[Luther et~al\mbox{.}(2015)]%
        {luther2015structuring}
\bibfield{author}{\bibinfo{person}{Kurt Luther}, \bibinfo{person}{Jari-Lee Tolentino}, \bibinfo{person}{Wei Wu}, \bibinfo{person}{Amy Pavel}, \bibinfo{person}{Brian~P Bailey}, \bibinfo{person}{Maneesh Agrawala}, \bibinfo{person}{Bj{\"o}rn Hartmann}, {and} \bibinfo{person}{Steven~P Dow}.} \bibinfo{year}{2015}\natexlab{}.
\newblock \showarticletitle{Structuring, aggregating, and evaluating crowdsourced design critique}. In \bibinfo{booktitle}{\emph{Proceedings of the 18th ACM conference on computer supported cooperative work \& social computing}}. \bibinfo{pages}{473--485}.
\newblock


\bibitem[Mack et~al\mbox{.}(2024)]%
        {mack2024they}
\bibfield{author}{\bibinfo{person}{Kelly~Avery Mack}, \bibinfo{person}{Rida Qadri}, \bibinfo{person}{Remi Denton}, \bibinfo{person}{Shaun~K Kane}, {and} \bibinfo{person}{Cynthia~L Bennett}.} \bibinfo{year}{2024}\natexlab{}.
\newblock \showarticletitle{“They only care to show us the wheelchair”: disability representation in text-to-image AI models}. In \bibinfo{booktitle}{\emph{Proceedings of the CHI Conference on Human Factors in Computing Systems}}. \bibinfo{pages}{1--23}.
\newblock


\bibitem[Madaio et~al\mbox{.}(2021)]%
        {madaio2021assessing}
\bibfield{author}{\bibinfo{person}{Michael Madaio}, \bibinfo{person}{Lisa Egede}, \bibinfo{person}{Hariharan Subramonyam}, \bibinfo{person}{Jennifer~Wortman Vaughan}, {and} \bibinfo{person}{Hanna Wallach}.} \bibinfo{year}{2021}\natexlab{}.
\newblock \showarticletitle{Assessing the Fairness of AI Systems: AI Practitioners' Processes, Challenges, and Needs for Support}.
\newblock \bibinfo{journal}{\emph{arXiv preprint arXiv:2112.05675}} (\bibinfo{year}{2021}).
\newblock


\bibitem[Madaio et~al\mbox{.}(2024)]%
        {madaio2024tinker}
\bibfield{author}{\bibinfo{person}{Michael~A Madaio}, \bibinfo{person}{Jingya Chen}, \bibinfo{person}{Hanna Wallach}, {and} \bibinfo{person}{Jennifer Wortman~Vaughan}.} \bibinfo{year}{2024}\natexlab{}.
\newblock \showarticletitle{Tinker, Tailor, Configure, Customize: The Articulation Work of Contextualizing an AI Fairness Checklist}.
\newblock \bibinfo{journal}{\emph{Proceedings of the ACM on Human-Computer Interaction}} \bibinfo{volume}{8}, \bibinfo{number}{CSCW1} (\bibinfo{year}{2024}), \bibinfo{pages}{1--20}.
\newblock


\bibitem[Madaio et~al\mbox{.}(2020)]%
        {madaio2020co}
\bibfield{author}{\bibinfo{person}{Michael~A Madaio}, \bibinfo{person}{Luke Stark}, \bibinfo{person}{Jennifer Wortman~Vaughan}, {and} \bibinfo{person}{Hanna Wallach}.} \bibinfo{year}{2020}\natexlab{}.
\newblock \showarticletitle{Co-designing checklists to understand organizational challenges and opportunities around fairness in ai}. In \bibinfo{booktitle}{\emph{Proceedings of the 2020 CHI Conference on Human Factors in Computing Systems}}. \bibinfo{pages}{1--14}.
\newblock


\bibitem[Maldaner et~al\mbox{.}(2024)]%
        {maldaner2024mirage}
\bibfield{author}{\bibinfo{person}{Matheus~Kunzler Maldaner}, \bibinfo{person}{Wesley~Hanwen Deng}, \bibinfo{person}{Jason Hong}, \bibinfo{person}{Ken Holstein}, {and} \bibinfo{person}{Motahhare Eslami}.} \bibinfo{year}{2024}\natexlab{}.
\newblock \showarticletitle{MIRAGE: Multi-model Interface for Reviewing and Auditing Generative Text-to-Image AI}.
\newblock \bibinfo{journal}{\emph{Demo of the AAAI Conference on Human Computation and Crowdsourcing}} (\bibinfo{year}{2024}).
\newblock


\bibitem[Metaxa et~al\mbox{.}(2021)]%
        {metaxa2021auditing}
\bibfield{author}{\bibinfo{person}{Dana{\"e} Metaxa}, \bibinfo{person}{Joon~Sung Park}, \bibinfo{person}{Ronald~E Robertson}, \bibinfo{person}{Karrie Karahalios}, \bibinfo{person}{Christo Wilson}, \bibinfo{person}{Jeff Hancock}, \bibinfo{person}{Christian Sandvig}, {et~al\mbox{.}}} \bibinfo{year}{2021}\natexlab{}.
\newblock \showarticletitle{Auditing algorithms: Understanding algorithmic systems from the outside in}.
\newblock \bibinfo{journal}{\emph{Foundations and Trends{\textregistered} in Human--Computer Interaction}} \bibinfo{volume}{14}, \bibinfo{number}{4} (\bibinfo{year}{2021}), \bibinfo{pages}{272--344}.
\newblock


\bibitem[Miceli and Posada(2021)]%
        {miceli2021wisdom}
\bibfield{author}{\bibinfo{person}{Milagros Miceli} {and} \bibinfo{person}{Julian Posada}.} \bibinfo{year}{2021}\natexlab{}.
\newblock \showarticletitle{Wisdom for the crowd: discoursive power in annotation instructions for computer vision}.
\newblock \bibinfo{journal}{\emph{arXiv preprint arXiv:2105.10990}} (\bibinfo{year}{2021}).
\newblock


\bibitem[Microsoft(2023)]%
        {mrbullwinkle_2023}
\bibfield{author}{\bibinfo{person}{Microsoft}.} \bibinfo{year}{2023}\natexlab{}.
\newblock \bibinfo{title}{Planning red teaming for large language models (LLMs) and their applications - Azure OpenAI Service}.
\newblock
\newblock
\urldef\tempurl%
\url{https://learn.microsoft.com/en-us/azure/ai-services/openai/concepts/red-teaming}
\showURL{%
\tempurl}


\bibitem[Mim et~al\mbox{.}(2024)]%
        {mim2024between}
\bibfield{author}{\bibinfo{person}{Nusrat~Jahan Mim}, \bibinfo{person}{Dipannita Nandi}, \bibinfo{person}{Sadaf~Sumyia Khan}, \bibinfo{person}{Arundhuti Dey}, {and} \bibinfo{person}{Syed~Ishtiaque Ahmed}.} \bibinfo{year}{2024}\natexlab{}.
\newblock \showarticletitle{In-Between Visuals and Visible: The Impacts of Text-to-Image Generative AI Tools on Digital Image-making Practices in the Global South}. In \bibinfo{booktitle}{\emph{Proceedings of the CHI Conference on Human Factors in Computing Systems}}. \bibinfo{pages}{1--18}.
\newblock


\bibitem[Morris and Horvitz(2007)]%
        {morris2007searchtogether}
\bibfield{author}{\bibinfo{person}{Meredith~Ringel Morris} {and} \bibinfo{person}{Eric Horvitz}.} \bibinfo{year}{2007}\natexlab{}.
\newblock \showarticletitle{SearchTogether: an interface for collaborative web search}. In \bibinfo{booktitle}{\emph{Proceedings of the 20th annual ACM symposium on User interface software and technology}}. \bibinfo{pages}{3--12}.
\newblock


\bibitem[Mun et~al\mbox{.}(2024)]%
        {mun2024particip}
\bibfield{author}{\bibinfo{person}{Jimin Mun}, \bibinfo{person}{Liwei Jiang}, \bibinfo{person}{Jenny Liang}, \bibinfo{person}{Inyoung Cheong}, \bibinfo{person}{Nicole DeCairo}, \bibinfo{person}{Yejin Choi}, \bibinfo{person}{Tadayoshi Kohno}, {and} \bibinfo{person}{Maarten Sap}.} \bibinfo{year}{2024}\natexlab{}.
\newblock \showarticletitle{Particip-ai: A democratic surveying framework for anticipating future ai use cases, harms and benefits}. In \bibinfo{booktitle}{\emph{Proceedings of the AAAI/ACM Conference on AI, Ethics, and Society}}, Vol.~\bibinfo{volume}{7}. \bibinfo{pages}{997--1010}.
\newblock


\bibitem[Naik and Nushi(2023)]%
        {naik2023social}
\bibfield{author}{\bibinfo{person}{Ranjita Naik} {and} \bibinfo{person}{Besmira Nushi}.} \bibinfo{year}{2023}\natexlab{}.
\newblock \showarticletitle{Social biases through the text-to-image generation lens}. In \bibinfo{booktitle}{\emph{Proceedings of the 2023 AAAI/ACM Conference on AI, Ethics, and Society}}. \bibinfo{pages}{786--808}.
\newblock


\bibitem[NIST(2023)]%
        {AIRiskManageF}
\bibfield{author}{\bibinfo{person}{NIST}.} \bibinfo{year}{2023}\natexlab{}.
\newblock \bibinfo{title}{Artificial Intelligence Risk Management Framework}.
\newblock
\newblock
\urldef\tempurl%
\url{https://www.nist.gov/itl/ai-risk-management-framework}
\showURL{%
\tempurl}


\bibitem[Noble(2018)]%
        {noble2018algorithms}
\bibfield{author}{\bibinfo{person}{Safiya~Umoja Noble}.} \bibinfo{year}{2018}\natexlab{}.
\newblock \bibinfo{booktitle}{\emph{Algorithms of oppression: How search engines reinforce racism}}.
\newblock \bibinfo{publisher}{NYU Press}.
\newblock


\bibitem[Nushi et~al\mbox{.}(2018)]%
        {nushi2018towards}
\bibfield{author}{\bibinfo{person}{Besmira Nushi}, \bibinfo{person}{Ece Kamar}, {and} \bibinfo{person}{Eric Horvitz}.} \bibinfo{year}{2018}\natexlab{}.
\newblock \showarticletitle{Towards accountable ai: Hybrid human-machine analyses for characterizing system failure}. In \bibinfo{booktitle}{\emph{Proceedings of the AAAI Conference on Human Computation and Crowdsourcing}}, Vol.~\bibinfo{volume}{6}. \bibinfo{pages}{126--135}.
\newblock


\bibitem[Ochigame and Ye(2021)]%
        {ochigame2021search}
\bibfield{author}{\bibinfo{person}{Rodrigo Ochigame} {and} \bibinfo{person}{Katherine Ye}.} \bibinfo{year}{2021}\natexlab{}.
\newblock \showarticletitle{Search Atlas: Visualizing Divergent Search Results Across Geopolitical Borders}. In \bibinfo{booktitle}{\emph{Designing Interactive Systems Conference 2021}}. \bibinfo{pages}{1970--1983}.
\newblock


\bibitem[Ojewale et~al\mbox{.}(2024)]%
        {ojewale2024towards}
\bibfield{author}{\bibinfo{person}{Victor Ojewale}, \bibinfo{person}{Ryan Steed}, \bibinfo{person}{Briana Vecchione}, \bibinfo{person}{Abeba Birhane}, {and} \bibinfo{person}{Inioluwa~Deborah Raji}.} \bibinfo{year}{2024}\natexlab{}.
\newblock \showarticletitle{Towards AI Accountability Infrastructure: Gaps and Opportunities in AI Audit Tooling}.
\newblock \bibinfo{journal}{\emph{arXiv preprint arXiv:2402.17861}} (\bibinfo{year}{2024}).
\newblock


\bibitem[OpenAI(2022a)]%
        {ChatGPT}
\bibfield{author}{\bibinfo{person}{OpenAI}.} \bibinfo{year}{2022}\natexlab{a}.
\newblock \bibinfo{title}{Open AI, ChatGPT}.
\newblock
\newblock
\urldef\tempurl%
\url{https://openai.com/blog/chatgpt/}
\showURL{%
\tempurl}


\bibitem[OpenAI(2022b)]%
        {openAI}
\bibfield{author}{\bibinfo{person}{OpenAI}.} \bibinfo{year}{2022}\natexlab{b}.
\newblock \bibinfo{title}{OpenAI: Our approach to alignment research}.
\newblock
\newblock
\urldef\tempurl%
\url{https://openai.com/blog/our-approach-to-alignment-research/}
\showURL{%
\tempurl}


\bibitem[OpenAI(2023)]%
        {openai2023gpt4card}
\bibfield{author}{\bibinfo{person}{OpenAI}.} \bibinfo{year}{2023}\natexlab{}.
\newblock
\newblock
\urldef\tempurl%
\url{https://cdn.openai.com/papers/gpt-4-system-card.pdf}
\showURL{%
\tempurl}


\bibitem[Pang and Reinecke(2023)]%
        {pang2023anticipating}
\bibfield{author}{\bibinfo{person}{Rock~Yuren Pang} {and} \bibinfo{person}{Katharina Reinecke}.} \bibinfo{year}{2023}\natexlab{}.
\newblock \showarticletitle{Anticipating Unintended Consequences of Technology Using Insights from Creativity Support Tools}.
\newblock \bibinfo{journal}{\emph{arXiv preprint arXiv:2304.05687}} (\bibinfo{year}{2023}).
\newblock


\bibitem[Pang et~al\mbox{.}(2024)]%
        {pang2024blip}
\bibfield{author}{\bibinfo{person}{Rock~Yuren Pang}, \bibinfo{person}{Sebastin Santy}, \bibinfo{person}{Ren{\'e} Just}, {and} \bibinfo{person}{Katharina Reinecke}.} \bibinfo{year}{2024}\natexlab{}.
\newblock \showarticletitle{BLIP: Facilitating the Exploration of Undesirable Consequences of Digital Technologies}. In \bibinfo{booktitle}{\emph{Proceedings of the CHI Conference on Human Factors in Computing Systems}}. \bibinfo{pages}{1--18}.
\newblock


\bibitem[Park et~al\mbox{.}(2022)]%
        {park2022social}
\bibfield{author}{\bibinfo{person}{Joon~Sung Park}, \bibinfo{person}{Lindsay Popowski}, \bibinfo{person}{Carrie Cai}, \bibinfo{person}{Meredith~Ringel Morris}, \bibinfo{person}{Percy Liang}, {and} \bibinfo{person}{Michael~S Bernstein}.} \bibinfo{year}{2022}\natexlab{}.
\newblock \showarticletitle{Social simulacra: Creating populated prototypes for social computing systems}. In \bibinfo{booktitle}{\emph{Proceedings of the 35th Annual ACM Symposium on User Interface Software and Technology}}. \bibinfo{pages}{1--18}.
\newblock


\bibitem[Park({[n.\,d.]})]%
        {park2024stakeholder}
\bibfield{author}{\bibinfo{person}{Tina Park}.} \bibinfo{year}{[n.\,d.]}\natexlab{}.
\newblock \bibinfo{title}{Stakeholder Engagement for Responsible AI: Introducing PAI’s Guidelines for Participatory and Inclusive AI}.
\newblock
\newblock
\urldef\tempurl%
\url{https://partnershiponai.org/stakeholder-engagement-for-responsible-ai-introducing-pais-guidelines-for-participatory-and-inclusive-ai/}
\showURL{%
\tempurl}


\bibitem[Passi and Jackson(2018)]%
        {passi2018trust}
\bibfield{author}{\bibinfo{person}{Samir Passi} {and} \bibinfo{person}{Steven~J Jackson}.} \bibinfo{year}{2018}\natexlab{}.
\newblock \showarticletitle{Trust in data science: Collaboration, translation, and accountability in corporate data science projects}.
\newblock \bibinfo{journal}{\emph{Proceedings of the ACM on Human-Computer Interaction}} \bibinfo{volume}{2}, \bibinfo{number}{CSCW} (\bibinfo{year}{2018}), \bibinfo{pages}{1--28}.
\newblock


\bibitem[Pierre et~al\mbox{.}(2021)]%
        {pierre2021getting}
\bibfield{author}{\bibinfo{person}{Jennifer Pierre}, \bibinfo{person}{Roderic Crooks}, \bibinfo{person}{Morgan Currie}, \bibinfo{person}{Britt Paris}, {and} \bibinfo{person}{Irene Pasquetto}.} \bibinfo{year}{2021}\natexlab{}.
\newblock \showarticletitle{Getting Ourselves Together: Data-centered participatory design research \& epistemic burden}. In \bibinfo{booktitle}{\emph{Proceedings of the 2021 CHI Conference on Human Factors in Computing Systems}}. \bibinfo{pages}{1--11}.
\newblock


\bibitem[Pirolli and Card(1999)]%
        {pirolli1999information}
\bibfield{author}{\bibinfo{person}{Peter Pirolli} {and} \bibinfo{person}{Stuart Card}.} \bibinfo{year}{1999}\natexlab{}.
\newblock \showarticletitle{Information foraging.}
\newblock \bibinfo{journal}{\emph{Psychological review}} \bibinfo{volume}{106}, \bibinfo{number}{4} (\bibinfo{year}{1999}), \bibinfo{pages}{643}.
\newblock


\bibitem[Pistilli(2022)]%
        {HuggingFace}
\bibfield{author}{\bibinfo{person}{Giada Pistilli}.} \bibinfo{year}{2022}\natexlab{}.
\newblock \bibinfo{title}{HuggingFace announcedthe new feature to flag any Model, Dataset, or Space on the Hub}.
\newblock
\newblock
\urldef\tempurl%
\url{https://twitter.com/GiadaPistilli/status/1571865167092396033?s=20&t=LRhhEu63s6ftPmtZdfz8Cw}
\showURL{%
\tempurl}


\bibitem[Prates et~al\mbox{.}(2020)]%
        {prates2020assessing}
\bibfield{author}{\bibinfo{person}{Marcelo~OR Prates}, \bibinfo{person}{Pedro~H Avelar}, {and} \bibinfo{person}{Lu{\'\i}s~C Lamb}.} \bibinfo{year}{2020}\natexlab{}.
\newblock \showarticletitle{Assessing gender bias in machine translation: a case study with google translate}.
\newblock \bibinfo{journal}{\emph{Neural Computing and Applications}} \bibinfo{volume}{32}, \bibinfo{number}{10} (\bibinfo{year}{2020}), \bibinfo{pages}{6363--6381}.
\newblock


\bibitem[Radharapu et~al\mbox{.}(2023)]%
        {radharapu2023aart}
\bibfield{author}{\bibinfo{person}{Bhaktipriya Radharapu}, \bibinfo{person}{Kevin Robinson}, \bibinfo{person}{Lora Aroyo}, {and} \bibinfo{person}{Preethi Lahoti}.} \bibinfo{year}{2023}\natexlab{}.
\newblock \showarticletitle{Aart: Ai-assisted red-teaming with diverse data generation for new llm-powered applications}.
\newblock \bibinfo{journal}{\emph{arXiv preprint arXiv:2311.08592}} (\bibinfo{year}{2023}).
\newblock


\bibitem[Raji et~al\mbox{.}(2020)]%
        {raji2020closing}
\bibfield{author}{\bibinfo{person}{Inioluwa~Deborah Raji}, \bibinfo{person}{Andrew Smart}, \bibinfo{person}{Rebecca~N White}, \bibinfo{person}{Margaret Mitchell}, \bibinfo{person}{Timnit Gebru}, \bibinfo{person}{Ben Hutchinson}, \bibinfo{person}{Jamila Smith-Loud}, \bibinfo{person}{Daniel Theron}, {and} \bibinfo{person}{Parker Barnes}.} \bibinfo{year}{2020}\natexlab{}.
\newblock \showarticletitle{Closing the AI accountability gap: Defining an end-to-end framework for internal algorithmic auditing}. In \bibinfo{booktitle}{\emph{Proceedings of the 2020 conference on fairness, accountability, and transparency}}. \bibinfo{pages}{33--44}.
\newblock


\bibitem[Rakova et~al\mbox{.}(2021)]%
        {rakova2021responsible}
\bibfield{author}{\bibinfo{person}{Bogdana Rakova}, \bibinfo{person}{Jingying Yang}, \bibinfo{person}{Henriette Cramer}, {and} \bibinfo{person}{Rumman Chowdhury}.} \bibinfo{year}{2021}\natexlab{}.
\newblock \showarticletitle{Where responsible AI meets reality: Practitioner perspectives on enablers for shifting organizational practices}.
\newblock \bibinfo{journal}{\emph{Proceedings of the ACM on Human-Computer Interaction}} \bibinfo{volume}{5}, \bibinfo{number}{CSCW1} (\bibinfo{year}{2021}), \bibinfo{pages}{1--23}.
\newblock


\bibitem[Rastogi et~al\mbox{.}(2023)]%
        {rastogi2023supporting}
\bibfield{author}{\bibinfo{person}{Charvi Rastogi}, \bibinfo{person}{Marco Tulio~Ribeiro}, \bibinfo{person}{Nicholas King}, \bibinfo{person}{Harsha Nori}, {and} \bibinfo{person}{Saleema Amershi}.} \bibinfo{year}{2023}\natexlab{}.
\newblock \showarticletitle{Supporting human-ai collaboration in auditing llms with llms}. In \bibinfo{booktitle}{\emph{Proceedings of the 2023 AAAI/ACM Conference on AI, Ethics, and Society}}. \bibinfo{pages}{913--926}.
\newblock


\bibitem[Reynante et~al\mbox{.}(2021)]%
        {reynante2021framework}
\bibfield{author}{\bibinfo{person}{Brandon Reynante}, \bibinfo{person}{Steven~P Dow}, {and} \bibinfo{person}{Narges Mahyar}.} \bibinfo{year}{2021}\natexlab{}.
\newblock \showarticletitle{A framework for open civic design: Integrating public participation, crowdsourcing, and design thinking}.
\newblock \bibinfo{journal}{\emph{Digital Government: Research and Practice}} \bibinfo{volume}{2}, \bibinfo{number}{4} (\bibinfo{year}{2021}), \bibinfo{pages}{1--22}.
\newblock


\bibitem[Salehi et~al\mbox{.}(2015)]%
        {salehi2015we}
\bibfield{author}{\bibinfo{person}{Niloufar Salehi}, \bibinfo{person}{Lilly~C Irani}, \bibinfo{person}{Michael~S Bernstein}, \bibinfo{person}{Ali Alkhatib}, \bibinfo{person}{Eva Ogbe}, {and} \bibinfo{person}{Kristy Milland}.} \bibinfo{year}{2015}\natexlab{}.
\newblock \showarticletitle{We are dynamo: Overcoming stalling and friction in collective action for crowd workers}. In \bibinfo{booktitle}{\emph{Proceedings of the 33rd annual ACM conference on human factors in computing systems}}. \bibinfo{pages}{1621--1630}.
\newblock


\bibitem[Sandvig et~al\mbox{.}(2014)]%
        {sandvig2014auditing}
\bibfield{author}{\bibinfo{person}{Christian Sandvig}, \bibinfo{person}{Kevin Hamilton}, \bibinfo{person}{Karrie Karahalios}, {and} \bibinfo{person}{Cedric Langbort}.} \bibinfo{year}{2014}\natexlab{}.
\newblock \showarticletitle{Auditing algorithms: Research methods for detecting discrimination on internet platforms}.
\newblock \bibinfo{journal}{\emph{Data and discrimination: converting critical concerns into productive inquiry}} \bibinfo{volume}{22}, \bibinfo{number}{2014} (\bibinfo{year}{2014}), \bibinfo{pages}{4349--4357}.
\newblock


\bibitem[Shelby et~al\mbox{.}(2023)]%
        {shelby2023sociotechnical}
\bibfield{author}{\bibinfo{person}{Renee Shelby}, \bibinfo{person}{Shalaleh Rismani}, \bibinfo{person}{Kathryn Henne}, \bibinfo{person}{AJung Moon}, \bibinfo{person}{Negar Rostamzadeh}, \bibinfo{person}{Paul Nicholas}, \bibinfo{person}{N'Mah Yilla-Akbari}, \bibinfo{person}{Jess Gallegos}, \bibinfo{person}{Andrew Smart}, \bibinfo{person}{Emilio Garcia}, {et~al\mbox{.}}} \bibinfo{year}{2023}\natexlab{}.
\newblock \showarticletitle{Sociotechnical harms of algorithmic systems: Scoping a taxonomy for harm reduction}. In \bibinfo{booktitle}{\emph{Proceedings of the 2023 AAAI/ACM Conference on AI, Ethics, and Society}}. \bibinfo{pages}{723--741}.
\newblock


\bibitem[Shelby et~al\mbox{.}(2024)]%
        {shelby2024generative}
\bibfield{author}{\bibinfo{person}{Renee Shelby}, \bibinfo{person}{Shalaleh Rismani}, {and} \bibinfo{person}{Negar Rostamzadeh}.} \bibinfo{year}{2024}\natexlab{}.
\newblock \showarticletitle{Generative AI in Creative Practice: ML-Artist Folk Theories of T2I Use, Harm, and Harm-Reduction}. In \bibinfo{booktitle}{\emph{Proceedings of the CHI Conference on Human Factors in Computing Systems}}. \bibinfo{pages}{1--17}.
\newblock


\bibitem[Shen et~al\mbox{.}(2021a)]%
        {shen2021value}
\bibfield{author}{\bibinfo{person}{Hong Shen}, \bibinfo{person}{Wesley~H Deng}, \bibinfo{person}{Aditi Chattopadhyay}, \bibinfo{person}{Zhiwei~Steven Wu}, \bibinfo{person}{Xu Wang}, {and} \bibinfo{person}{Haiyi Zhu}.} \bibinfo{year}{2021}\natexlab{a}.
\newblock \showarticletitle{Value cards: An educational toolkit for teaching social impacts of machine learning through deliberation}. In \bibinfo{booktitle}{\emph{Proceedings of the 2021 ACM conference on fairness, accountability, and transparency}}. \bibinfo{pages}{850--861}.
\newblock


\bibitem[Shen et~al\mbox{.}(2021b)]%
        {shen2021everyday}
\bibfield{author}{\bibinfo{person}{Hong Shen}, \bibinfo{person}{Alicia DeVos}, \bibinfo{person}{Motahhare Eslami}, {and} \bibinfo{person}{Kenneth Holstein}.} \bibinfo{year}{2021}\natexlab{b}.
\newblock \showarticletitle{Everyday algorithm auditing: Understanding the power of everyday users in surfacing harmful algorithmic behaviors}.
\newblock \bibinfo{journal}{\emph{Proceedings of the ACM on Human-Computer Interaction}} \bibinfo{volume}{5}, \bibinfo{number}{CSCW2} (\bibinfo{year}{2021}), \bibinfo{pages}{1--29}.
\newblock


\bibitem[Siangliulue et~al\mbox{.}(2015)]%
        {siangliulue2015toward}
\bibfield{author}{\bibinfo{person}{Pao Siangliulue}, \bibinfo{person}{Kenneth~C Arnold}, \bibinfo{person}{Krzysztof~Z Gajos}, {and} \bibinfo{person}{Steven~P Dow}.} \bibinfo{year}{2015}\natexlab{}.
\newblock \showarticletitle{Toward collaborative ideation at scale: Leveraging ideas from others to generate more creative and diverse ideas}. In \bibinfo{booktitle}{\emph{Proceedings of the 18th ACM Conference on Computer Supported Cooperative Work \& Social Computing}}. \bibinfo{pages}{937--945}.
\newblock


\bibitem[Sindlinger(2010)]%
        {Sindlinger2010CrowdsourcingWT}
\bibfield{author}{\bibinfo{person}{Ted~S. Sindlinger}.} \bibinfo{year}{2010}\natexlab{}.
\newblock \showarticletitle{Crowdsourcing: Why the Power of the Crowd is Driving the Future of Business}.
\newblock \bibinfo{journal}{\emph{American Journal of Health-system Pharmacy}}  \bibinfo{volume}{67} (\bibinfo{year}{2010}), \bibinfo{pages}{1565--1566}.
\newblock


\bibitem[Solyst et~al\mbox{.}(2023)]%
        {solyst2023potential}
\bibfield{author}{\bibinfo{person}{Jaemarie Solyst}, \bibinfo{person}{Ellia Yang}, \bibinfo{person}{Shixian Xie}, \bibinfo{person}{Amy Ogan}, \bibinfo{person}{Jessica Hammer}, {and} \bibinfo{person}{Motahhare Eslami}.} \bibinfo{year}{2023}\natexlab{}.
\newblock \showarticletitle{The Potential of Diverse Youth as Stakeholders in Identifying and Mitigating Algorithmic Bias for a Future of Fairer AI}.
\newblock \bibinfo{journal}{\emph{Proceedings of the ACM on Human-Computer Interaction}} \bibinfo{volume}{7}, \bibinfo{number}{CSCW2} (\bibinfo{year}{2023}), \bibinfo{pages}{1--27}.
\newblock


\bibitem[Steiger et~al\mbox{.}(2021)]%
        {steiger2021psychological}
\bibfield{author}{\bibinfo{person}{Miriah Steiger}, \bibinfo{person}{Timir~J Bharucha}, \bibinfo{person}{Sukrit Venkatagiri}, \bibinfo{person}{Martin~J Riedl}, {and} \bibinfo{person}{Matthew Lease}.} \bibinfo{year}{2021}\natexlab{}.
\newblock \showarticletitle{The psychological well-being of content moderators: the emotional labor of commercial moderation and avenues for improving support}. In \bibinfo{booktitle}{\emph{Proceedings of the 2021 CHI conference on human factors in computing systems}}. \bibinfo{pages}{1--14}.
\newblock


\bibitem[Sweeney(2013)]%
        {sweeney2013discrimination}
\bibfield{author}{\bibinfo{person}{Latanya Sweeney}.} \bibinfo{year}{2013}\natexlab{}.
\newblock \showarticletitle{Discrimination in online ad delivery}.
\newblock \bibinfo{journal}{\emph{Queue}} \bibinfo{volume}{11}, \bibinfo{number}{3} (\bibinfo{year}{2013}), \bibinfo{pages}{10--29}.
\newblock


\bibitem[{The White House}(2023)]%
        {House_2023}
\bibfield{author}{\bibinfo{person}{{The White House}}.} \bibinfo{year}{2023}\natexlab{}.
\newblock \bibinfo{title}{Executive Order on the Safe, Secure, and Trustworthy Development and Use of Artificial Intelligence}.
\newblock
\newblock
\urldef\tempurl%
\url{https://www.whitehouse.gov/briefing-room/presidential-actions/2023/10/30/executive-order-on-the-safe-secure-and-trustworthy-development-and-use-of-artificial-intelligence/}
\showURL{%
\tempurl}


\bibitem[Vaughan(2017)]%
        {vaughan2017making}
\bibfield{author}{\bibinfo{person}{Jennifer~Wortman Vaughan}.} \bibinfo{year}{2017}\natexlab{}.
\newblock \showarticletitle{Making Better Use of the Crowd: How Crowdsourcing Can Advance Machine Learning Research.}
\newblock \bibinfo{journal}{\emph{J. Mach. Learn. Res.}} \bibinfo{volume}{18}, \bibinfo{number}{1} (\bibinfo{year}{2017}), \bibinfo{pages}{7026--7071}.
\newblock


\bibitem[Vincent et~al\mbox{.}(2021)]%
        {vincent2021data}
\bibfield{author}{\bibinfo{person}{Nicholas Vincent}, \bibinfo{person}{Hanlin Li}, \bibinfo{person}{Nicole Tilly}, \bibinfo{person}{Stevie Chancellor}, {and} \bibinfo{person}{Brent Hecht}.} \bibinfo{year}{2021}\natexlab{}.
\newblock \showarticletitle{Data leverage: A framework for empowering the public in its relationship with technology companies}. In \bibinfo{booktitle}{\emph{Proceedings of the 2021 ACM Conference on Fairness, Accountability, and Transparency}}. \bibinfo{pages}{215--227}.
\newblock


\bibitem[Wang et~al\mbox{.}(2023)]%
        {wang2023designing}
\bibfield{author}{\bibinfo{person}{Qiaosi Wang}, \bibinfo{person}{Michael Madaio}, \bibinfo{person}{Shaun Kane}, \bibinfo{person}{Shivani Kapania}, \bibinfo{person}{Michael Terry}, {and} \bibinfo{person}{Lauren Wilcox}.} \bibinfo{year}{2023}\natexlab{}.
\newblock \showarticletitle{Designing responsible ai: Adaptations of ux practice to meet responsible ai challenges}. In \bibinfo{booktitle}{\emph{Proceedings of the 2023 CHI Conference on Human Factors in Computing Systems}}. \bibinfo{pages}{1--16}.
\newblock


\bibitem[Wang et~al\mbox{.}(2024)]%
        {wang2024farsight}
\bibfield{author}{\bibinfo{person}{Zijie~J. Wang}, \bibinfo{person}{Chinmay Kulkarni}, \bibinfo{person}{Lauren Wilcox}, \bibinfo{person}{Michael Terry}, {and} \bibinfo{person}{Michael Madaio}.} \bibinfo{year}{2024}\natexlab{}.
\newblock \showarticletitle{Farsight: Fostering Responsible AI Awareness During AI Application Prototyping}. In \bibinfo{booktitle}{\emph{Proceedings of the CHI Conference on Human Factors in Computing Systems}}. \bibinfo{pages}{1--40}.
\newblock


\bibitem[Warkentin and Woodward(2022)]%
        {AItest}
\bibfield{author}{\bibinfo{person}{Tris Warkentin} {and} \bibinfo{person}{Josh Woodward}.} \bibinfo{year}{2022}\natexlab{}.
\newblock \bibinfo{title}{AI Test Kitchen}.
\newblock
\newblock
\urldef\tempurl%
\url{https://blog.google/technology/ai/join-us-in-the-ai-test-kitchen/}
\showURL{%
\tempurl}


\bibitem[Weidinger et~al\mbox{.}(2024)]%
        {weidinger2024star}
\bibfield{author}{\bibinfo{person}{Laura Weidinger}, \bibinfo{person}{John Mellor}, \bibinfo{person}{Bernat~Guillen Pegueroles}, \bibinfo{person}{Nahema Marchal}, \bibinfo{person}{Ravin Kumar}, \bibinfo{person}{Kristian Lum}, \bibinfo{person}{Canfer Akbulut}, \bibinfo{person}{Mark Diaz}, \bibinfo{person}{Stevie Bergman}, \bibinfo{person}{Mikel Rodriguez}, {et~al\mbox{.}}} \bibinfo{year}{2024}\natexlab{}.
\newblock \showarticletitle{STAR: SocioTechnical Approach to Red Teaming Language Models}.
\newblock \bibinfo{journal}{\emph{arXiv preprint arXiv:2406.11757}} (\bibinfo{year}{2024}).
\newblock


\bibitem[Weidinger et~al\mbox{.}(2022)]%
        {weidinger2022taxonomy}
\bibfield{author}{\bibinfo{person}{Laura Weidinger}, \bibinfo{person}{Jonathan Uesato}, \bibinfo{person}{Maribeth Rauh}, \bibinfo{person}{Conor Griffin}, \bibinfo{person}{Po-Sen Huang}, \bibinfo{person}{John Mellor}, \bibinfo{person}{Amelia Glaese}, \bibinfo{person}{Myra Cheng}, \bibinfo{person}{Borja Balle}, \bibinfo{person}{Atoosa Kasirzadeh}, {et~al\mbox{.}}} \bibinfo{year}{2022}\natexlab{}.
\newblock \showarticletitle{Taxonomy of risks posed by language models}. In \bibinfo{booktitle}{\emph{Proceedings of the 2022 ACM Conference on Fairness, Accountability, and Transparency}}. \bibinfo{pages}{214--229}.
\newblock


\bibitem[Widder et~al\mbox{.}(2024)]%
        {widder2024power}
\bibfield{author}{\bibinfo{person}{David~Gray Widder}, \bibinfo{person}{Laura Dabbish}, \bibinfo{person}{James~D Herbsleb}, {and} \bibinfo{person}{Nikolas Martelaro}.} \bibinfo{year}{2024}\natexlab{}.
\newblock \showarticletitle{Power and Play: Investigating" License to Critique" in Teams' AI Ethics Discussions}.
\newblock \bibinfo{journal}{\emph{Proceedings of the ACM on Human-Computer Interaction}} \bibinfo{volume}{8}, \bibinfo{number}{CSCW2} (\bibinfo{year}{2024}), \bibinfo{pages}{1--23}.
\newblock


\bibitem[Wu et~al\mbox{.}(2019)]%
        {wu2019errudite}
\bibfield{author}{\bibinfo{person}{Tongshuang Wu}, \bibinfo{person}{Marco~Tulio Ribeiro}, \bibinfo{person}{Jeffrey Heer}, {and} \bibinfo{person}{Daniel~S Weld}.} \bibinfo{year}{2019}\natexlab{}.
\newblock \showarticletitle{Errudite: Scalable, reproducible, and testable error analysis}. In \bibinfo{booktitle}{\emph{Proceedings of the 57th Annual Meeting of the Association for Computational Linguistics}}. \bibinfo{pages}{747--763}.
\newblock


\bibitem[Young et~al\mbox{.}(2019)]%
        {young2019toward}
\bibfield{author}{\bibinfo{person}{Meg Young}, \bibinfo{person}{Lassana Magassa}, {and} \bibinfo{person}{Batya Friedman}.} \bibinfo{year}{2019}\natexlab{}.
\newblock \showarticletitle{Toward inclusive tech policy design: a method for underrepresented voices to strengthen tech policy documents}.
\newblock \bibinfo{journal}{\emph{Ethics and Information Technology}} \bibinfo{volume}{21}, \bibinfo{number}{2} (\bibinfo{year}{2019}), \bibinfo{pages}{89--103}.
\newblock


\bibitem[Zaydi and Maleh(2024)]%
        {zaydi2024empowering}
\bibfield{author}{\bibinfo{person}{Mounia Zaydi} {and} \bibinfo{person}{Yassine Maleh}.} \bibinfo{year}{2024}\natexlab{}.
\newblock \showarticletitle{EMPOWERING RED TEAMS WITH GENERATIVE AI: TRANSFORMING PENETRATION TESTING THROUGH ADAPTIVE INTELLIGENCE}.
\newblock \bibinfo{journal}{\emph{EDPACS}} (\bibinfo{year}{2024}), \bibinfo{pages}{1--26}.
\newblock


\bibitem[Zhang et~al\mbox{.}(2024)]%
        {zhang2024partiality}
\bibfield{author}{\bibinfo{person}{Lili Zhang}, \bibinfo{person}{Xi Liao}, \bibinfo{person}{Zaijia Yang}, \bibinfo{person}{Baihang Gao}, \bibinfo{person}{Chunjie Wang}, \bibinfo{person}{Qiuling Yang}, {and} \bibinfo{person}{Deshun Li}.} \bibinfo{year}{2024}\natexlab{}.
\newblock \showarticletitle{Partiality and Misconception: Investigating Cultural Representativeness in Text-to-Image Models}. In \bibinfo{booktitle}{\emph{Proceedings of the CHI Conference on Human Factors in Computing Systems}}. \bibinfo{pages}{1--25}.
\newblock


\end{thebibliography}

\appendix

\section{Appendix}

\subsection{Demographic Information of Participants} \label{Appendix: demo}

\begin{table}[h!]
\centering
\begin{tabular}{|c|c|c|c|c|}
\hline
\textbf{Participant} & \textbf{Age} & \textbf{Gender} & \textbf{Educational Level} & \textbf{Ethnicity} \\
\hline \hline
U01 & 22 & Female & Bachelor's degree & Asian \\ \hline 
U02 & 18 & Female & High school or lower & Asian \\ \hline 
U03 & 27 & Female & Bachelor's degree & White \\ \hline 
U04 & 37 & Female & Bachelor's degree & Hispanic \\ \hline 
U05 & 80 & Female & Doctoral degree (Ph.D.) & White \\ \hline 
U06 & 78 & Female & Some college, no degree & White \\ \hline 
U07 & 68 & Female & Bachelor's degree & N/A \\ \hline 
U08 & 50 & Man & Bachelor's degree & White \\ \hline 
U09 & 24 & Female & Associate's degree & N/A \\ \hline 
U10 & 46 & Female & Master's degree & White \\ \hline 
U11 & 72 & Female & Bachelor's degree & Caucasian \\
\hline \hline
\end{tabular}
\caption{Formative study participant Demographics. Among all participants, only U10 reported reported having previous experiences with text-to-image AI systems; other participants reported no experience with text-to-image AI.}
\label{tab:formative user participants}
\end{table}

\begin{table}[h!]
    \centering
    \begin{tabular}{p{4.0 cm} p{3.5 cm} p{2.5cm} p{3cm}}
        \toprule
        \textbf{Major / Area of Study} & \textbf{Race} & \textbf{Gender} & \textbf{Current Program} \\ 
        \midrule
     Computer Science (13) &  Asian (31) & Female (30) & Undergraduate (22) \\ 
     Engineering (11) &  White (9) & Male (13) & Master (18)  \\ 
     Design (10) &  African American (3) & Non-binary (2) & Ph.D. (5) \\ 
     Social Science (4) &  Prefer not to answer (2) &  & \\ 
     Public Policy (2) &   &  &  \\
     Philosophy (2) &   &  &  \\
     Math (2) &   &  & \\
     English (1) &  &  & \\

     \bottomrule
    \end{tabular}
    \caption{Overall demographics and background of our 45 user auditors from a U.S. university. Next to each demographic information, we include the number of the participants within that demographic group in parenthesis. Following our IRB protocol, we only collect minimum demographic data.}
    \label{tab: user auditors}
\end{table}

\begin{table*}[]
\centering
\resizebox{\textwidth}{!}{%
\begin{tabular}
    {
    | >{\centering\arraybackslash}p{0.05\linewidth}
    | >{\centering\arraybackslash}p{0.1\linewidth}
    | >{\centering\arraybackslash}p{0.4\linewidth}
    | >{\centering\arraybackslash}p{0.1\linewidth}
    | >{\centering\arraybackslash}p{0.1\linewidth}
    | >{\centering\arraybackslash}p{0.1\linewidth}
    |
    }
\hline
 &
  
  \textbf{Job title} &
  \textbf{Work experience relevant to our study} &
  \textbf{Years of experience} &
  \textbf{In Formative Study} &
  \textbf{In Evaluation Interview}\\ \hline
\hline
P01 & UX Researcher & Building a web-based application to engage users in flagging the potential biased and harmful behavior in a conversational agent & 3 - 10 & \cellcolor{green!30}Yes & \cellcolor{green!30}Yes \\
\hline
P02 & ML Engineer & Engaging end users in rating the risk of representation harms of LLM applications & 3 - 10 & \cellcolor{green!30}Yes & \cellcolor{gray!30}No \\
\hline
P03 & UX Researcher & Building an application to engage users in auditing potential problematic behaviors of LLM models that have not yet been incorporated into products & 3 - 10 & \cellcolor{green!30}Yes & \cellcolor{gray!30}No \\
\hline
P04 & UX Researcher & Organizing focus groups to engage end users in testing their AI products and services & 0 - 3 & \cellcolor{green!30}Yes & \cellcolor{gray!30}No \\
\hline
P05 & Product Lead & Building an internal crowdsourcing tool for GenAI auditing & 10+ & \cellcolor{green!30}Yes & \cellcolor{green!30}Yes \\
\hline
P06 & Technical Lead & Engaging users in auditing a range of AI products built by their customers & 10+ & \cellcolor{green!30}Yes & \cellcolor{gray!30}No \\
\hline
P07 & Data Scientist & Incorporating end users’ feedback in measuring the representational harms of LLM applications & 3 - 10 & \cellcolor{green!30}Yes & \cellcolor{green!30}Yes \\
\hline
P08 & Research Scientist & Human-centered evaluation for multimodal generative AI systems & 0 - 3 & \cellcolor{gray!30}No & \cellcolor{green!30}Yes \\
\hline
P09 & Research Engineer & Building an internal tool for engaging external domain experts in red-teaming generative AI & 0 - 3 & \cellcolor{gray!30}No & \cellcolor{green!30}Yes \\
\hline
P10 & Software Engineer & Building an internal evaluation pipeline for text-to-image generative AI & 3 - 10 & \cellcolor{gray!30}No & \cellcolor{green!30}Yes \\
\hline
P11 & Research Scientist & Training data curation for pre-trained large language and vision models & 0 - 3 & \cellcolor{gray!30}No & \cellcolor{green!30}Yes \\
\hline
P12 & Research Scientist & Human-centered evaluation for multimodal generative AI systems & 10+ & \cellcolor{gray!30}No & \cellcolor{green!30}Yes \\
\hline
P13 & Data Scientist & Red teaming large language and vision models & 3 - 10 & \cellcolor{gray!30}No & \cellcolor{green!30}Yes \\
\hline
P14 & Software Engineer & Fairness testing for large language and vision models & 0 - 3 & \cellcolor{gray!30}No & \cellcolor{green!30}Yes \\
\hline
\end{tabular}%
}

\Description[]{}
\caption{Summary of industry practitioner participants' backgrounds and relevant experience. All but one industry practitioner (P13, start up company) participants are from large technology companies.}
\label{tab:industry participants}
\end{table*}

  \subsection{Formative Study Intervention} 

\begin{figure*}[h]
  \centering
  \includegraphics[width=0.5\linewidth]{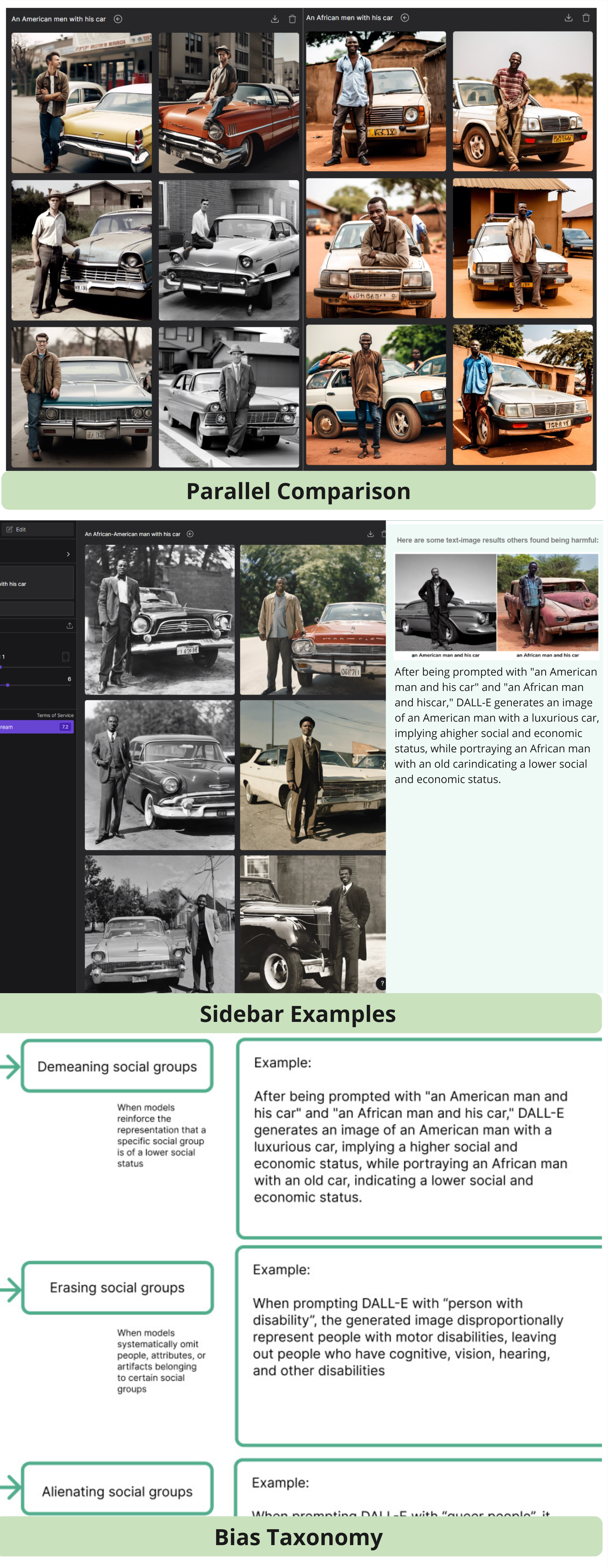}
  \caption{
   Screenshot of three low-fidelity prototypes we used in our formative study.
}
  \Description{TBA}
  \label{fig:formative prototypes}
\end{figure*}

\end{document}